\documentclass[%
 reprint,
superscriptaddress,
 amsmath,amssymb,
 aps,
]{revtex4-2}

\usepackage{graphicx}
\usepackage{dcolumn}
\usepackage{bm}

\usepackage{amsmath}
\usepackage{xspace} 
\usepackage{hyperref}
\newcommand{\harm}{\textsc{Harm3d}\xspace}
\newcommand{\harmnuc}{\textsc{HARM3D+NUC}\xspace}
\newcommand{\igm}{\textsc{IllinoisGRMHD}\xspace}

\newcommand{\handoff}{\textsc{HandOff}\xspace}
\newcommand{\BNSSRtz}{\texttt{BNS\_SmallRout\_t0}\xspace}
\newcommand{\BNSSRto}{\texttt{BNS\_SmallRout\_t1}\xspace}
\newcommand{\BNSLRto}{\texttt{BNS\_LargeRout\_t1}\xspace}
\newcommand{\Msun}{M_{\odot}}
\providecommand{\e}[1]{\ensuremath{\times 10^{#1}}}
\providecommand{\eqref}[1]{(\ref{#1})}

\begin{document}

\preprint{APS/123-QED}

\title{Handing off the outcome of binary neutron star mergers for accurate and long-term postmerger simulations}%

\author{Federico G. Lopez Armengol}
\affiliation{Center for Computational Relativity and Gravitation, 
Rochester Institute of Technology, 
Rochester, New York 14623, USA}

\author{Zachariah B. Etienne}
\affiliation{Department of Physics, University of Idaho, Moscow, ID 83843, USA}
\affiliation{Department of Physics and Astronomy, West Virginia University, Morgantown, WV 26506}
\affiliation{Center for Gravitational Waves and Cosmology, West Virginia University, Chestnut Ridge Research Building, Morgantown, WV 26505}

\author{Scott C. Noble}
\affiliation{Gravitational Astrophysics Lab, NASA Goddard Space Flight Center, Greenbelt, MD 20771, USA}

\author{Bernard J. Kelly}
\affiliation{Department of Physics, University of Maryland Baltimore County, 1000 Hilltop Circle Baltimore, MD 21250, USA}
\affiliation{Gravitational Astrophysics Lab, NASA Goddard Space Flight Center, Greenbelt, MD 20771, USA}
\affiliation{Center for Research and Exploration in Space Science and Technology, NASA Goddard Space Flight Center, Greenbelt, MD 20771, USA}

\author{Leonardo R. Werneck}
\affiliation{Department of Physics, University of Idaho, Moscow, ID 83843, USA}
\affiliation{Department of Physics and Astronomy, West Virginia University, Morgantown, WV 26506}
\affiliation{Center for Gravitational Waves and Cosmology, West Virginia University, Chestnut Ridge Research Building, Morgantown, WV 26505}

\author{Brendan Drachler}
\affiliation{Center for Computational Relativity and Gravitation, 
Rochester Institute of Technology, 
Rochester, New York 14623, USA}
\affiliation{School of Physics and Astronomy, Rochester Institute of Technology, Rochester, New York 14623, USA}

\author{Manuela Campanelli}
\affiliation{Center for Computational Relativity and Gravitation, 
Rochester Institute of Technology, 
Rochester, New York 14623, USA}
\affiliation{School of Mathematical Sciences, Rochester Institute of Technology, Rochester, New York 14623, USA}
\affiliation{School of Physics and Astronomy, Rochester Institute of Technology, Rochester, New York 14623, USA}

\author{Federico Cipolletta}
\affiliation{Center for Computational Relativity and Gravitation, 
Rochester Institute of Technology, 
Rochester, New York 14623, USA}
\affiliation{Leonardo Corporate LABS - via Raffaele Pieragostini 80, 16149 Genova GE - Italy}

\author{Yosef Zlochower}
\affiliation{Center for Computational Relativity and Gravitation, 
Rochester Institute of Technology, 
Rochester, New York 14623, USA}
\affiliation{School of Mathematical Sciences, Rochester Institute of Technology, Rochester, New York 14623, USA}
\affiliation{School of Physics and Astronomy, Rochester Institute of Technology, Rochester, New York 14623, USA}

\author{Ariadna~Murguia-Berthier}
\affiliation{Department of Astronomy and Astrophysics, University of California, Santa Cruz, CA 95064, USA}
\affiliation{Center for Interdisciplinary Exploration and Research in Astrophysics (CIERA), 1800 Sherman Ave., Evanston, IL 60201, USA}
\affiliation{NASA Einstein Fellow}

\author{Lorenzo Ennoggi}
\affiliation{Center for Computational Relativity and Gravitation, 
Rochester Institute of Technology, 
Rochester, New York 14623, USA}

\author{Mark Avara}
\affiliation{Center for Computational Relativity and Gravitation, 
Rochester Institute of Technology, 
Rochester, New York 14623, USA}

\author{Riccardo Ciolfi}
\affiliation{INAF, Osservatorio Astronomico di Padova, Vicolo dell'Osservatorio 5, I-35122 Padova, Italy}
\affiliation{INFN, Sezione di Padova, Via Francesco Marzolo 8, I-35131 Padova, Italy}

\author{Joshua Faber}
\affiliation{Center for Computational Relativity and Gravitation, 
Rochester Institute of Technology, 
Rochester, New York 14623, USA}
\affiliation{School of Mathematical Sciences, Rochester Institute of Technology, Rochester, New York 14623, USA}
\affiliation{School of Physics and Astronomy, Rochester Institute of Technology, Rochester, New York 14623, USA}

\author{Grace Fiacco}
\affiliation{Center for Computational Relativity and Gravitation, Rochester Institute of Technology, Rochester, New York 14623, USA}
\affiliation{School of Physics and Astronomy, Rochester Institute of Technology, Rochester, New York 14623, USA}
\affiliation{Department of Physics, Montana State University, Bozeman, Montana 59717, USA}

\author{Bruno Giacomazzo}
\affiliation{Universit\`{a} degli Studi di Milano - Bicocca, Dipartimento di Fisica G. Occhialini, Piazza della Scienza 3, I-20126 Milano, Italy}
\affiliation{INFN, Sezione di Milano-Bicocca, Piazza della Scienza 3, I-20126 Milano, Italy}
\affiliation{INAF, Osservatorio Astronomico di Brera, via E. Bianchi 46, I-23807 Merate (LC), Italy}

\author{Tanmayee Gupte}
\affiliation{Center for Computational Relativity and Gravitation, Rochester Institute of Technology, Rochester, New York 14623, USA}
\affiliation{School of Physics and Astronomy, Rochester Institute of Technology, Rochester, New York 14623, USA}

\author{Trung Ha}
\affiliation{Center for Computational Relativity and Gravitation, Rochester Institute of Technology, Rochester, New York 14623, USA}
\affiliation{Department of Physics, University of North Texas, Denton, Texas 76203, USA}

\author{Julian H. Krolik}
\affiliation{Physics and Astronomy Department, Johns Hopkins University, Baltimore, MD 21218, USA}

\author{Vassilios Mewes}
\affiliation{National Center for Computational Sciences, Oak Ridge National Laboratory, P.O. Box 2008, Oak Ridge, TN 37831-6164, USA}

\author{Richard O'Shaughnessy}
\affiliation{Center for Computational Relativity and Gravitation, 
Rochester Institute of Technology, 
Rochester, New York 14623, USA}
\affiliation{School of Mathematical Sciences, Rochester Institute of Technology, Rochester, New York 14623, USA}
\affiliation{School of Physics and Astronomy, Rochester Institute of Technology, Rochester, New York 14623, USA}

\author{Jes\'{u}s M. Rueda-Becerril}
\affiliation{Center for Computational Relativity and Gravitation, 
Rochester Institute of Technology, 
Rochester, New York 14623, USA}

\author{Jeremy Schnittman}
\affiliation{Gravitational Astrophysics Lab, NASA Goddard Space Flight Center, Greenbelt, MD 20771, USA}

\date{\today}

\begin{abstract}
We perform binary neutron star (BNS) merger simulations in full dynamical general relativity with \igm, on a Cartesian grid with adaptive-mesh refinement. 
After the remnant black hole has become nearly stationary, the evolution of the surrounding accretion disk on Cartesian grids over long  timescales (${\sim}$1s) is suboptimal, as Cartesian coordinates
over-resolve the angular coordinates at large
distances, and  the accreting plasma flows obliquely across coordinate lines
dissipating angular momentum artificially from the disk. 
To address this, we present the \handoff, a set of computational tools that enables the transfer of general relativistic magnetohydrodynamic (GRMHD) and spacetime data from \igm to \harm, a GRMHD code that specializes in modeling black hole accretion disks in static spacetimes over long timescales, making use of general coordinate systems with spherical topology. We demonstrate that the \handoff allows for a smooth and reliable transition of GRMHD fields and spacetime data, enabling us to efficiently and reliably evolve BNS dynamics well beyond merger. We also discuss future plans, which involve incorporating advanced equations of state and neutrino physics into BNS simulations using the \handoff approach.
\end{abstract}

\maketitle


\section{Introduction} \label{sec:intro}
    The detection of the gravitational wave (GW) GW170817 
	and its electromagnetic counterparts confirms
	longstanding theses
	about the outcomes of binary neutron star (BNS) mergers. 
	Although this event remains unique, the current high rate of GW detections suggests that similar events
	will follow \citep{LIGOScientific:2020ibl}.
	The signal GW170817 is compatible with a BNS merger 
	of total mass
	${\sim}2.73~\Msun$ and mass ratio $0.73-1.0$ 
	(low spins assumed), 
	in the galaxy NGC~4993 \citep{Coulter+2017} at a distance of  
	${\sim}40~\mathrm{Mpc}$ \citep{Abbott+2017a, Abbott+2019}.

	The observation of the coincident short 
	$\gamma$-ray burst (SGRB) GRB~170817A after $1.7s$ 
	from the merger \citep{Goldstein+2017} 
	supports earlier theoretical  
	connections between SGRB and BNS mergers 
	\citep{Eichler+1989, Narayan+1992, Lee+2007, Nakar2007, Berger2014}.
	In the standard picture, the merger remnant
	powers a mildly relativistic jet that drills 
	through the postmerger medium and generates a hot
	cocoon. The jet-cocoon system eventually breaks out
	from the dense ejecta and releases 
	$\gamma$-rays
	over wide angles
	\citep{Gottlieb2018b,Lazzati+2018, Bromberg+2018, Gottlieb+2018}.

	Kilonova emission detected in
	UV-optical-IR bands within the first hours and weeks after merger
	is consistent with radioactive heating from neutron-rich
        elements synthetized in the ejecta that decay
        either via $\beta$-decay, $\alpha$-decay, or spontaneous fission
	\citep{SoaresSantos+2017, Drout+2017, Cowperthwaite+2017, Nicholl+2017, Chornock+2017, Pian+2017}.
	This proves that BNS mergers are a propitious environment
	for the rapid neutron-capture process (r-process) and play
	an important role in the nucleosynthesis of the Universe 
	\citep{Li+1998, Metzger+2010, Roberts+2011, 
	Goriely+2011, Metzger+2012}.	
	More specifically, an early ``blue'' component in the
	kilonova spectrum, and a later ``red'' component, reveal
	multi-component ejecta, 
	the former being lanthanide-poor,
	with lower opacity, 
	plausibily ejected from the polar region of the remnant, 
	and the latter being lanthanide-rich, with higher opacity, 
	and tidally ejected around the equator of the system
	\citep[see, for instance, Ref. ][]{Kilpatrick+2017}.
	
	Moreover, later light curves in radio \citep{Hallinan+2017}
	and X-rays \citep{Troja+2017}
	rise together in time as
	${\sim}t^{0.8}$ \citep{Mooley+2018a},
	consistent with a single power-law of index $-0.6$ 
	for synchrotron radiation, emitted by 
	accelerated electrons in the shocked 
	interstellar medium (ISM)
	\citep{Margutti+2018}.
	The quick decline of these light curves 
	as ${\sim}t^{-2.2}$ after their peaks
	at 150 days after merger
	\citep{Troja+2018,Mooley+2018b},
	and the measurement of apparent superluminal motion with 
	very long-baseline interferometry 
	\citep{Mooley+2018c} 
	proves that this outflow is powered by
	an anisotropic and mildly relativistic outflow, 
	viewed off axis by $15^{\circ}-30^{\circ}$, 
	consistent with the picture of the  
	jet-cocoon breakout
	\citep[see also Refs.][]{Kasliwal+2017, Ghirlanda+2019}.

	Although this physical model is in 
	agreement with the available data, 
	many questions remain unanswered
	\citep[see the reviews ][]{Baiotti+2017,
	Shibata+2019, Ciolfi2020c, Sarin+2021, Dietrich+2021}.
	The equation of state (EOS) for the 
	neutron stars (NSs) has been 
	constrained by the GW signal and electromagnetic
	(EM) counterparts 
	\citep{Abbott+2019, Radice+2018a},
	but remains degenerate
	\citep[see][and references therein]{Raithel+2019}.
	While the remnant compact object seems to have collapsed 
	to a black hole (BH), the time of collapse remains
	uncertain, and the central engine for the jet
	could be both a rotating BH 
	or a \textit{long-lived} hypermassive neutron
	star (HMNS)  \citep[see][]{Margalit+2017,
	Ciolfi2018, MurguiaBerthier+2021a}.
	Furthermore, the dependence of the total mass, 
	composition, and magnetization
	of the ejecta on the 
	properties of the binary,
	remains to be fully understood 
	\citep[see][]{Hotokezaka+2013, Radice+2018b}.

	Numerical simulations 
	are key to answering these questions but,
	despite the remarkable progress in the last 
	decades \citep{Davies1994,Rosswog1999,Oohara+1999, Shibata+2000, Shibata+2003, 
	Anderson+2008, Baiotti+2008, Liu+2008, Giacomazzo+2009,
	Giacomazzo+2011, Rezzolla+2011, Palenzuela+2013,
	Foucart+2015,
	Kiuchi+2014, Kiuchi+2015, Endrizzi+2016, Ruiz+2016,
	Fernandez+2017,
	Kawamura+2016, Ciolfi+2017, Ruiz+2017, Ciolfi+2019,
	Ruiz+2019, Tsokaros+2019, Ciolfi2020a, Ruiz+2020, Ciolfi2020b},
	there are still computational limitations to address.
	The key ingredients for a realistic simulation
	of a BNS merger and postmerger 
	are, at least, \textit{Numerical Relativity} (NR)
	for the evolution of the spacetime metric during the
	inspiral and merger, \textit{General-Relativistic Magnetohydrodynamics}
	(GRMHD) for the evolution of the matter fields,
	realistic EOS as tabulated from nuclear interactions,
	and consistent emission, absorption and transport of neutrinos
	\citep[see the reviews ][]{Baiotti+2017, Shibata+2019, Ciolfi2020b}.
	However, numerical codes that take NR into account 
	usually make use of Cartesian coordinates 
	in a hierarchy of inset boxes with different resolutions 
	(Adaptive Mesh Refinement, AMR)
	and adopt a finest resolution of ${\sim}200~\mathrm{m}$ \footnote{Some exceptions being \cite{Kiuchi+2014, Kiuchi+2015,
	Kiuchi+2018, Mosta+2020}.}.
	Such a grid topology and resolution are insufficient to 
	resolve in detail the length scales of the relevant mechanisms
	of the postmerger, like the magneto-rotational 
	instability \citep[MRI; see Ref. ][]{Kiuchi+2018}.
	Moreover, since Cartesian coordinates over-resolve the angles 
	at large distances, the outer boundary of such domains cannot
	be placed far enough without making the simulation
	prohibitively expensive. For this reason, 
	outer boundaries are usually set at 
	${\sim}5000~\mathrm{km}$, preventing long-term simulations
	(${\sim}1~\mathrm{s}$)
	that can follow the propagation of outflows
	\citep[see Refs. ][]{Hotokezaka+2013, Ciolfi+2020}.
	Another drawback of such grid structure
	is that the approximate symmetries of the system change after
	merger and Cartesian coordinates with AMR introduce numerical
	dissipation in the postmerger disk that can dominate the
	long-term accretion.
	We will discuss about this issue later in
	the manuscript.
	See Ref. \cite{Ciolfi2018} for further comments on 
	current computational limitations, 
	and see Ref. \cite{Mewes+2020} for NR simulations
	in spherical coordinates.

	In this work we solve these computational problems 
	by transitioning
	a BNS postmerger simulation 
	from the code \igm~\citep{Etienne+2015} that uses Cartesian AMR grids, to
	the code \harm~\citep{Noble+2009, MurguiaBerthier+2021b} that adopts a grid adapted
	to the requirements of the postmerger.
	\harm's grid uses spherical-like coordinates for better conservation of angular momentum; the grid
	has higher resolution in the polar coordinate 
	towards the equator
	if close to the BH, to resolve the disk, but
	higher resolution towards the 
	polar axis if farther
	away, to resolve the jet-cocoon system. Further its outer boundary is far enough (${\sim}10^5~\mathrm{km}$)
	to include the region of the 
	jet breakout.
    Finally, we use novel boundary conditions
    at the polar axis that 
	allow us to fully resolve
	the funnel region.
	Transitioning between \igm
	and \harm
	is possible because they
	rest on the same formalism for describing the GRMHD fields. However the numerical infrastructures of these
	codes are very different. 
	Our new code \handoff consists of the set of
	routines needed to 
	translate the state of a 
	BNS postmerger
	from \igm to \harm, and
	the description and validation of 
	this package entails the main purpose
	of this work.

	Our work is organized as follows.
	In Sec.~\ref{sec:methods} we describe the formalism
	and numerical methods adopted by 
	the numerical codes \igm and \harm, and we
	describe in detail the set of routines that
	compose the \handoff. 
	In Sec.~\ref{sec:validation} we validate the 
	\handoff in the well-known system of a magnetized
	torus around a rotating BH.
	Then in Sec.~\ref{sec:results} we apply
	the \handoff to a BNS postmerger, and show
	the continued evolution in \harm gives the
	results expected.
	Finally, in Sec.~\ref{sec:conclusions} we
	present final remarks and our future work.

\section{The \handoff package} \label{sec:methods}

    \begin{figure}[htb!]
    	\centering
    	\includegraphics[width=.8\columnwidth]{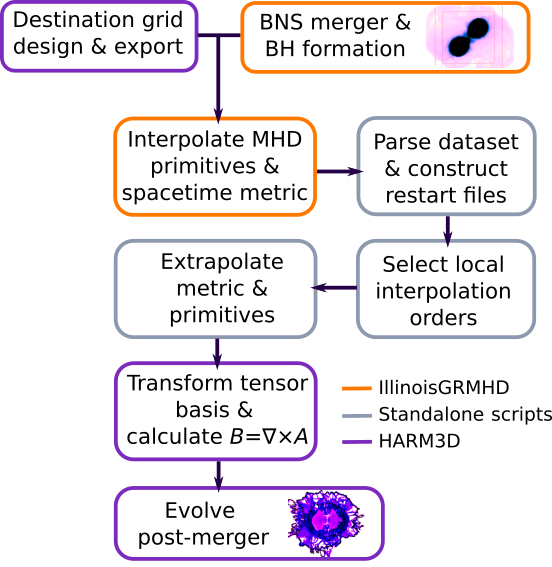}  
    	\caption{\handoff package workflow. The colors of the borders shows the code
    	responsible for the task, with \textit{orange} for \igm, \textit{gray}
    	for standalone scripts, and \textit{purple} for \harm.}
    	\label{f:sketch}
    \end{figure}

    We first describe the GRMHD formalism adopted by
    both \igm and \harm, as well as the steps performed by the
    \handoff package that allow us to transition a BNS postmerger 
    simulation from one code to the other.
    In Fig.~\ref{f:sketch} we show a sketch of the 
    workflow around the \handoff:
    We perform a BNS merger simulation until BH formation
    with \igm. At the same time,
    we design and export the numerical grid that we will use
    to continue the postmerger evolution in \harm.
    We interpolate the MHD primitives and spacetime metric
    of the postmerger onto this grid, 
    and use these results to construct restart files
    readable by \harm.
    Then we select results from lower or higher orders of 
    interpolation, depending on the local smoothness
    of each grid function.
    If the destination grid is larger than the original grid
    of \igm, then we extrapolate the primitives and spacetime metric
    to populate the complementary cells.
    Finally, we transform the tensor basis from Cartesian coordinates
    to the new coordinate system, calculate the magnetic field
    from the curl of the magnetic vector potential,
    and continue the postmerger evolution in the new grid with
    the usual methods of \harm.

    \subsection{GRMHD formalism}
    The GRMHD formalisms on which \igm and \harm
    are based are equivalent, but the
    formulation and conventions adopted by each code differ.
    Below we present the equations of motion as implemented in \harm, 
    and we will describe the differences with \igm when presenting the
    methods adopted for the BNS merger.
    
	The evolution of the MHD fields follows from the integration 
	of the general relativistic equations of motion for
	a perfect fluid with infinite conductivity (ideal MHD).
	These are the continuity equation, the 
	local conservation of energy 
	and momentum, and Maxwell's equations
	\cite[see, for instance, Refs. ][]{Noble+2009, Etienne+2015, Cipolletta+2020, MurguiaBerthier+2021b}. 
	In flux-conservative form, they read:

        \begin{equation}
        	\label{consEqs}
                \partial_t \mathbf{U}(\mathbf{P}) = - \partial_i \mathbf{F}^{i} + \mathbf{S} (\mathbf{P}) \,,
        \end{equation}
        where $\mathbf{P}$ is the vector of \textit{primitive} variables,
        $\mathbf{U}$ the vector of \textit{conserved} variables,
        $\mathbf{F}$ the \textit{fluxes}, 
        and $\mathbf{S}$ the \textit{sources}:
        \begin{equation}
        	\mathbf{P} = \left[\rho, p, \tilde{v}^k, B^k\right]^{\mathrm{T}},
        \end{equation}
        \begin{equation}
        	\mathbf{U}(\mathbf{P}) = \sqrt{-g} \left[\rho u^t, {T^t}_t + \rho u^t, {T^t}_j, B^k\right]^{\mathrm{T}},
        \end{equation}
        \begin{equation}
        	\mathbf{F}^i(\mathbf{P}) = \sqrt{-g} \left[ \rho u^i, {T^i}_t + \rho u^{i},\right. \left.{T^i}_j, \left(b^i u^k - b^k u^i \right) \right]^{\mathrm{T}},
        \end{equation}
        \begin{equation}
        \label{sourceFuncs}
                \mathbf{S}(\mathbf{P}) = \sqrt{-g} \left[ 0, {T^{\kappa}}_{\lambda} {\Gamma^{\lambda}}_{t\kappa}, \right. \left. {T^{\kappa}}_{\lambda} {\Gamma^{\lambda}}_{j\kappa}, 0^k \right]^{\mathrm{T}},
        \end{equation}
        where $g$ denotes the determinant of the metric, 
        $\rho$ is the rest mass density,
        $p$ is the fluid pressure,
        $u^{\mu}$ is the fluid four-velocity, 
        and $\tilde{v}^{k}$ is the fluid velocity as measured 
    	by normal observers with four-velocity \mbox{$n_{\mu} = (-\alpha,\vec{0})$}, $\alpha=\sqrt{-1/g^{tt}}$,
    	with $g^{\mu \nu}$ ($g_{\mu \nu}$) the contravariant (covariant) components of the
    	spacetime metric.
        The magnetic field is represented by 
        $B^k ={}^*F^{kt}$, where
    	${}^*F^{\mu \nu}$ is the dual of the Maxwell tensor times $1/\sqrt{4\pi}$,
        $b^{\mu} = \left(\delta^{\mu}_{\nu} + u^{\mu} u_{\nu} \right) B^{\nu}/u^t$
        is the projection of the magnetic field into the fluid's 
        comoving frame.  In addition, ${\Gamma^{\lambda}}_{\mu \nu}$
        is the affine connection and  ${T^{\mu}}_{\nu}$ is the sum of the
        stress-energy tensor of a perfect fluid and the EM 
        stress energy tensor, defined as:
        \begin{equation}
                T_{\mu \nu} = \left(\rho h + 2 {p_{\mathrm{m}}} \right) u_{\mu} u_{\nu} + \left(p + {p_{\mathrm{m}}} \right) g_{\mu \nu} - b_{\mu} b_{\nu} \,,
        \end{equation}
        where 
        $h = 1 + \epsilon + p / \rho$ denotes
        the specific enthalpy,
        $\epsilon$ the specific internal energy, 
        and ${p_{\mathrm{m}}} = b^{\mu} b_{\mu} / 2$ the magnetic pressure.
        The internal energy is $u=\rho \epsilon$, and
        we assume an adiabatic $\Gamma$-law equation of state:
        $p=(\Gamma - 1) u$.
        
        The \handoff package can be extended to deal
        with more general formalisms.
        For instance, if using a 
        realistic and 
        finite-temperature equation of state, 
        we can also include 
        the electron fraction and temperature of 
        the plasma. 
        We will describe such extensions to the
        \handoff package in an forthcoming 
        article, where we will transition a simulation from \igm to the numerical code \harmnuc \citep{MurguiaBerthier+2021b}.
    
	\subsection{BNS merger}
	\label{ss:merger}

	We simulate the BNS merger with the numerical code
	\igm \citep{Etienne+2015}. 
	This code is implemented as a set of modules, or ``thorns'', built upon the
	\texttt{Cactus} \footnote{See \url{http://www.cactuscode.org/}.} and \texttt{Carpet} infrastructure 
     \citep{Goodale+2002, Schnetter+2003}, within the \texttt{Einstein Toolkit} 
	framework \citep{Loffler:2011ay,Haas+2020}. \texttt{Carpet} enables \igm to sample the physical fields on Cartesian AMR numerical grids.

	The integration in time of the conservation equations
	for the primitives $\rho, \ p,$ and $\tilde{v}^k$ 
	follows from high-resolution shock-capturing (HRSC) schemes.
	To be precise, \igm adopts the coordinate velocity $v^k = u^k/u^0$ as 
	a primitive instead of $\tilde{v}^k$.
	HRSC schemes, in a nutshell, reconstruct the	
	primitive variables at the cell 
	interfaces, solve for the fluxes with an approximate Riemann solver, 
	and integrate in time with the Method of Lines (MoL).
	The standard methods adopted are Piecewise Parabolic 
	Method (PPM) \citep{Colella+1984} for the reconstruction of primitives, 
	Harten-Lax-van Leer (HLL)
	for the approximate Riemann solver,
	and fourth-order Runge-Kutta (RK4) for the MoL
	integration.

	The integration of the induction equation for $B^k$
	requires further care because the propagation of truncation errors can violate
	the solenoidal constraint $\partial_i \left( \sqrt{-g} B^i \right)=0$.
	Constrained-transport schemes for $B^k$ have proven to avoid these violations
	by finite-differencing the derivatives in 
	the induction equation with specific stencils 
	\citep[see, for instance, ][]{Toth2000},
	but their scope is limited to uniform grids, unless special steps or conditions are implemented.
	\igm instead evolves the vector potential $A_k$ directly at staggered locations on a given cell, ensuring that results match those adopting a standard constrained-transport algorithm \citep{Evans+1988}.
	The algorithm is summarized as follows. The induction equation is recast as an evolution equation for $A_k$,
	which is integrated with HRSC methods, 
	and then $B^k$ is 
	obtained from the curl of $A_k$, satisfying the solenoidal constraint
	to roundoff error \footnote{The definition of the magnetic field 
	primitives in \igm
	differs from the definition introduced previously in this manuscript.
	Specifically $B^k = B^k_{\mathrm{IGM}} / (\alpha \sqrt{4 \pi})$.}.
	
    After each time step, the primitive variables need to be recovered from the conserved variables in the conservative-to-primitive step. \igm adopts a Newton-Raphson-based 2D recovery scheme, which is also available in \harm~\citep{Noble+2006}. This step can fail, however, if the conserved variables become invalid during the evolution. This happens most often in regions where high accuracy is difficult to maintain: the low density ``atmosphere'', inside the BH horizon, or at AMR refinement boundaries. \igm enforces MHD inequalities that the conserved variables must satisfy in order to mitigate the number of recovery failures. If a failure still occurs, a backup primitive recovery method is used that is guaranteed to succeed, as described in Appendix~A of \cite{Etienne:2011ea}.
	
	The conservative-to-primitive step is prone to failure
	for lower densities of the artificial atmosphere.
    A typical value that results in robust, stable evolutions with small numbers of recovery failures is \mbox{$\rho^{\mathrm{Cactus}}_{\mathrm{atm}}\sim10^{-8}\rho_{\mathrm{max}}$}, where $\rho_{\mathrm{max}}$ is the maximum initial baryonic density. The minimum allowed value for the conserved energy used by \igm,
	\begin{equation}
	    \tau \equiv \sqrt{-g}\left(\alpha T^{tt} - u^{t}\rho\right),
	\end{equation}
    is obtained by assuming $\rho=\rho^{\mathrm{Cactus}}_{\rm atm}$, flat space, zero velocities, and zero magnetic fields, yielding
    \begin{equation}
        \tau_{\rm min} = \epsilon^{\mathrm{Cactus}}_{\rm atm}\rho^{\mathrm{Cactus}}_{\rm atm},
        \label{eq:taumindef}
    \end{equation}
    where the ``atmosphere'' energy $\epsilon^{\mathrm{Cactus}}_{\rm atm} = \epsilon(\rho_{\rm atm})$ is computed using the gamma-law EOS with $\Gamma=2$ and $K = 0.0332 \rho_{\rm nuc} c^{2}/n_{\rm nuc}^{\Gamma} = 123.6$ (in code units), where $\rho_{\rm nuc}$ and $n_{\rm nuc}$ are the nuclear rest-mass and number densities, respectively. Empirically, we have determined that choosing an ``atmosphere" value
    $\tau^{\mathrm{Cactus}}_{\rm atm}$ up to ${\sim}10^{2} \tau_{\rm min}$ (see  Eq.~\eqref{eq:taumindef})
    leads to stable evolutions and accurate data transfers to \harm, while choosing values a few orders of magnitude larger can lead to an unstable transition.

	During the BNS merger, the spacetime is highly dynamical and, besides integrating 
	the equations of motion for the plasma, we need 
        to integrate the equations of motion for the metric components $g_{\mu \nu}$.
	These equations are the Einstein field equations, written in
	the standard 3+1 Arnowitt-Deser-Misner (ADM) formalism \citep{Arnowitt+1962},
	in the BSSNOK formulation \citep{Nakamura+1987, Shibata+1995, Baumgarte+1998}.
	In order to integrate these equations numerically, 
	we use the thorn 
	\texttt{McLachlan} \footnote{See \url{https://www.cct.lsu.edu/~eschnett/McLachlan/}.}
	\citep{Brown+2008}, which has been written with the 
	\texttt{Kranc} \footnote{See \url{http://kranccode.org/}.} application.

	\subsection{Grid design and export} 
	Numerical errors rapidly sap angular momentum from fluid flows that obliquely cross coordinate lines. Thus continuing to model the dynamics of a postmerger BH accretion disk over timescales far longer than the inspiral/merger timescale with moderate-resolution Cartesian AMR grids would be ill advised. Thus we choose a post-handoff grid that samples the remnant accretion disk in a spherical coordinate system designed specifically for modeling black hole accretion. The post-handoff grid is implemented within the same code adopted for post-handoff evolution: \harm.
 	
	To begin the handoff, \harm specifies the Cartesian 
	coordinate locations
	of all cell centers, lower faces, and corners, for both physical and 
	ghost cells on its spherical grid.
	The number of coordinates dumped, then, is 
	$\left(N^r + 2 N_{\mathrm{G}} \right) \times \left(N^{\theta} + 2 N_{\mathrm{G}} \right)
	\times \left(N^{\phi} + 2 N_{\mathrm{G}} \right) \times 3 \times 5$, 
	where $N^r$, $N^{\theta}$, and $N^{\phi}$ are the number of cells in each dimension, 
	$N_{\mathrm{G}}$ is the number of ghost cells beyond each 
	coordinate boundary, 
	$\times 3$ stands for
	each spatial dimension, and $\times 5$ for each cell location.
	
	In case the numerical codes assume different values for
	the unity of mass in geometrical units, we normalize 
	the exported coordinates. For instance, 
	if the unit of length of the destination grid equals the mass of the black hole remnant $M_{\mathrm{BH}}$, but the unit
	of length during the merger equals $\Msun$, then we multiply
	the exported coordinates of the destination grid by $M_{\mathrm{BH}} / M_{\odot}$.

	\subsection{Interpolation routine}
	In order to restart a GRMHD simulation in \harm we need
	two sets of grid functions at a given time:
	the MHD primitives $\mathbf{P}_{\mathrm{HandOff}}$, and the geometry 
	of spacetime $\mathbf{g}_{\mathrm{HandOff}}$.
 	We have implemented a thorn \footnote{Publicly available at 
	\url{https://github.com/zachetienne/nrpytutorial/blob/master/Tutorial-ETK_thorn-Interpolation_to_Arbitrary_Grids_multi_order.ipynb}} within the \textsc{Cactus} infrastructure
	that reads in the Cartesian coordinates of the cell positions of the 
	destination grid
	and proceeds to interpolate both sets of grid functions into them. 

	Specifically, regarding the first set $\mathbf{P}_{\mathrm{HandOff}}$, 
	we interpolate the following grid functions:
	\begin{equation}
		\mathbf{P}_{\mathrm{HandOff}} = \left\{ \rho, p,
		\tilde{v}^{r}, \tilde{v}^{\theta}, \tilde{v}^{\phi}, A_{x}, A_{y}, A_{z} \right\}.
	\end{equation}
	We require most of the values of $\mathbf{P}_{\mathrm{HandOff}}$ only at 
	the cell centers 
	and, since their behavior can be smooth or discontinuous,
	we use three different orders of Lagrangian interpolation: first, 
	second, and fourth.
	The vector potential components $A_k$ are an exception:
	To be consistent with the algorithms that calculate the curl of 
	$A_k$ in \harm,
	we interpolate the components of $A_k$ into the corners of the 
	destination cells.
	Further, we use third-order Hermite interpolation for $A_k$ 
	since it ensures continuity in the first derivatives of the interpolant
	function.
	Notice that, although we will transform the basis of tensorial quantities 
	later in the \handoff, we transform
	the Cartesian velocities $\left\{v^{x}, v^{y}, v^{z}\right\}$ to a 
	spherical basis
	$\left\{v^{r}, v^{\theta}, v^{\phi}\right\}$ before interpolation.
	In this way, we avoid the propagation of truncation errors of dominant 
	components---usually $\tilde{v}^{\phi}$---into other components during the basis transformation.	

	Regarding the second set,  $\mathbf{g}_{\mathrm{HandOff}}$, consider that \harm does not 
	evolve the spacetime metric, so the \handoff is limited to 
	stationary geometries.
	Then $\mathbf{g}_{\mathrm{HandOff}}$ is simply given by 
	the components of the four-dimensional metric in Cartesian coordinates:
	\begin{equation}
		\mathbf{g}_{\mathrm{HandOff}} = \left\{ g_{\mu \nu}^{\mathrm{Cart}}, 
		\ \mu,\nu=t,x,y,z \right\}.
	\end{equation}
	We interpolate $\mathbf{g}_{\mathrm{HandOff}}$ to every 
	cell position and, since
	the metric components will be differentiated for the 
	calculation of the affine connections, we use third-order
	Hermite interpolation that, as mentioned, ensures continuity 
	in the first derivatives of the interpolant function.

	Even in the equal-mass case, the collapsed BH after a 
	BNS merger might have a gauge-induced velocity $\tilde{v}^i_{\mathrm{BH}}$.
	We correct for this effect by applying the interpolation
	routine in the frame of the BH.
	Specifically, we shift 
	$x^i \rightarrow x^i - x^i_{\mathrm{BH}}$ and
	$\beta^i \rightarrow \beta^i + \tilde{v}^i_{\mathrm{BH}}$
	before calculating and interpolating 
	$\mathbf{P}_{\mathrm{HandOff}}$ and $\mathbf{g}_{\mathrm{HandOff}}$.

	If the radial extent of the destination grid exceeds the boundaries 
	of \igm 's grid, then
	we set the grid functions to an undefined value, or \texttt{NaN}, 
	at the corresponding cells.
	Marking cells in this way ensures that they are easily found and their values overwritten with extrapolated data later in the \handoff procedure.

    The interpolation routine returns a unique file in binary format where
    the values of each grid function in the destination grid are stored in 
	contiguous memory locations.

	\subsection{Parse dataset}
	With a standalone script, we parse the binary file that results from 
	the interpolation routine,
	and construct restart files, readable by \harm.
	Specifically, we read each grid function and distribute it in 
	an array of dimension 
	$(N^r + 2 N_{\mathrm{G}}, N^{\theta} + 2 N_{\mathrm{G}}, N^{\phi} + 2 N_{\mathrm{G}})$.
	We discard the values of the primitives at the ghost cells as we will 
	refill them at runtime 
	based on the boundary conditions selected for the continued evolution. 
	The vector potential $A_k$ entails an exception: We need 
	to keep its interpolated values 
	at the ghosts cells in order to calculate its curl when initializing 
	the continued run.
	We also keep the interpolated values of the metric at the ghost cells, 
	since these will be required at initialization
	for calculating the spacetime connections in the physical cells next 
	to the coordinate boundaries. 

	We dump these reordered arrays in three restart files, 
	each containing a different interpolation order for the primitives 
	$\mathbf{P}_{\mathrm{HandOff}}$.

	\subsection{Selective interpolation orders}
	
	The truncation errors in an interpolation scheme over a smooth function 
	are proportional to $\Delta x^{n+1}$, where $\Delta x$
	denotes a measure of the grid resolution and $n$ the order of the 
	algorithm. 
	Higher orders are preferred for smooth functions to minimize the truncation error. 
	MHD primitives, however, might present strong shocks and 
	discontinuities that induce Gibbs phenomena when using high interpolation orders. 
	Our strategy in the \handoff is to keep higher-order results 
	where the solution is smooth, but lower-order results where it is sharp.
	To that end, for each primitive, we take the results from the 
	first-order interpolation, 
	which is free of Gibbs phenomena, and we compare its local values 
	to the averages
	over its neighboring cells ($\pm2$ cells in each dimension). If the relative error between these 
	quantities is lower than 
	$1\%$, then we take the local result from the fourth-order 
	interpolation; if it is higher than
	$10\%$, then we take the first-order result;  and if it is in between, then 
	we take the second-order result.
	The final outcome of this step is a unique restart file that mixes
	first-, second-, and fourth-order results for the primitives $\mathbf{P}_{\mathrm{HandOff}}$.
	
	This step is crucial for the robustness and precision of the \handoff.
	During our testing stage, we found that higher-order methods can introduce Gibbs 
	phenomena in the MHD primitives at the
	boundaries of handed-off accretion disks, 
	leading to severe MHD perturbations and shocks that were
	not in the original simulation, and preventing
	a continuous transition between codes.
	On the other hand, using solely lower-order methods compromised
	the accuracy of the transition.
	
	\subsection{Metric and primitives extrapolation}
	\label{sec:extrapolation}
	In case the outer boundary of the destination grid exceeds the
	outer boundary of the original grid, we 
	extrapolate the geometry $\mathbf{g}_{\mathrm{HandOff}}$ and primitives 
	$\mathbf{P}_{\mathrm{HandOff}}$ into the complementary cells.

	Regarding the geometry $\mathbf{g}_{\mathrm{HandOff}}$, we
	treat each of the ten independent components of the 
	four-metric individually. 
	Unlike the primitives $\mathbf{P}_{\mathrm{HandOff}}$, the metric components 
	will generally take nonzero values everywhere in the 
	computational domain, falling off as inverse polynomials of radius.
	To extrapolate each metric component $g_{\mu \nu}$ out to the 
	larger computational domain, we follow this procedure:

	\begin{itemize}
        \item Perform a low-order (up to $(l,m) = (4,\pm 4)$) 
			spherical harmonic mode decomposition of the 
			component at each radius $r$ in the destination grid 
			The result is a radial array of mode coefficients 
			$g_{\mu \nu}^{(l,m)}(r) \equiv \oint g_{\mu \nu}(r,\theta,\phi) Y_{lm}(\theta,\phi) d\Omega$.

		\item In a ``trusted window'' $r \in \{R_1,R_2\}$, 
			perform a least-squares fit of each mode coefficient $g_{\mu \nu}^{(l,m)}(r)$ 
			to a power-law function $a_0 + a_1/r^n$.

        \item Construct a new radial array of coefficients that 
			transitions continuously from the original numerical data 
			to the power-law fit over the course of the trusted window:

            \[
            \bar{g}_{\mu \nu}^{(l,m)}(r) = (1 - \lambda(r)) g_{\mu \nu}^{(l,m)}(r) + \lambda(r) ( a_0 + a_1/r^n ),
            \]

            where

            \[ 
            \lambda(r) \equiv \begin{cases}
            0 & r < R_1\\
            \frac{r-R_1}{R_2-R_1} & R_1 \leq r \leq R_2 \\
            1 & r > R_2
            \end{cases}
            \]

		\item Use the power-law fit to generate $g_{\mu \nu}^{(l,m)}(r)$ values 
			for all $r>R_2$, extending to fill the full \harm domain.

		\item Finally, reconstruct  $g_{\mu \nu}(r,\theta,\phi)$  at every point 
			in the larger domain using the new extended $g_{\mu \nu}^{(l,m)}(r)$.

	\end{itemize}

	For some individual modes, the power-law functional least-squares fit may fail to converge---generally a symptom of that mode being dominated by low-amplitude noise. 
	In these cases, we keep the original mode data out to $R_2$, and use 
	the value at $r=R_2$ to fill the remainder of the domain---equivalent to setting $a_0 = g_{\mu \nu}^{(l,m)}(R_2), a_1 = 0$.

	After extrapolating the metric components, we proceed to extrapolate
	the primitives $\mathbf{P}_{\mathrm{HandOff}}$.
	We fill the complementary cells with the numerical 
	atmosphere of \harm, $\rho=\rho_{\mathrm{atm}}$, $u=u_{\mathrm{atm}}$,
	where $\rho_{\mathrm{atm}},u_{\mathrm{atm}}$ are defined below, 
	and we set the rest of the primitives to zero.

	To avoid a sharp transition between the numerical atmospheres
	of \igm and \harm at the the boundary between
	the handed-off
	and extrapolated primitives, we also adjust the handed-off atmosphere.
	If $\rho<1.1 \times \rho_{\mathrm{atm}}^{\mathrm{Cactus}}$, we set 
	$\rho=\rho_{\mathrm{atm}}$, $u=u_{\mathrm{atm}}$, and keep the
	rest of the primitives unchanged.

	\subsection{Continued evolution}
	At this stage we have a restart file in a format readable 
	by \harm that contains the 
	MHD primitives at the physical cell centers, the vector potential $A^k$ at the  corners of physical and ghost cells, and the metric at all cell positions for all cells.
	Using existing restart routines, 
	we read in and distribute the MHD primitives among
	different processors in \harm. 
	The vector potential and metric components require specific 
	routines that take into account the ghost cells and 
	different cell positions when reading in and distributing its values.
	
	We transform the metric and vector potential components 
	from the Cartesian basis 
	to the numerical basis of the	destination grid:
	 \begin{equation}
         	\label{eq:metric_transformation}
         	g_{\mu' \nu'} = \frac{dx^{\alpha}_{\mathrm{Cart}}}{dx^{\mu'}}
         	\frac{dx^{\beta}_{\mathrm{Cart}}}{dx^{\nu'}} g_{\alpha \beta}^{\mathrm{Cart}} \,,
	 \end{equation}
	 \begin{equation}
         	\label{eq:Amu_transformation}
         	A_{\mu'} = \frac{dx^{\alpha}_{\mathrm{Cart}}}{dx^{\mu'}} A_{\alpha}^{\mathrm{Cart}} \,,
	 \end{equation}
	 and we transform the velocity from the spherical basis to numerical:
	 \begin{equation}
         	\label{eq:velocity_transformation}
		\tilde{v}^{\mu'} = \frac{dx^{\mu'}}{dx^{\alpha}_{\mathrm{sph}}} \tilde{v}^{\alpha}_{\mathrm{sph}} \,.
	 \end{equation}
	
	We calculate the curl of the vector potential $A_k$ with standard
	routines in \harm and obtain the magnetic field 
	$B^k$ at the cell centers, 
	maintaining the solenoidal constraint \citep{Zilhao+2014}.
	We also calculate the affine connections for the numerical metric 
	with standard routines of \harm.
	These use fourth-order (second-order) 
	finite differences for spatial (temporal) 
	derivatives of the metric to obtain the Christ\"{o}ffel
	symbols; cell extents are used as the discrete 
	spacing in the spatial finite differences, while a much
	smaller time spacing than the evolution's Courant-limited
	time step is used in order to keep its truncation error
	smaller than that of the spatial difference. We typically
	limit the simulations from handed-off initial data (ID) to instances when the metric is nearly stationary, so the continued
	evolution is assumed to have a time-independent
	spacetime.
	For this reason during \handoff we simply use the same metric field data for all nominal time slices fed into the \harm affine connection routine, guaranteeing that all temporal derivatives will be zero.

	We evolve the handed-off primitives over the numerical 
	spacetime in the destination grid with the well-tested
	methods of \harm \citep{Gammie+2003,Noble+2009,MurguiaBerthier+2021b}.
	We integrate the equations of motion, Eq.~(\ref{consEqs}), with HRSC
    schemes.
	We reconstruct the primitive variables to the
    cell interfaces through piecewise parabolic interpolation, 
	we apply the Lax-Friedrichs formula to compute
    the local fluxes, and the
    method of lines for time integration with a second-order Runge-Kutta (RK2) method.  The primitive variables are recovered from the conserved variables with the scheme described in \cite{Noble+2006, MurguiaBerthier+2021b}. 
	If the updates of $\rho$ or $u$ go
        below the corresponding atmosphere values 
	$\rho_{\mathrm{atm}} = 2 \times 10^{-10} (r/M)^{-3/2}$,
	$u_{\mathrm{atm}} = 2 \times 10^{-12} (r/M)^{-5/2}$
        they are reset to the latter.
	We use the constrained transport algorithm 
	\texttt{FluxCT} \citep{Toth2000}
	to evolve the magnetic field $B^k$ and
	maintain the solenoidal constraint.
	For more details, see Ref. \cite{Noble+2009}.

\section{Code Verification: Fishbone-Moncrief disk} \label{sec:validation}

	To verify the \handoff we take the case of a
	magnetized Fishbone-Moncrief (FM) disk \cite{Fishbone+1976}
	around a rotating BH.
	We evolve the system 
	with \igm and \harm independently, and we use the \handoff 
	to perform several transitions between the codes at
	different stages of the simulation.

    \begin{figure}[htb!]
        \centering
        \includegraphics[width=.8\columnwidth]{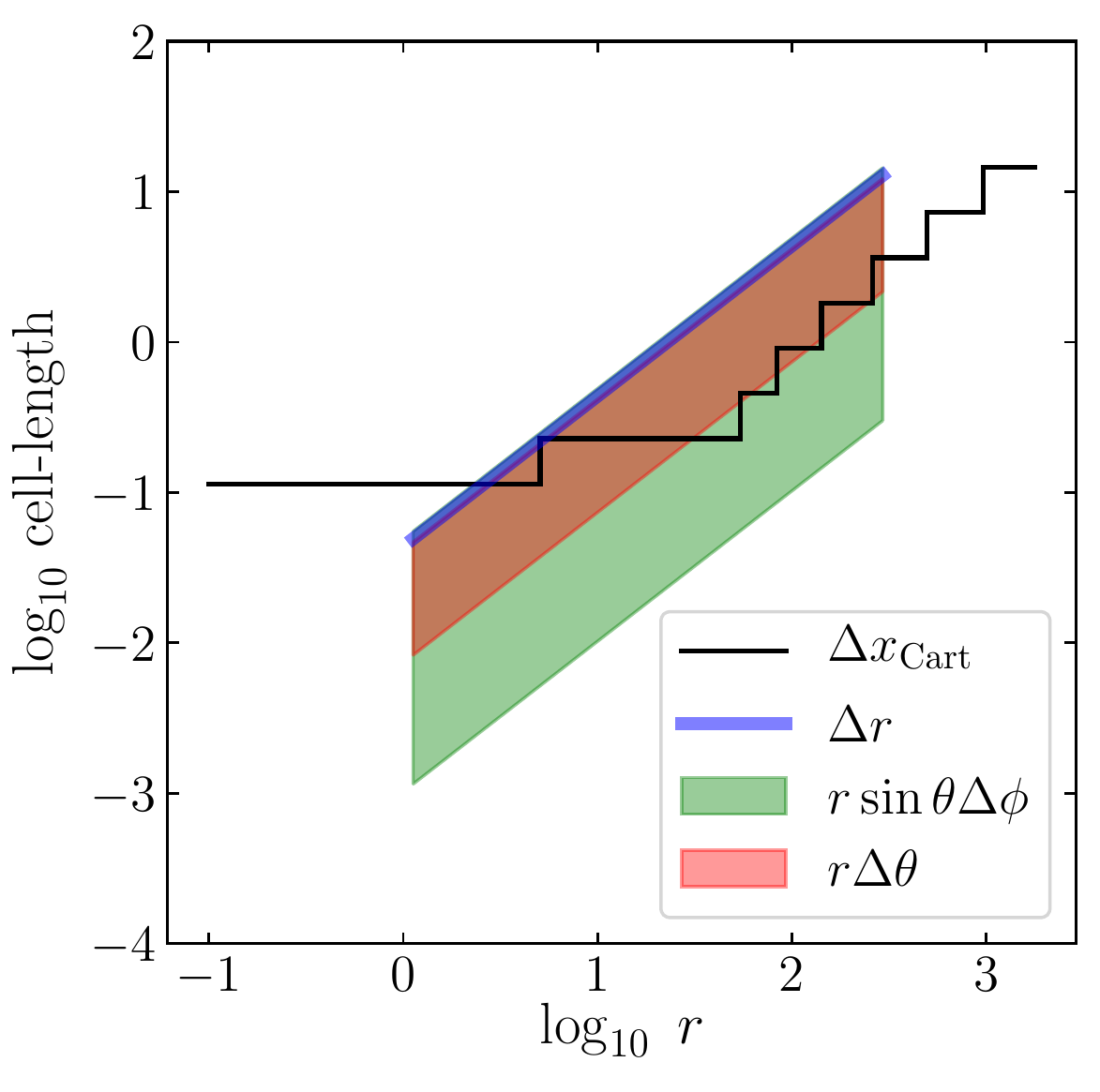}  
        \caption{Cell lengths as a function of radii for the FM simulations in \igm (\textit{black}) and \harm (\textit{green}, \textit{red},
		and \textit{blue}). Shaded regions show how the cell lengths change over the full span of $\theta$.
		Note that regions of overlap between \textit{green} and \textit{red} appear brown in the plot.}
        \label{fig:FM_e_resolutions} 
    \end{figure}

	We begin by setting the same initial data in \igm and \harm
	for a magnetized
	FM disk around a BH of mass $M$ with specific angular momentum $a/M=0.9375$, 
	as dictated by Ref. \cite{Porth+2019}. 
	We evolve this data with \igm and \harm up to  
	$t=10^{4}M$.
	The grid in \igm consists of a Cartesian hierarchy of eight
	refinement
	levels, as described in Ref. \cite{Porth+2019}.
	The grid in \harm has a spherical topology and
	consists of $128^3$ cells, 
	parametrized by numerical coordinates
	that refine the resolution towards the equator and towards
	the BH, with the functional dependence used by Ref.
	\cite{McKinney+2004} 
	(we use $h=0.3$ in Eq.~(8) of this reference). 
	We make use of novel boundary conditions
	at the 
	coordinate boundaries of $\theta$, i.e.
	at the polar axis,
	which allow us to fully extend $\theta\in(0,\pi)$.
	In Appendix~\ref{sec:BC} we describe
	the details of these boundary conditions.
	The coordinate boundaries
	of $\phi$ are internal to the domain, and
	the coordinate boundaries of $r\in\left[1.1M,300M\right]$
	are physical boundaries, where we impose
	outflow boundary conditions.
	
	In Fig.~\ref{fig:FM_e_resolutions} we plot the cell lengths
	in both \igm and \harm as a function of radii and for $\theta \in (0,\pi)$, and conclude the resolutions
	are comparable.
	The cell length $\Delta r$ grows as a function of radius because the radial resolution has an 
	exponential dependence with radius.
	The cell length $r \sin \theta \Delta \phi$ grows as a function of radius because of its explicit radial dependence, and also spans different values at each radius because of its $\theta$ dependence.
	Finally the cell length $r\Delta \theta$ grows as a 
	function of radius because of its explicit radial 
	dependence and has different values for a given
	radius because $\Delta \theta$ is not uniform, but progressively \emph{smaller} toward the equator.
	Such flexibility of the numerical grids in \harm to
	adapt optimally to the relevant physical processes is
	one of the main motivations behind the \handoff. In
	this case, the numerical grid in \harm has less than $\times 80$ cells than the numerical grid in \igm and will bring
	comparable results.
	
	As can be seen from Fig.~\ref{fig:FM_e_resolutions}, 
	the numerical domain in \harm is contained in the domain
	in \igm, therefore we do not 
	need to extrapolate the primitives or  metric
	(see Sec.~\ref{sec:extrapolation}).
	We will validate the extrapolation procedure 
	in the next section, where we apply the \handoff
	to a BNS postmerger.

	We do the first transition at $t=0$, i.e. we hand off
	the initial data (ID) from \igm to \harm.
	By comparing the resulting dataset with the
	ID constructed in \harm, 
	we can measure the truncation error introduced in  
	the \handoff.
	Based on the resolution of \igm in the region of 
	the disk $\Delta x_{\mathrm{IGM}}\approx 0.2274$, 
	and the interpolation orders considered $n=1,2,3,4$, 
	we expect the following values for the truncation errors
	\begin{equation}
		\label{eq:epsilon_estimates}
		\epsilon \sim \Delta x_{\mathrm{IGM}}^{n+1} \sim 
		\begin{cases}
  			5 \times 10^{-2}, \ n=1 \\ 
  			1 \times 10^{-2}, \ n=2 \\
  			2 \times 10^{-3}, \ n=3 \\
  			6 \times 10^{-4}, \ n=4  
	    \end{cases} \, .
	\end{equation}

\begin{figure*}[htb!]
	\centering
	\includegraphics[width=.9\columnwidth]{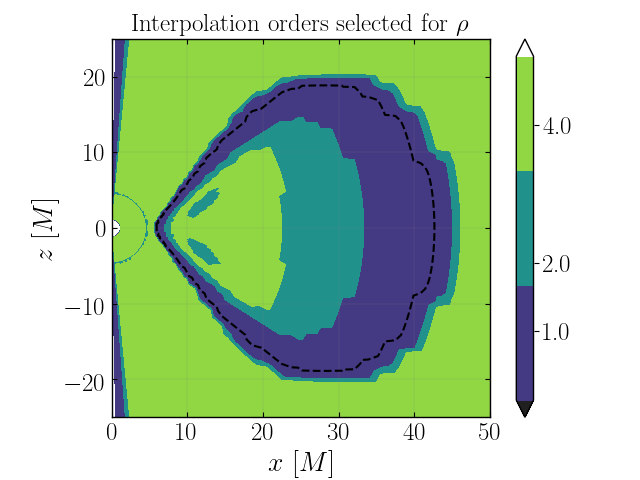}  
	\includegraphics[width=.9\columnwidth]{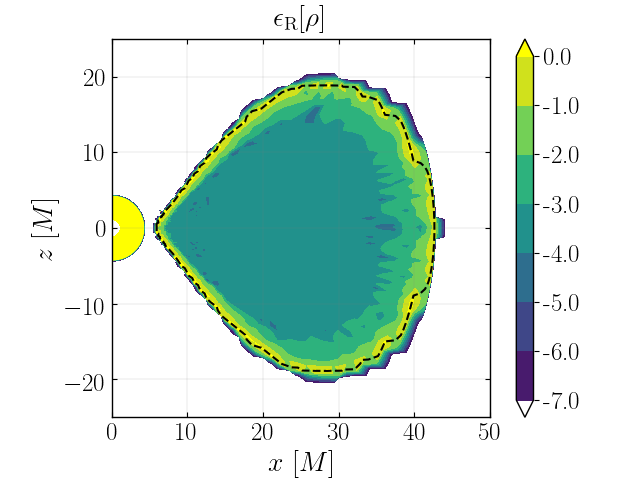}  
	\caption{Interpolation orders selected (\textit{left}) and the 
	correspoding relative errors (\textit{right}) for $\rho$, 
	when handing-off ID for a magnetized
	FM from \igm to \harm.
	Fourth order interpolation is selected at the core of the
	disk where $\rho$ is smooth, and first order interpolation is selected
	at the surface of the disk where there is a jump
	to the numerical floor. There is also a transition
	region where second order interpolation is selected.
	The numerical errors follow the expected values from the
	resolution and interpolation orders (see 
	Eq.~\eqref{eq:epsilon_estimates}).}
 	\label{fig:rho_int_error} 
\end{figure*}

	In Fig.~\ref{fig:rho_int_error} (\textit{left})
	we plot the interpolation
	orders selected for the primitive $\rho$, in the $xz$ plane.
	The distribution follows the expected behavior:
	At the core of the disk, around the position
	of the maximum of pressure at $r=12M$, fourth order
	interpolation is selected. In contrast, at the
	surface of the disk, where the density jumps to the 
	value of the numerical floor, first order is selected.
	There is also a transition region in the body
	of the disk where second order is selected.

	In Fig.~\ref{fig:rho_int_error} (\textit{right})
	we plot $\epsilon_{\mathrm{R}}[\rho]$
	in the $xz$ plane,
	where
	\begin{equation}
		\label{eq:epsilon_R}
		\epsilon_{\mathrm{R}}[P] = \frac{ \left| P_{\mathrm{harm3d}} - P_{\mathrm{HandOff}} \right|}{\frac{1}{2}\left( \left| P_{\mathrm{harm3d}} \right| +  \left| P_{\mathrm{HandOff}} \right| \right)}
	\end{equation}
	is the relative error between the handed-off primitive 
	and the value initialized in \harm. 
	As expected from the estimates in Eq.~\eqref{eq:epsilon_estimates},
	the body of the disk presents
	$\epsilon_{\mathrm{R}}[\rho]\sim 1\e{-4}$ 
	and transitions to $\epsilon_{\mathrm{R}}[\rho]\sim 1\e{-2}$ 
	near the surface of the disk.
	The region around the BH with higher errors
	happens because \igm sets a tenuous atmosphere
	with $\rho>1.1\rho_{\mathrm{atm}}^{\mathrm{Cactus}}$
	around the BH and the \handoff does not reset it to
	the \harm's atmosphere.
	This is still a low-density region, where $\rho<1\e{-5}$
	in code units,
	and it does not affect our results.	
	In global terms, the density-weighted relative error 
	$\left< \epsilon_{\mathrm{R}}[\rho] \right>_{\rho}$, where
	\begin{equation}
		\label{eq:epsilon_R_weighted}
		\left< X \right>_{\rho} = \frac{ \int X  \rho \sqrt{-g} \, dV}{ \int \rho \sqrt{-g} \, dV},
	\end{equation}
	is $\left< \epsilon_{\mathrm{R}}[\rho] \right>_{\rho}=2\e{-4}$.

	For the remaining primitives, we find equivalent distributions
	for the interpolation orders selected, but
	different values for the integrated errors.
	We find a higher integrated
	error for the internal energy
	$\left< \epsilon_{\mathrm{R}}[u] \right>_{\rho}=1\e{-2}$
	but this is dominated by random perturbations, 
	introduced artificially
	to trigger accretion.
	We find lower errors for nonvanishing components
	of the velocity 
	$\left< \epsilon_{\mathrm{R}}[v^{(1)}] \right>_{\rho}=
	8\e{-6}$ and 
	$\left< \epsilon_{\mathrm{R}}[v^{(3)}] \right>_{\rho} = 
	1\e{-5}$.
	These functions are smoother than $\rho$, and
	monotonically decrease in $r$ within the disk,
	explaining why the interpolation is more accurate.
	We find higher errors for the nonvanishing
	components of the magnetic field 
	$\left< \epsilon_{\mathrm{R}}[B^{(1)}] \right>_{\rho}=
	\left< \epsilon_{\mathrm{R}}[B^{(2)}] \right>_{\rho}=1\e{-1}$
	that come from the propagation of 
	truncation errors in $A_k^{\mathrm{Cart}}$
	 after the basis transformation and 
	curl calculation.
	Regarding the primitives that are initialized to zero, we
	define the relative error by
	\begin{equation}
		\label{eq:epsilon_R_zero}
		\tilde{\epsilon}_{\mathrm{R}}[P] = \left| P_{\mathrm{HandOff}} \right|
	\end{equation}
	and find $\left< \tilde{\epsilon}_{\mathrm{R}}[v^{(2)}] \right>_{\rho} = 5\e{-19}$ 
	and $\left< \tilde{\epsilon}_{\mathrm{R}}[B^{(3)}] \right>_{\rho} = 4 \e{-8}$.
	Comparing $\left< \tilde{\epsilon}_{\mathrm{R}}[v^{(2)}] \right>_{\rho}$
	with $\left< \tilde{\epsilon}_{\mathrm{R}}[B^{(3)}] \right>_{\rho}$ 
	we notice the convenience of transforming to the spherical basis
	before interpolation and the effect of error propagation 
	by the curl calculation. 
	In the case of the BNS postmerger, however, we expect these errors on the 
	magnetic field components to be smaller, since
	the different components, and therefore the truncation errors, will be more comparable in magnitude.

	In this validation test, we can also measure the truncation errors
	of the handed-off metric components
	, since	we know their analytical values in advance.
	Overall, we find lower errors for the
	metric components than for the MHD primitives,
	$\left< \epsilon_{\mathrm{R}}[g_{\mu\nu}] \right>_{\rho} < 1\e{-7}$.
	These lower errors in the metric components were expected
	because these are differentiable functions, and because 
	we use third-order interpolation in every cell.

	After confirming that the handed-off ID and spacetime metric
	are consistent with our expectations, we evolve the dataset
	within \harm.
	We denote this simulation \texttt{HandOff\_0}.
	
\begin{figure}[htb!]
    \centering
    \includegraphics[width=\columnwidth]{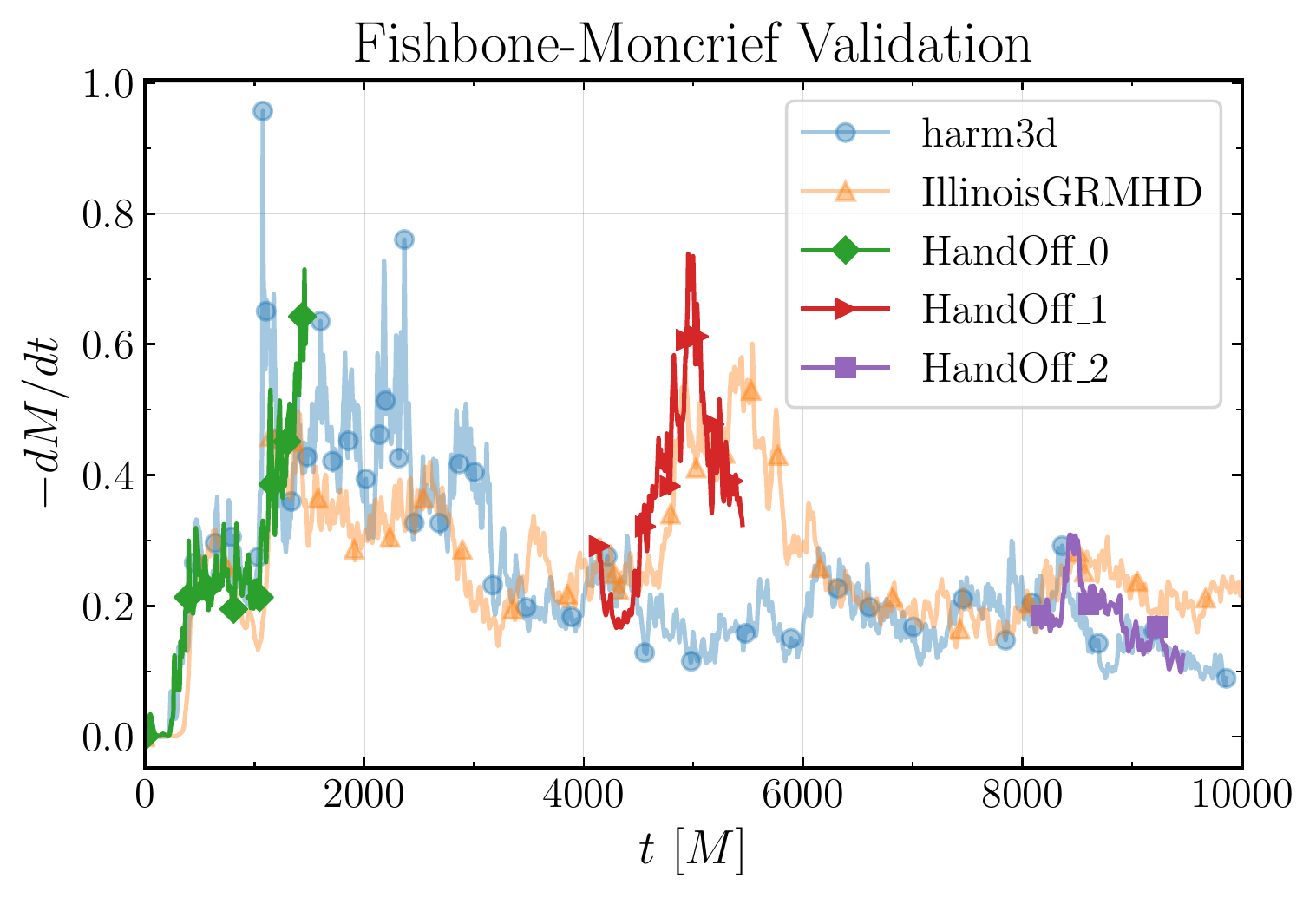}
    \includegraphics[width=\columnwidth]{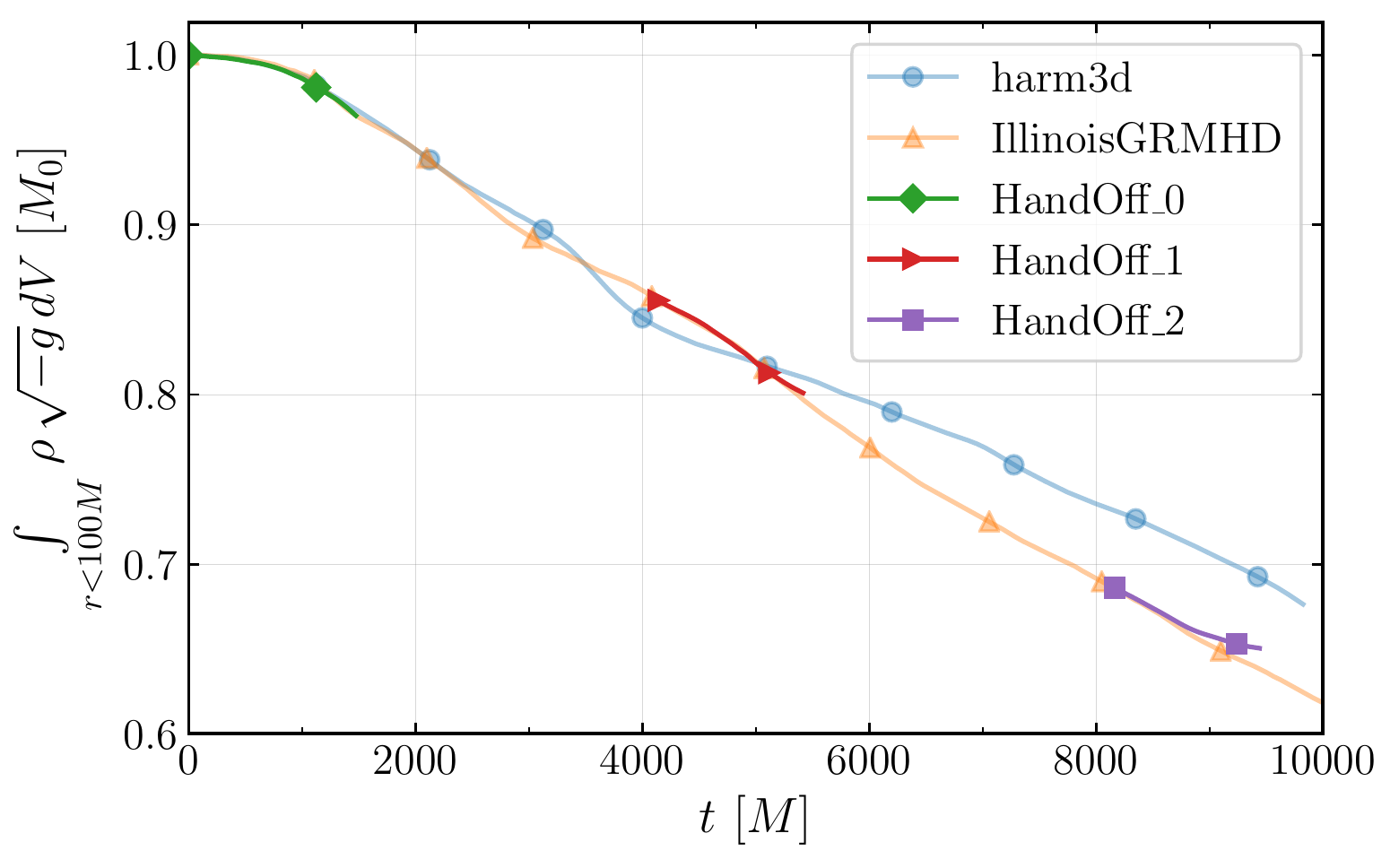}
    \caption{Accretion rate at the horizon (\textit{top}) 
    and total mass within $r=100M$ (\textit{bottom}) 
    as a function of time for the FM validation case.
    The curves \texttt{harm3d}
	and \igm stand for the fiducial  
	simulations with these codes.
	The curves \texttt{HandOff\_0}, \texttt{HandOff\_1} 
	and \texttt{HandOff\_2} represent handed-off 
	runs from \igm to
	\harm, at times $t=0, \ 4.1\e{3}, \ 8.1\e{3} M$,
	respectively.
	In every case, the transition is continuous and the 
	evolved system follows the expected behavior from
	the fiducial runs.}
    \centering
    \label{fig:FM_t}
\end{figure}

	Defining the accretion rate as
	\begin{equation}
		\label{eq:mdot}
		-\frac{dM}{dt}(r) = \int \rho u^r \sqrt{-g} \, d\theta \, d\phi,
	\end{equation}
	in Fig.~\ref{fig:FM_t} (\textit{top}) we plot this quantity
	at the BH horizon for \texttt{HandOff\_0} (\textit{green})
	and for the fiducial run in \harm (\textit{light blue}).
	We also include the accretion rate for the fiducial run
	in \igm (\textit{orange}).
	We notice the evolution of \texttt{HandOff\_0} is dynamically
	equivalent to the fiducial run in \harm, capturing 
	the MRI growth, saturation, and relaxation.

	To test the \handoff in a more realistic and turbulent
	scenario, we transition the same simulation from \igm to
	\harm at $t=4.1\e{3}M$. 
	We denote this continued evolution \texttt{HandOff\_1}.
	In Fig.~\ref{fig:FM_t} (\textit{top, red}) we plot
	the resulting accretion rate and, 
	again, we find the \handoff succesfully captures 
	the dynamical state of the disk, 
	reproducing the value of the accretion rate at the 
	time of transition, and the spike in the accretion rate
	at $t\sim 5\e{3}M$.

	Finally, we perform a transition at $t=8.1\e{3}M$
	and we denote this continued evolution \texttt{HandOff\_2}.
    In Fig.~\ref{fig:FM_t} (\textit{top, violet}) 
	we plot the resulting accretion rate and
	we notice its value at the transition time
	matches the value in \igm, proving a continuous transition.
	The accretion rate then converges to 
	the results of \harm.
	Indeed, while the state of the disk 
	at the transition time
	is determined by \igm, the continued evolution
	is determined by the numerical methods in \harm.

    Beyond the local measure of the accretion rate, 
    we analyze
    the continuity of global quantities
    after the \handoff.
    In Fig.~\ref{fig:FM_t} (\textit{bottom}) we plot 
    the integrated mass of the disk within $r=100M$,
    normalized by the initial mass $M_0$.
    We find the curves of \texttt{HandOff\_0}, \texttt{HandOff\_1} and
    \texttt{HandOff\_2} match the results from \igm at the time
    of transition, proving the high fidelity of the \handoff.
    
	We conclude the \handoff succesfully translates
	the GRMHD state of a magnetized torus, and
	the spacetime metric, from a Cartesian-AMR
	grid within \igm to a more flexible grid
	within \harm. In the next section we will apply
	the \handoff to the interesting case of a BNS postmerger.

\section{Results: BNS postmerger} \label{sec:results}

	In this section we evolve a BNS merger with
	\igm and apply the 
	\handoff to continue the postmerger
	evolution in \harm.
	The results presented in this section
	are well known from
	the literature, but serve to demonstrate the
	consistency and validity of the continued
	evolution from the \handoff.

	\subsection{Merger proper}
	\label{ss:merger-proper}
	The ID for the BNS is similar to that of
	\cite{Rezzolla+2011}
	\citep[see also Refs. ][for similar settings]
	{Ruiz+2016, Kawamura+2016}: 
	Two NSs in a circular orbit, separated
	by ${\sim}45~\mathrm{km}$, each with
	a gravitational mass of $1.5~\Msun$ and
	an equatorial radius of $13.6~\mathrm{km}$.
	We also initialize two poloidal
	magnetic fields, confined in each star, but
	with a maximum strength of
	${\sim}10^{15}~\mathrm{G}$, three orders of
	magnitude higher than in Ref. 
	\cite{Rezzolla+2011}.
	We use a Cartesian-AMR grid centered in the
	center of mass of the system, with outer 
	boundaries at ${\sim}5700~\mathrm{km}$, and
	seven refinement boundaries with a finest 
	resolution of ${\sim}180~\mathrm{m}$.
	We model the fluid as an ideal gas, with
	adiabatic index $\Gamma = 2$.

	We evolve this ID with \igm
	and, consistent with Ref. \cite{Rezzolla+2011},
	we find the NSs inspiral towards the center of mass
	of the system as they transfer energy and angular
	momentum to gravitational waves (GWs), until they 
	merge after $8~\mathrm{ms}$ (${\sim}3$ orbits).
	After merger, a HMNS forms 
	but, given the large mass of the system,
	it promptly collapses to a BH 
	(${\sim}8~\mathrm{ms}$ after merger).
	If we set the ID at time
	$t=0.0~\mathrm{s}$, then the BH forms 
	at $t_{\mathrm{BH}} = 0.017~\mathrm{s}$.
	We use the thorn \texttt{AHFinderDirect}
	\citep{Thornburg2003} to locate the apparent horizon;
	the resulting BH has a mass of $2.911~\Msun$,
	a specific angular momentum of $0.82$, and 
	a drift velocity 
	$v_{\mathrm{BH}}\sim 1\e{-3} c$.
	At $t=t_{\mathrm{BH}}$, we add 
	two refinement levels centered at
	the remnant, yielding a 
	finest resolution after merger 
	of ${\sim}50~\mathrm{m}$. 
	The remnant is surrounded by an orbiting and
	magnetized torus, 
	with mass $0.076~\Msun$.

	\subsection{Handing off the postmerger: Destination grids and boundary conditions}
	
		\begin{figure}[htb!]
    	\centering
        \includegraphics[width=\columnwidth]{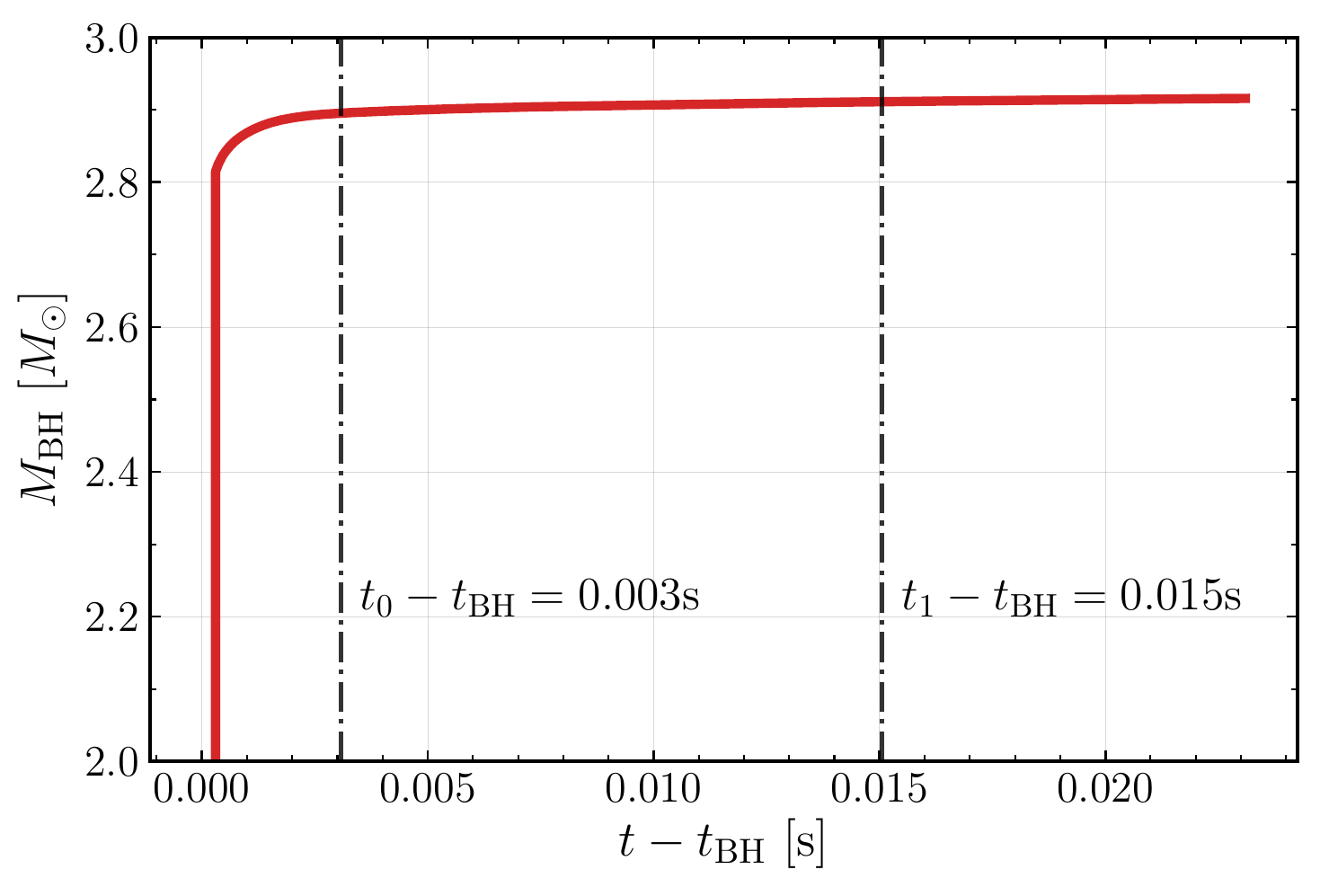}  
        \caption{Mass of BH remnant from the BNS merger, 
        as a function of time. \textit{Dashed-dotted} lines
        represent the times at which we apply the
        \handoff.
        For our fiducial run, $M_{\mathrm{BH}}(t_1)=2.911\Msun$. }
        \label{fig:BNS_mass_BH} 
    \end{figure}
	
	\begin{figure*}[htb!]
	\centering        \includegraphics[width=.8\columnwidth]{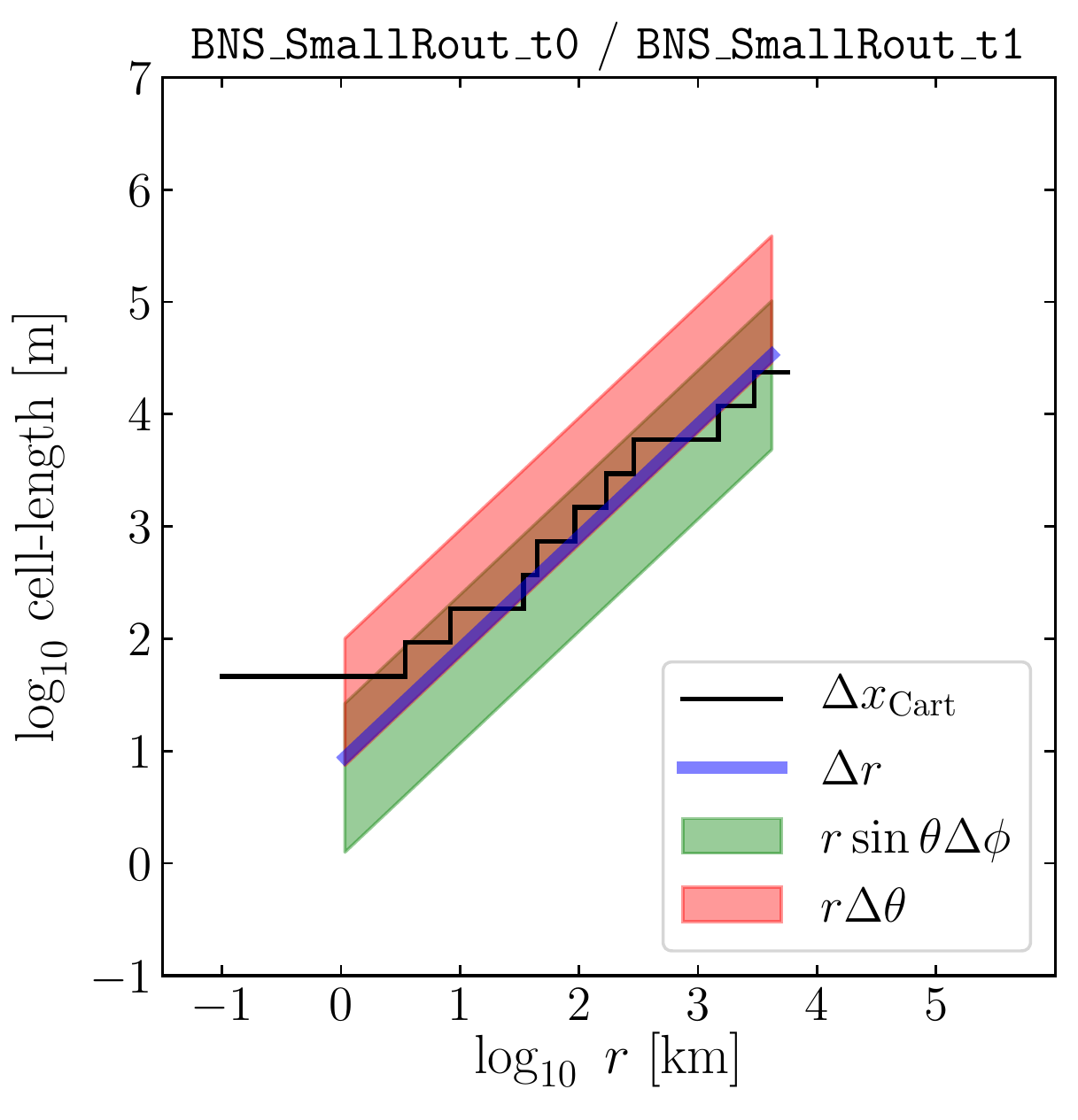}
    \includegraphics[width=.8\columnwidth]{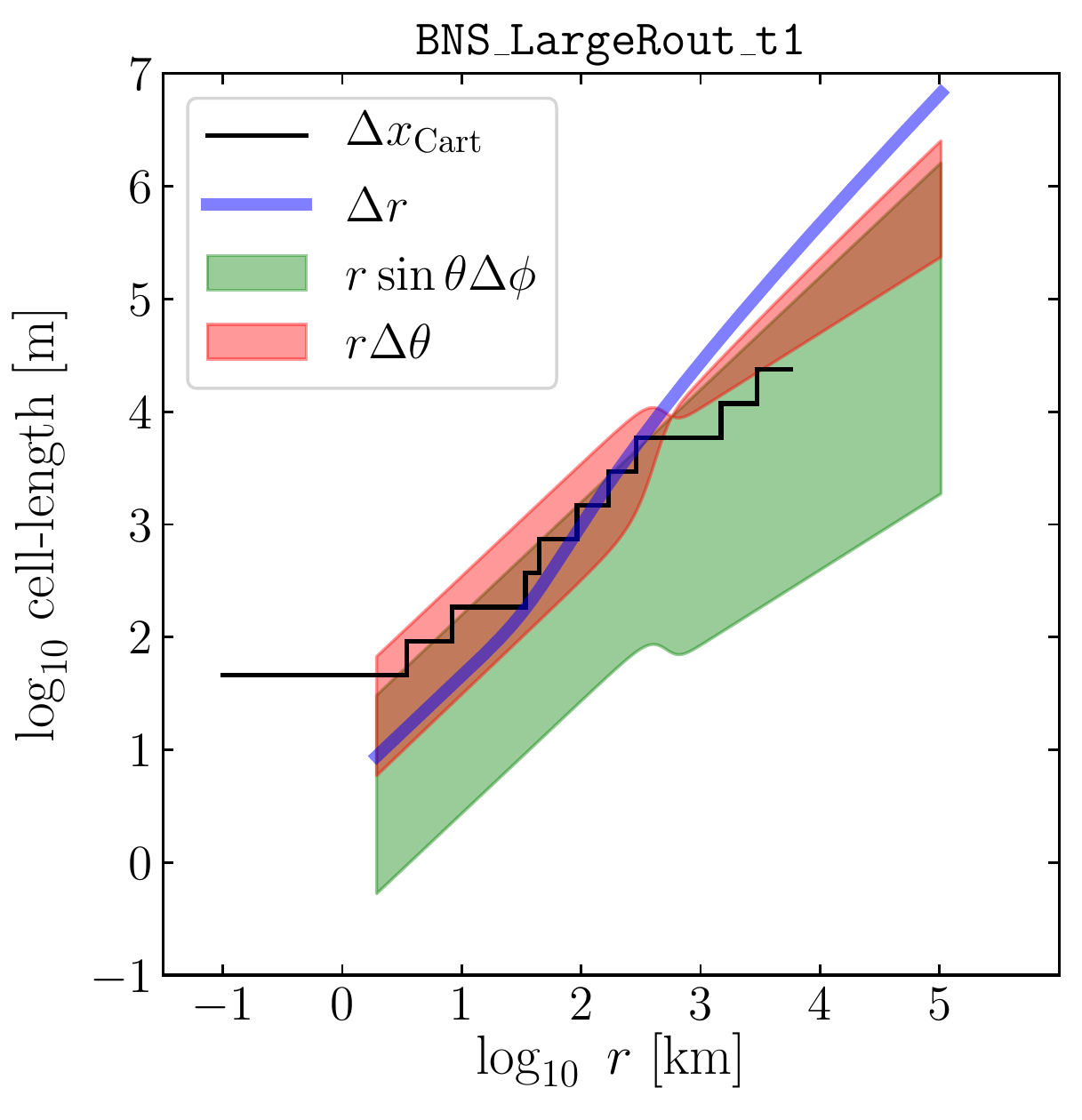}
    \caption{cell lengths for BNS postmerger simulations as a function of radii. The \textit{black} line 
	    represent the resolution of the 
	    AMR-levels used during merger. 
	    \textit{Green}, \textit{red}, and \textit{blue}
	    points represent the grid lengths 
	    for spherical-like cells used during the continued evolution
	    in \BNSSRtz (\textit{left}), \BNSSRto (\textit{left}),
	    and \BNSLRto (\textit{right}).}
    \label{fig:BNS_resolutions}
	\end{figure*}
	
    After BH formation, we apply
    the \handoff  
    and transition the BH-torus system to \harm.
	For consistency checks, we do the transition
	at different times and to different grids.

	We do the first \handoff at $t_0=0.020~\mathrm{s}$, 
	$0.003~\mathrm{s}$ after BH formation,
	and use a grid equivalent to that of Ref. 
	\cite{Noble+2012}, but with $\xi = 0.65$, 
	$\theta_{\mathrm{c}}\sim1\e{-14}$, and $n=7$
	(see Eq.~(22) of that reference), 
	we set the radial extent $r\in(1.07,4291.53)~\mathrm{km}$,
	and use $1024\times160\times256$ cells.
	This grid is contained in the original
	grid for the merger, so we do not need to 
	extrapolate the primitives or spacetime metric. 
	We denote this transition, and continued evolution,
	\BNSSRtz.
	Using this same grid, 
	we do a later transition, at 
	$t_1 = 0.032~\mathrm{s}$, denoted \BNSSRto.
	In Fig.~\ref{fig:BNS_mass_BH} we plot the
	mass of the BH remnant $M_{\mathrm{BH}}$ as a function of time,
	calculated with the thorn \texttt{QuasiLocalMeasures} \citep{Dreyer+2003}, 
	and notice that $M_{\mathrm{BH}}$ is in a converging regime at
	times $t_0$ and $t_1$. Specifically, $M_{\mathrm{BH}}(t_0)=2.895 \Msun$ 
	and $M_{\mathrm{BH}}(t_1)=2.911 \Msun$.

	Next, we do a transition at the same time
	$t_1$, but to a grid designed specifically
	for a BNS postmerger.
	As described in the Introduction, this new 
	grid has spherical topology, the outer boundary 
	is far enough (${\sim}10^5~\mathrm{km}$) to capture the  jet breakout,  
	it  has  higher resolution  in $\theta$ towards the equator 
	if close to the BH to resolve the disk, 
	but higher resolution towards the polar axis if farther away,
	to resolve the funnel region.
	The implementation details of this grid are described in
	Appendix~\ref{sec:grid}. 
	In this case, since the destination grid is much
	larger than the original grid for the merger, we need to 
	extrapolate the primitives and spacetime metric
	to the complementary cells.
	We denote this run \BNSLRto, and consider it
	our fiducial simulation.

	In Fig.~\ref{fig:BNS_resolutions} we include
	plots for the cell lengths of these grids
	as a function of radii and for $\theta \in (0,\pi)$, 
	compared with the resolution in \igm.
	We notice that, at the bulk of the disk
	($r<200~\mathrm{km}$),
	cells at the equator have
	higher resolution for both $\theta$
	and $r$ with respect to the original
	resolution
	in \igm, and comparable resolution
	for $\phi$.
	The description of Fig.~\ref{fig:BNS_resolutions}
	(\textit{left}) is similar to that 
	of Fig.~\ref{fig:FM_e_resolutions} in the previous
	section, for an exponential grid in $\Delta r$ with
	$\Delta \theta$ focused in the equator.
	The description of Fig.~\ref{fig:BNS_resolutions}
	(\textit{right}) is more subtle and is given
	in Appendix~\ref{sec:grid}.

	In every case, we apply novel boundary conditions
	of \harm 
	at the polar axis that refer to neighboring cells
	in the domain,
	allowing us to extend
	$\theta \in (0,\pi)$ and to fully resolve the funnel
	region. See Appendix~\ref{sec:BC} for the details
	on implementing these boundary conditions.
    The coordinate boundaries of $\phi$ are are 
    also internal to the domain, and the 
    coordinate boundaries of $r$ are the actual physical 
    boundaries, where we impose outflow
    boundary conditions.

	\subsection{Handing off the postmerger: Initial data}
	\begin{figure}[htb!]
        	\centering
        	\includegraphics[width=\columnwidth]{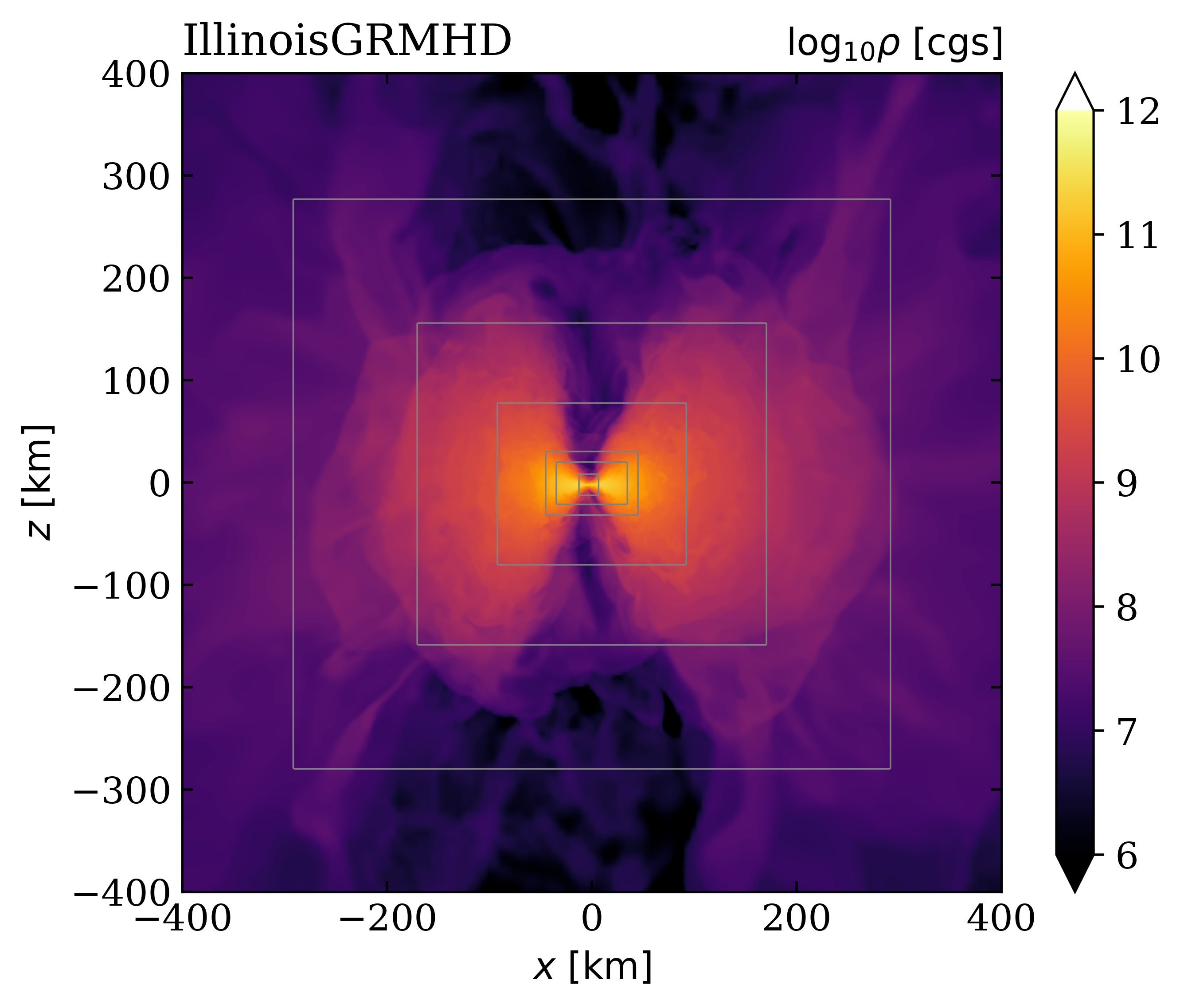}  
        	\includegraphics[width=\columnwidth]{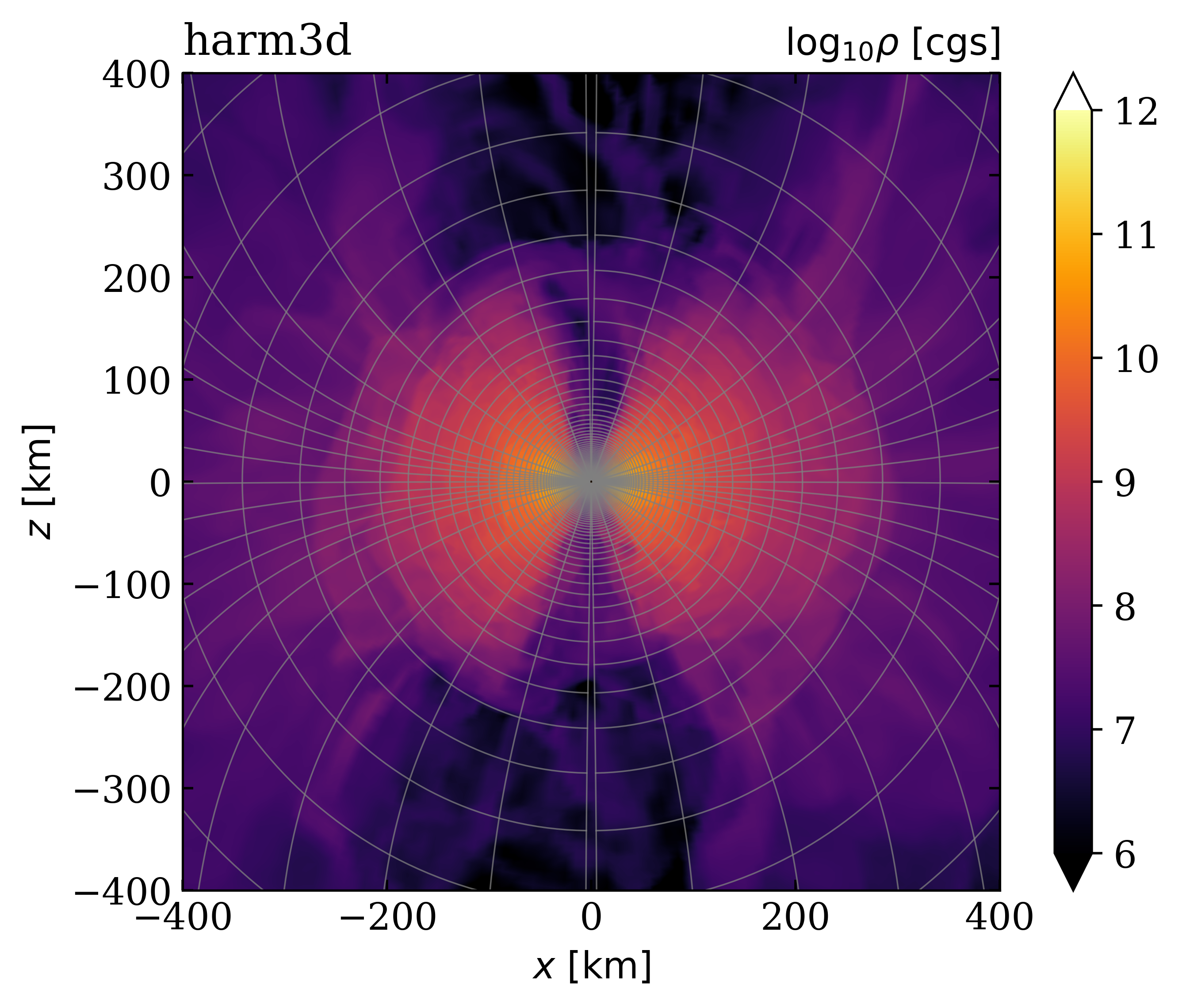}  
        	\caption{Baryonic density $\rho$ in the $xz$-plane
		for \BNSLRto before
		(\textit{top}) and after (\textit{bottom}) the
		hand-off procedure, in \igm and \harm, respectively.
		The gray lines in the plot for \igm represent the 
		AMR boundaries and, in the plot for \harm, 
		they represent 1 every 10 grid lines of the
		destination grid.}
         	\label{fig:BNS_rho_b} 
        \end{figure}
	
	In Fig.~\ref{fig:BNS_rho_b} we plot 
	the rest-mass density $\rho$ 
	in the $x-z$ plane at the time of transition
	for \BNSLRto (\textit{bottom}).
	We also plot the original data from \igm (\textit{top}), and we plot
	gray lines that represent 
	the grid topology for each case.
	As a demonstration of the precision of
	the \handoff, 
	we note the plots of $\rho$ are
	indistinguishable between the two codes.

	\begin{figure}[htb!]
        	\centering
        	\includegraphics[width=.7\columnwidth]{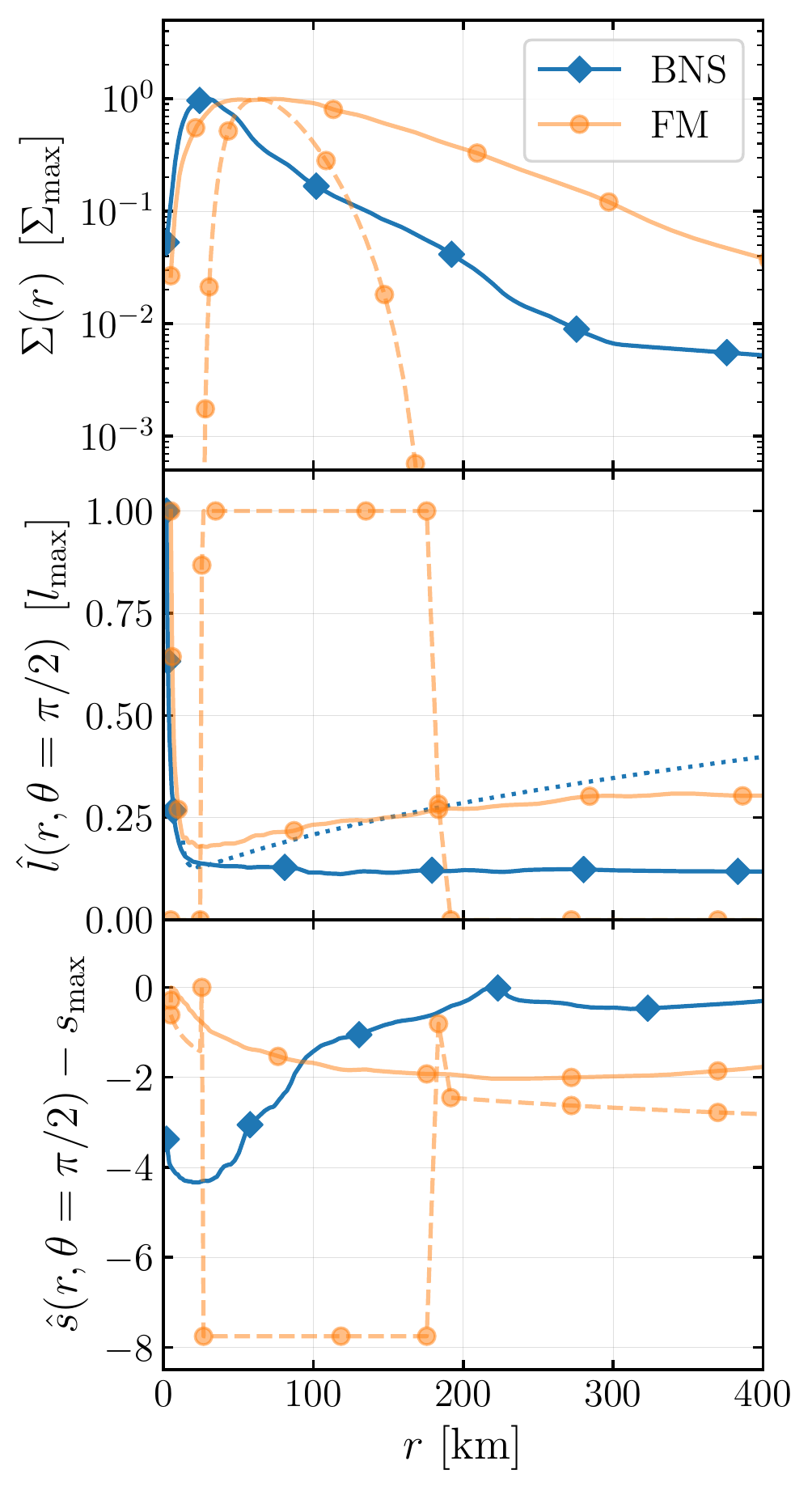}  
		\caption{Disk measures for \BNSLRto at the time
		of transition (\textit{solid, diamond, blue}), and
		for the FM disk evolved in Sec.~\ref{sec:validation} 
		at $t=0$ (\textit{dashed, circle, orange}) and $t=9000M$
		(\textit{solid, circle, orange}). From top to bottom, 
		we plot the $\theta$-integrated and $\phi$-averaged
		density $\Sigma(r)$, the specific angular momentum 
		at the equator $\hat{l}(r,\theta=\pi/2)$,
		and specific entropy at the equator
		$\hat{s}(r,\theta=\pi/2)$. The \textit{blue, dotted}
		line in the middle panel represent the Keplerian
		value for $\hat{l}(r,\theta=\pi/2)$, given the 
		mass and spin of the remnant BH.}
         	\label{fig:BNS_FM_r} 
        \end{figure}
	In Fig.~\ref{fig:BNS_FM_r} we focus on the 
	accretion disk around the remnant for \BNSLRto
	(\textit{solid, blue}),
	and plot
	the $\theta$-integrated and $\phi$-averaged
	density:
	\begin{equation}
		\Sigma(r) = \frac{ \int \sqrt{-g} \rho \, d\theta \, d\phi}{\int \sqrt{ g_{\phi \phi}(\theta=\pi/2)} \, d\phi},
	\end{equation}
	and the $\phi$-averaged specific angular momentum 
	$\hat{l}~(r,\theta=~\pi/2)$ and specific entropy 
	$\hat{s}~(r,\theta=~\pi/2)$ at the equator, where
	$l=u_{\phi}/u_{t}$, $s=\ln( p/\rho^{\Gamma})$, and
	\begin{equation}
		\hat{ X}(r,\theta) = \frac{\int X \sqrt{g_{\phi \phi}} \, d\phi}{\int \sqrt{g_{\phi \phi}} \, d\phi}.
	\end{equation}
	Since many references initialize
	the postmerger disk with an FM distribution 
	\citep[e.g.][]{Siegel+2017},
	or similar isentropic solutions with
	constant specific angular momentum, in 
	Fig.~\ref{fig:BNS_FM_r}
	we also plot the latter quantities 
	for the FM disk evolved
	in the previous section at $t=0$
	(\textit{dashed, orange})
	and $t=9000M$
	(\textit{solid, orange}).
	For the FM, we set the unit of length assuming 
	the central BH has a mass of $2.911~M_{\odot}$, 
	and set the units of MHD variables assuming
	the code units are the same as in the BNS case.
	Comparing the postmerger disk with the curves for
	the evolved FM disk, 
	we notice a steeper decay of $\Sigma(r)$ in the postmerger case. The angular momentum
	of the postmerger is rather
	constant at the bulk of
	the disk, but presents
	a peculiar decreasing trend.
	The specific entropy, on the
	other hand, presents the most
	remarkable difference.
	For the postmerger disk, it grows steeply outwards because of the shocked and
	ejected material from the merger.

	The accretion disk is fairly magnetized.
	At the time of transition, for \BNSLRto, we find the
	ratio of the $\rho$-weighted integrals of
	thermal and magnetic pressure to be:
	\begin{equation}
		\label{eq:int_beta_rho}
		\frac{ \int p \rho \sqrt{-g} \, dV}{ \int p_\mathrm{m} \rho \sqrt{-g} \, dV} \sim 140
	\end{equation}
	For comparison, we find this ratio 
	to be ${\sim}560$ and ${\sim}30$ for the the FM evolved 
	in the last section, at $t=0$ and $t=9000M$,
	respectively.
	Therefore, according to this parameter, 
	the magnetization of the relaxed FM is $\times 4-5$
	larger than the initial postmerger disk.
	In the next subsection we analyze the evolution
	of this parameter.
	
	We proceed to investigate the topology of the magnetic
	field.
	Following \cite{Fanci+2013}, the 
	decomposition of the magnetic field into
	toroidal $B_{||}$ and poloidal $B_{\perp}$ 
	components for
	nonaxysymmetric distributions of
	matter follows from 
	the projection of the field with respect
	to the fluid frame:
	\begin{equation}
		B^i = B_{||} \frac{\tilde{v}^i}{\sqrt{\gamma_{ij} \tilde{v}^i \tilde{v}^j}} + B_{\perp}^i.
	\end{equation}
	Then the magnetic energy can be decomposed
	into toroidal and poloidal components
	$E_{\mathrm{mag}} = E_{\mathrm{mag}}^{||} + E_{\mathrm{mag}}^{\perp}$,
	where these quantities can be computed 
	independently as
	\begin{equation}
		\label{eq:Emag}
		E_{\mathrm{mag}} = \int T^{00}_{\mathrm{EM}} \alpha \sqrt{-g} \, dV,
	\end{equation}
	\begin{equation}
		\label{eq:Emag_par}
		E_{\mathrm{mag}}^{||} = \int \frac{1}{2} B_{||} B_{||} \frac{\sqrt{-g}}{\alpha} \, dV,
	\end{equation}
	\begin{equation}
		\label{eq:Emag_perp}
		E_{\mathrm{mag}}^{\perp} = \int \frac{1}{2} g_{ij} B_{\perp}^i B_{\perp}^j \left( 1 + g_{i j} \tilde{v}^i \tilde{v}^j\right) \frac{\sqrt{-g}}{\alpha} \, dV.
	\end{equation}
	In Appendix~\ref{sec:bfield_decomp} we validate the interpretation of
	$E_{\mathrm{mag}}^{||}$ and $E_{\mathrm{mag}}^{\perp}$ 
	as toroidal and poloidal components of the magnetic energy
	for accretion disks.
	At the time of transition, for \BNSLRto,
	we find $E_{\mathrm{mag}}=3.136\e{48}~\mathrm{erg}$,
	with $E_{\mathrm{mag}}^{||}=2.417\e{48}~\mathrm{erg}$ and 
	$E_{\mathrm{mag}}^{\perp}=7.192\e{47}~\mathrm{erg}$
	, i.e. we find the toroidal
	field is dominant.
	As a comparison, for the FM case, at $t=0$
	and $t=9000M$, respectively,
	we find $E_{\mathrm{mag}}=1.197\e{53}~\mathrm{erg}, 
	\ 1.928\e{55}~\mathrm{erg}$,
	with $E_{\mathrm{mag}}^{||}=1.297\e{52}~\mathrm{erg}, 
	\ 9.497\e{54} \mathrm{erg}$ and
	$E_{\mathrm{mag}}^{\perp}=1.067\e{53}~\mathrm{erg},
	\ 9.791\e{54}~\mathrm{erg}$, 
	i.e. it transitions from poloidal dominance 
	to equipartition. 
	Recall we have freedom of scale over the system of units	in the FM disk, so
	only relative comparisons of $E_{\mathrm{mag}},
	E_{\mathrm{mag}}^{||}, E_{\mathrm{mag}}^{\perp}$
	within each run are meaningful.

	\begin{figure}[htb!]
        	\centering
        	\includegraphics[width=\columnwidth]{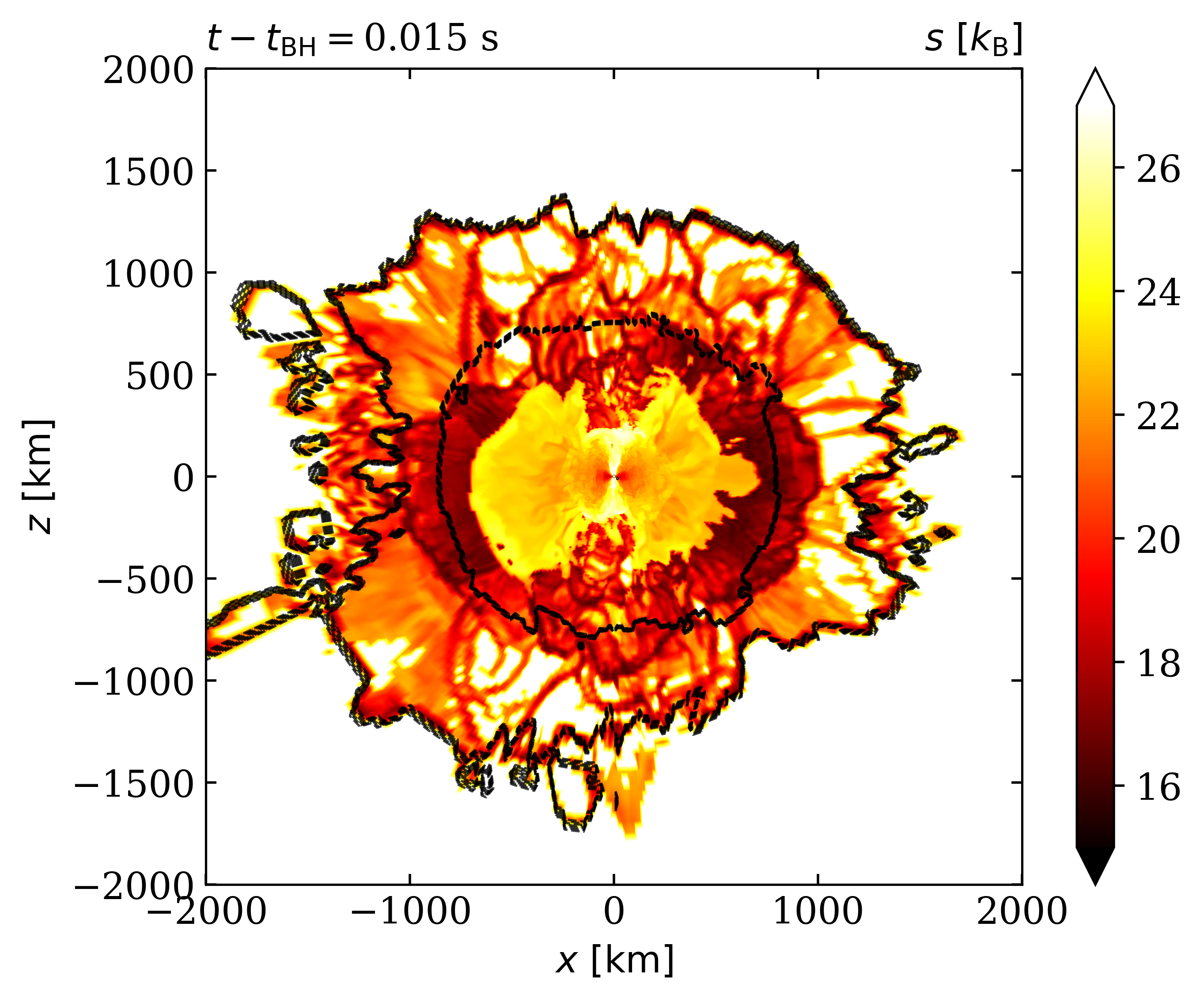}  
        	\caption{Specific entropy of dynamical ejecta
		for \BNSLRto.
		Unbound matter with outgoing radial velocity is enclosed
		within \textit{black, dashed} curves.}
        \label{fig:ejecta}

        \centering
        \includegraphics[width=\columnwidth]{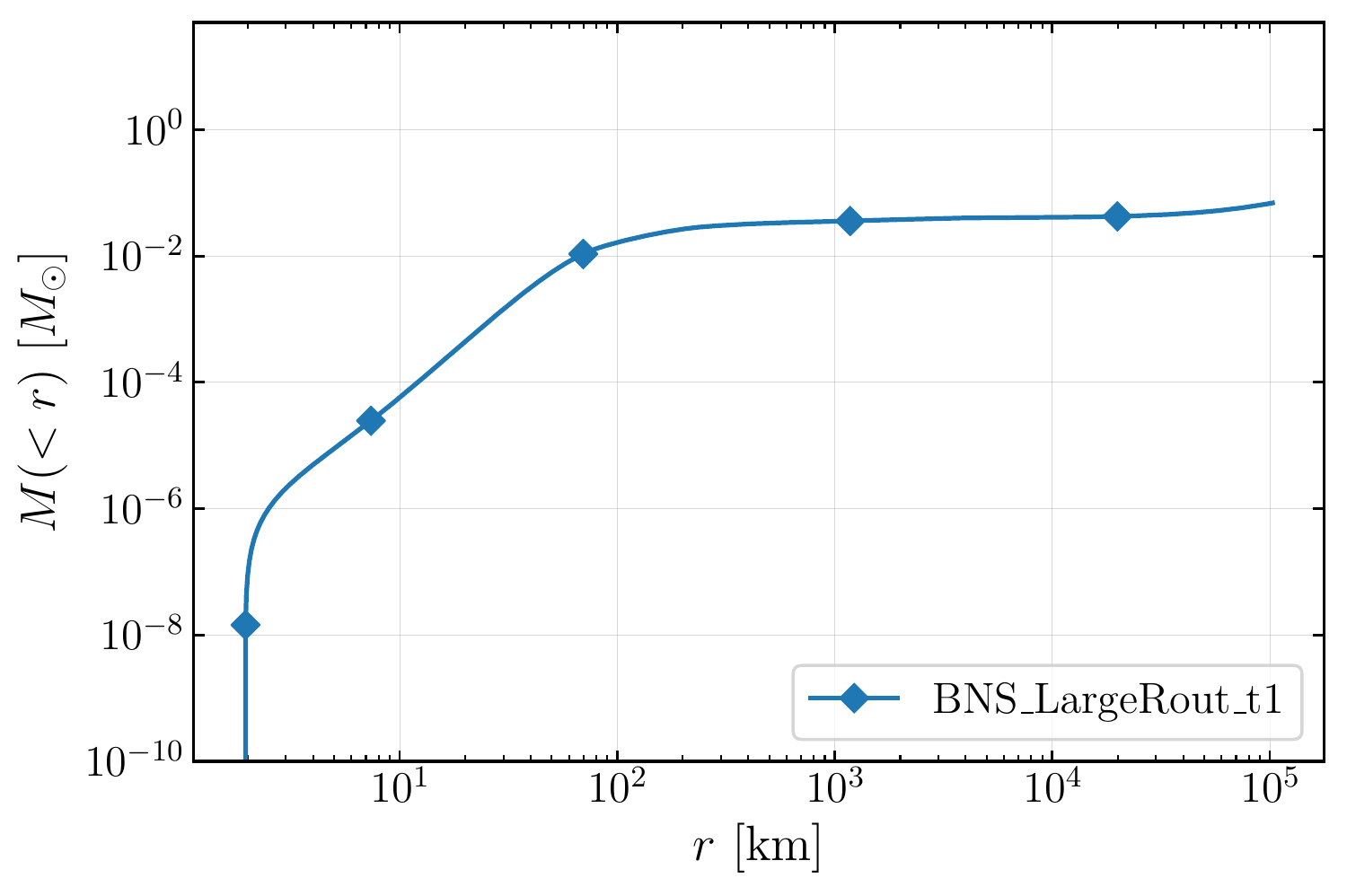}  
		\caption{Integrated mass profile (see Eq.~\ref{eq:mass_r})
		for \BNSLRto at the time of hand off. We notice that, 
		even though the domain is large, the integrated mass
		is dominated by the ejecta; the numerical atmosphere
		does not contribute significantly.}
        \label{fig:BNS_mass_r}
    \end{figure}
    
	In addition to the accretion disk, the ID obtained
	from the \handoff includes the unbound debris
	from the merger (i.e., the dynamical ejecta).
	There are two main mechanisms of mass ejection
	during the merger, tidal interactions and 
	shock heating \citep[see, for instance, Ref.][]{Sekiguchi+2015}.
	While the first mechanism predominantly expels
	cool material from the NSs to the orbital plane
	of the binary, the second ejects heated material
	quasiisotropically.
	In Fig.~\ref{fig:ejecta} we visualize the dynamical
	ejecta for \BNSLRto 
	by plotting the specific entropy, along
	with dashed
	lines that contain the ejected material.
	We identify the ejecta as the parcels of fluid that
	are unbound, satisfying $(h+b^2/\rho) u_t < -1$,
	and move outwards, satisfying $v^r > 0$.
	Indeed, we recognize regions with low entropy 
	around the equator, which we associate with
	the early tidal ejecta, and a quasispherical
	region with higher entropy, which we associate with
	the shocked-heated ejecta.
	We find the total ejecta have a mass of $0.002 M_{\odot}$,
	and an average radial velocity of $0.15 c$.
	These values are consistent with the ranges
	found in the literature \citep{Hotokezaka+2013, Sekiguchi+2015,
	Shibata+2019},
	supporting the validity of our methods.
	Regarding our specific model, the total mass of the ejecta
	is likely to be overestimated since we do not take
	neutrino cooling into account and this excess of internal energy
	increases the amount of unbound material \citep{Baiotti+2017}.
	In Fig.~\ref{fig:BNS_mass_r} we plot the
    mass profile
    \begin{equation}
        \label{eq:mass_r}
        M(<r) = \int^r \rho \sqrt{-g} \, dV
    \end{equation}
    for the radial extent of the domain and 
    confirm that the enclosed mass in
    the simulation is dominated by the ejecta--
    the numerical atmosphere has a negligible contribution, even for such a large domain.

	\begin{figure}[htb!]
        	\centering
        	\includegraphics[width=\columnwidth]{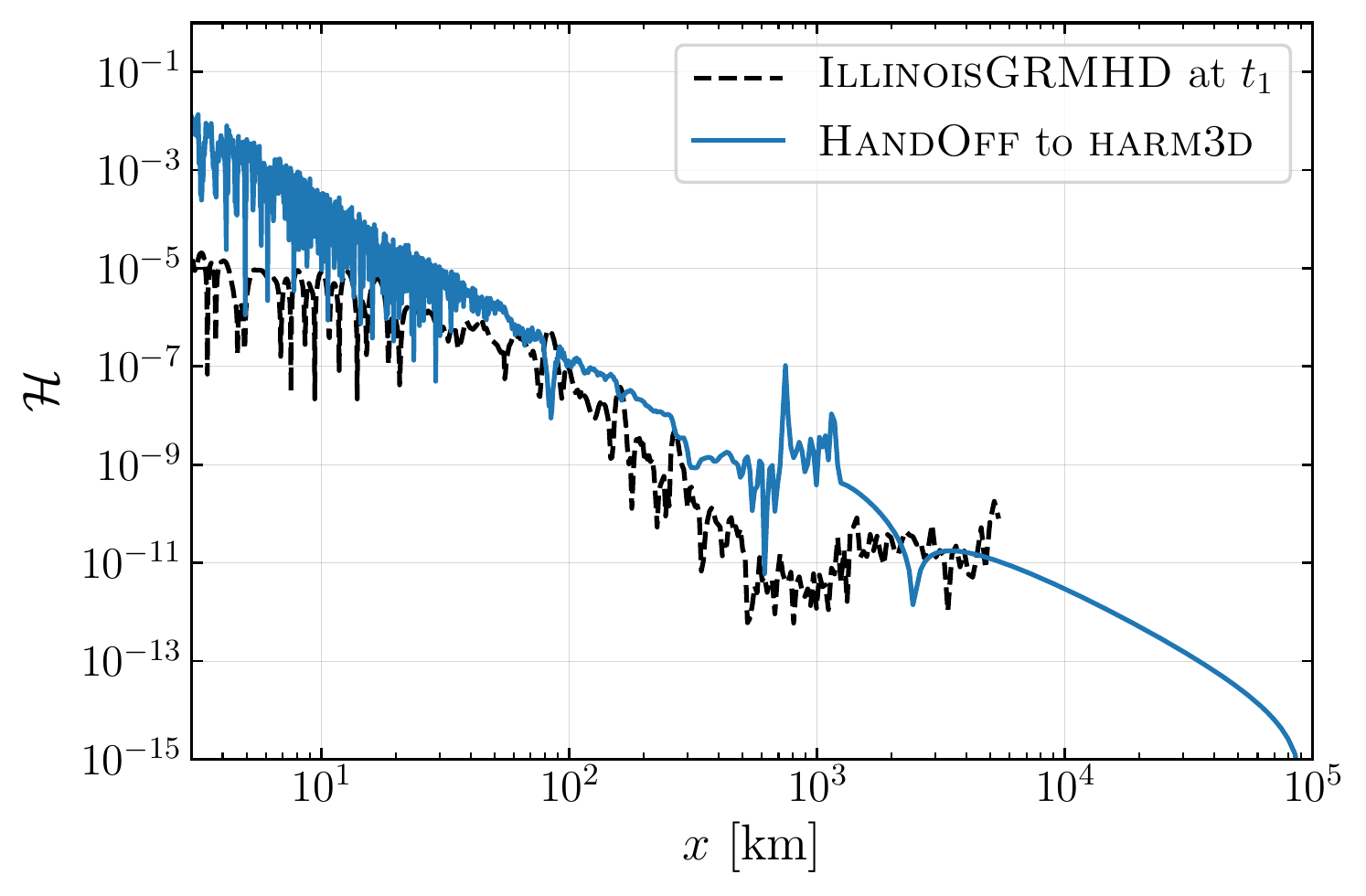}  
        	\caption{Hamiltonian constraint
        	$\mathcal{H}$
        	at the $x$-axis at the
        	time of transition for
        	\BNSLRto. We plot both the
        	constraints for the evolved metric
        	in \igm (\textit{dashed, black}),
        	and for the interpolated/extrapolated metric
        	in \harm after the \handoff
        	(\textit{solid, blue}).}
         	\label{fig:Hamiltonian} 
    \end{figure}
    As mentioned, the destination grid of \BNSLRto
	has its outer boundary
	much farther away than the original grid of \igm.
	Then, as described in Sec.~\ref{sec:extrapolation}, 
	we need to extrapolate the primitives and 
	spacetime metric.
	For the ``trusted window'' in the extrapolation 
	of the spacetime metric,
    we use $R1= 500\Msun$ ($\sim 750 \rm{km}$) and $R2= 800\Msun$ ($\sim 1200 \rm{km}$).
    
    To	test the validity of both the
	interpolated and extrapolated geometry, in
	Fig.~\ref{fig:Hamiltonian} we plot
	the Hamiltonian constraint along the $x$-axis
	for \igm at the time
	of transition, and for \harm after the \handoff.
	For the latter,
	we neglect the contribution of
	matter and assume the spacetime is static.
	In the extrapolated region, we find the values of $\mathcal{H}$ to be
	comparable between both codes, 
	although some deterioration in the transition region for \harm.
	In the accretion disk region, we find the values of $\mathcal{H}$
	to be comparable between both codes, ensuring the validity of the spacetime
	metric.
	However, in the region close to the BH, we find the values of $\mathcal{H}$ 
	to be worst for \harm after the \handoff. 
	A similar behavior was found for the momentum constraints.
	In the following
	we discuss why this is not a significant concern.
	
	\begin{figure}[htb!]
        	\centering
        	\includegraphics[width=\columnwidth]{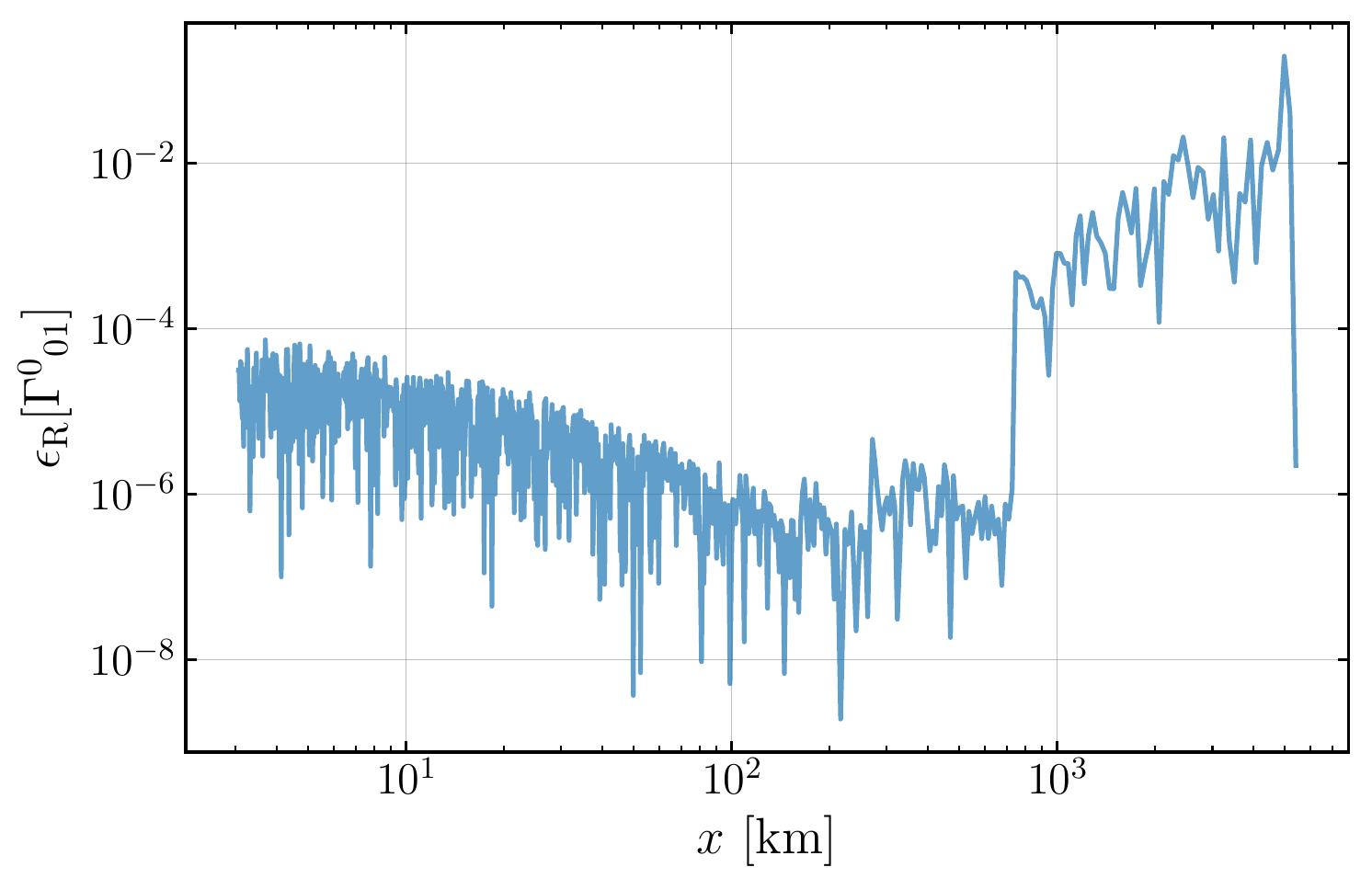}  
        	\caption{Relative error (see Eq.~\ref{eq:epsilon_R}) between the affine
        	connection ${\Gamma^0}_{01}$ in \igm and its values as calculated 
        	from the interpolated/extrapolated metric in \harm.}
         	\label{fig:connection} 
    \end{figure}
    
	Although the spacetime metric is transferred
	with high-fidelity, we noticed that in regions where the destination grid
	has higher resolution the interpolant function have regions
	of size $dx^{\mathrm{IGM}}$ in its domain where the functional
	form is that of a polynomial, and second order derivatives are not trustworthy.
	This deteriorates the values of the constraints near the BH.
	However, only first order derivatives of the spacetime metric play a role
	in the evolution equations, via the affine connections 
	${\Gamma^{\lambda}}_{\mu \nu}$.
	Comparing the relevant connections in \igm with those recalculated in \harm
	from the interpolated/extrapolated metric, we find they agree with a relative error
	of order $\sim{}10^{-2}$ or lower in the BH surroundings, ensuring an equivalent
	gravitational field after the \handoff.
    As an example of this, in Fig.~\ref{fig:connection} we plot the relative errors for the case of
    ${\Gamma^0}_{01}$ along the x-axis.
    Noticeably, the relative errors raise in the region where the numerical 
    metric is replaced by the extrapolated values, but the connections still remain
    comparable.
	Indeed, in the next subsection we will show 
	that the evolution of the MHD fields
	in the extrapolated spacetime 
	has the expected physical behavior, and
	matches the results from 
	\BNSSRto, where extrapolation
	is not required and the numerical metric
	is taken directly from interpolation.
	In particular, we will compare
	the radial velocity $v^r$ of the 
	expanding ejecta of \BNSLRto in the
	extrapolated region with those 
	results for \BNSSRto.
	
	To monitor the degree of axisymmetry of the
	transitioned spacetime metric we calculate 
	the relative deviations of each local element of volume 
	$\sqrt{-\left| g(r,\theta,\phi) \right|}$ from the corresponding
	$\phi$-average $\left<\sqrt{-\left| g(r,\theta,\phi)  \right|} \right>_{\phi}$
	for \BNSLRto,
	and find that:
	\begin{equation}
	    \frac{ \left| \sqrt{- \left| g(r,\theta,\phi)\right| }  - \left<\sqrt{-\left| g(r,\theta,\phi)\right|}\right>_{\phi} \right|}{\left| \left<\sqrt{-\left| g(r,\theta,\phi)\right|}\right>_{\phi} \right|} < 3\e{-3}.
	\end{equation}
	Furthermore, to monitor the degree of stationarity of the
	transitioned spacetime metric we calculate the relative
	deviations of each local element of volume for 
	\BNSSRtz and \BNSSRto which make use of the same destination
	grid but transitioned at the different times $t_0$ and $t_1$,
	respectively. We find that:
	\begin{equation}
	    \frac{ 2 \left| \sqrt{- \left| g(t_0,r,\theta,\phi)\right| }  - 
	    \sqrt{-\left| g(t_1,r,\theta,\phi)\right|} \right|}{\left| \sqrt{-\left| g(t_0,r,\theta,\phi)\right|} \right| + \left| \sqrt{-\left| g(t_1,r,\theta,\phi)\right|} \right|} < 2\e{-2}.
	\end{equation}
	In this sense, we argue that the spacetime is
	approximately axisymmetric and static for the continued evolution
	in \harm.

	\subsection{Handing off the postmerger: Continued evolution}
    We evolve \BNSLRto
	in \harm for $0.1~\mathrm{s}$, i.e.
	up to $0.115~\mathrm{s}$ after BH formation.
	As we describe below,
	during this period we find the BH accretes $0.023~M_{\odot}$
	(${\sim}27\%$ of the initial mass of the torus), the accretion
	rate converges to ${\sim}0.1~M_{\odot}s^{-1}$
	, the magnetic energy
	increases by ${\sim}50\%$, and the ejecta expand freely.
	We do not find significant magnetic
	outflows from the ergosphere of the
	remnant.
	We could have evolved the system for longer,
	but were limited by computational resources.
	We used 
	229 nodes of the the supercomputer 
	\texttt{Frontera} \footnote{Visit https://frontera-portal.tacc.utexas.edu/},
	for approximately 5 days, 
	implying a system usage of almost
	30k~SUs (System Units).
	Considering that the number of cells is 
	$1024\times200\times400$
	(see Appendix~\ref{sec:grid}), and that each
	node has $56$ processors, our code
	runs at a pace of almost
	70k cell updates per second per processor.
	For consistency checks, we also evolve 
	\BNSSRtz and \BNSSRto
	for approximately $0.04~\mathrm{s}$.

\begin{figure}[htb!]
  \centering
  \includegraphics[width=\columnwidth]{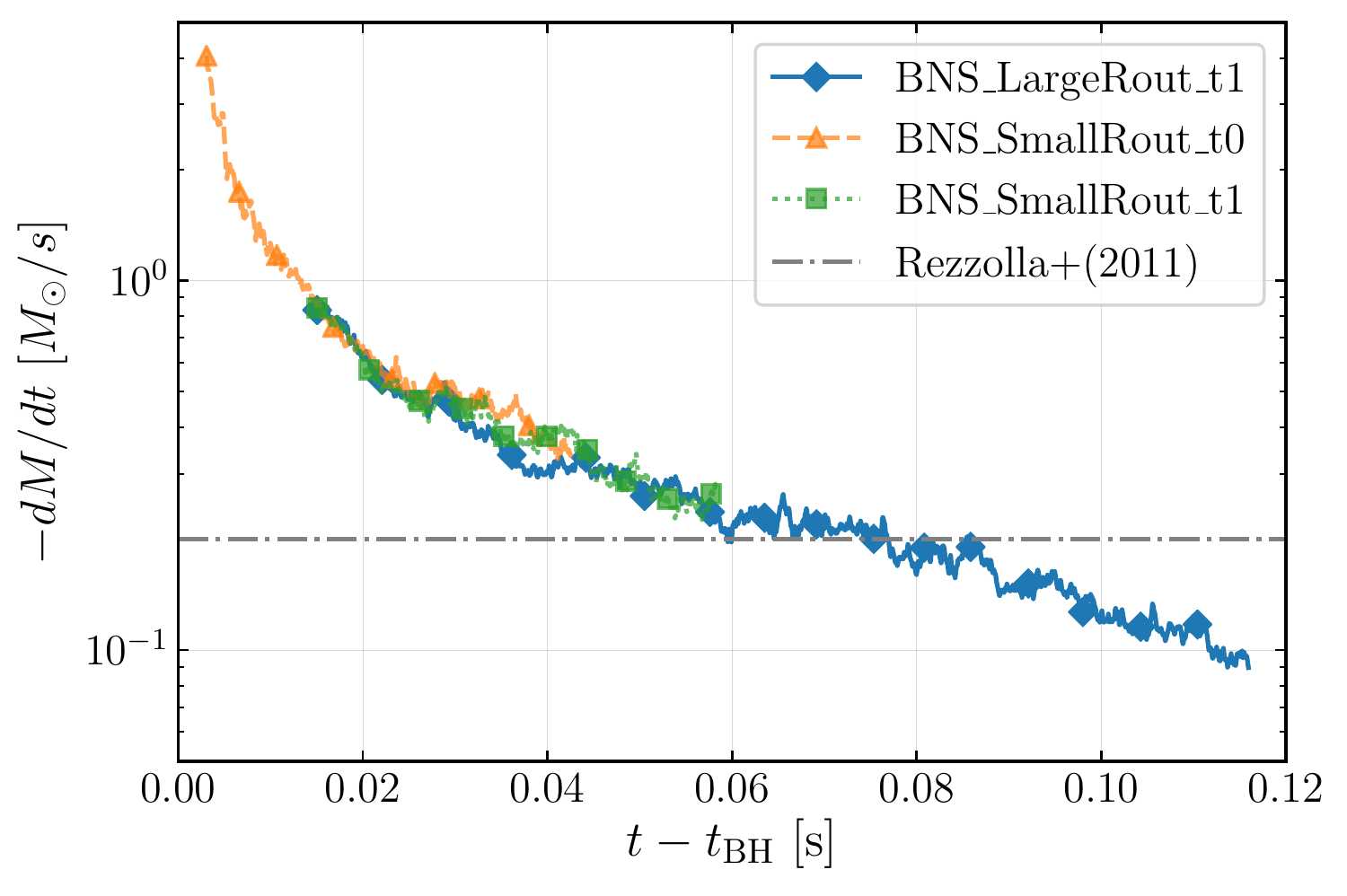}
  \caption{Accretion rate at the horizon 
	for the BNS postmerger simulations, and the reference
	value of~\cite{Rezzolla+2011}.
	In every case, the transition is continuous,
	proving the robustness of the \handoff.}
  \label{fig:BNS_mdot_t}
\end{figure}

	In Fig.~\ref{fig:BNS_mdot_t} we plot the accretion
	rate at the horizon (${\sim}2.65~\mathrm{km}$)
	for \BNSSRtz, \BNSSRto and \BNSLRto.
	The plot demonstrates the robustness of the \handoff in
	the following senses.
	First, the initial values of the accretion rate
	for the later transitions at $t_1$,
	\BNSSRto and \BNSLRto,
	match the accretion rate of \BNSSRtz
	at time $t_1$, proving both that the \handoff
	does not introduce unphysical transients and
	that the evolution of \BNSSRtz follows 
	the continuing run in \igm.
	Second, the plot shows that the extrapolation of
	primitives and spacetime metric for \BNSLRto
	does not lead to unphysical results but
	matches the results from runs where 
	extrapolation was
	not needed.
	Third, the accretion rate approximates to the
	reference value
	${\sim}0.2~\Msun s^{-1}$ \citep{Rezzolla+2011}.
\begin{figure}[htb!]
  \centering
  \includegraphics[width=\columnwidth]{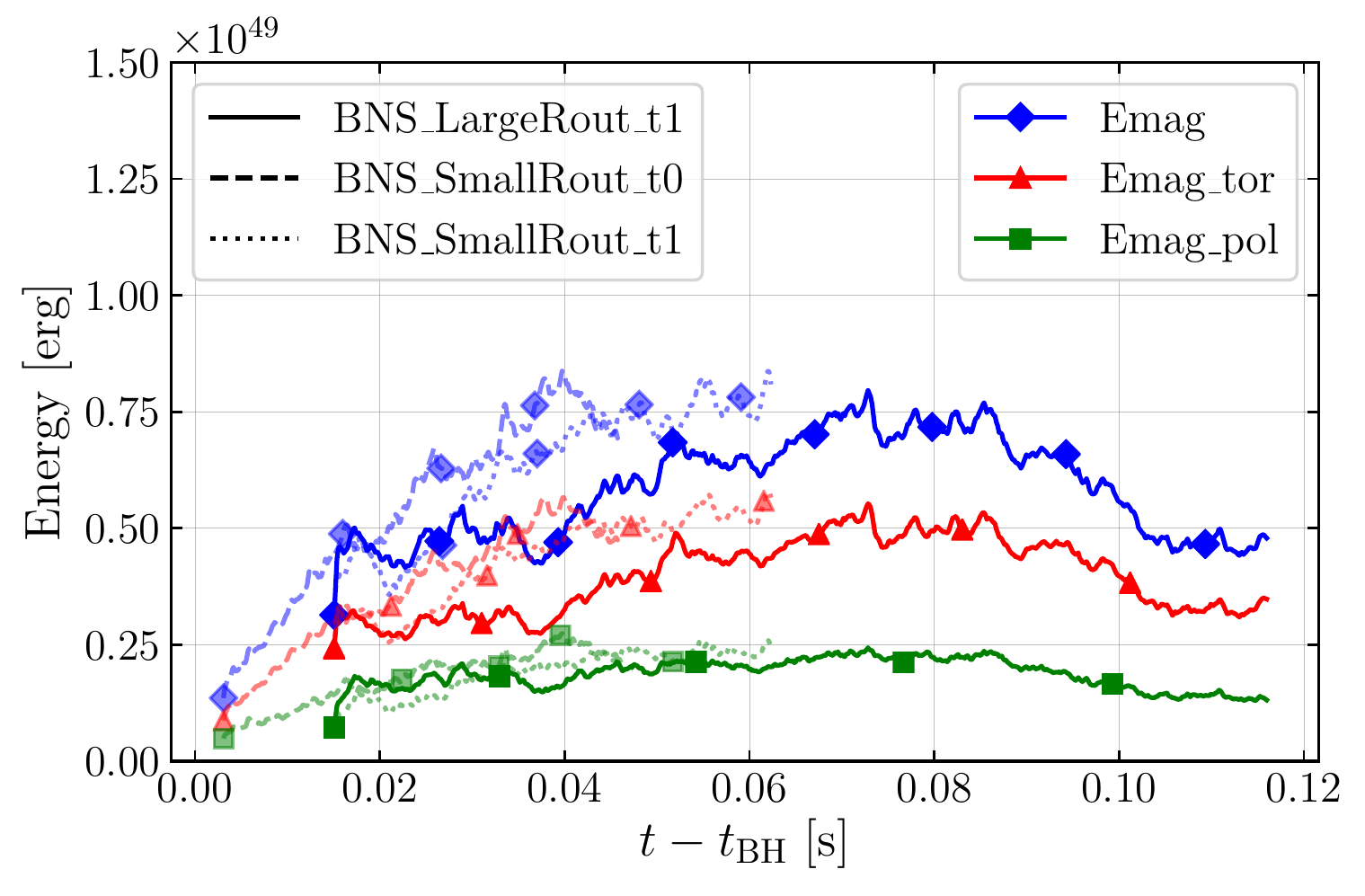}
  \includegraphics[width=\columnwidth]{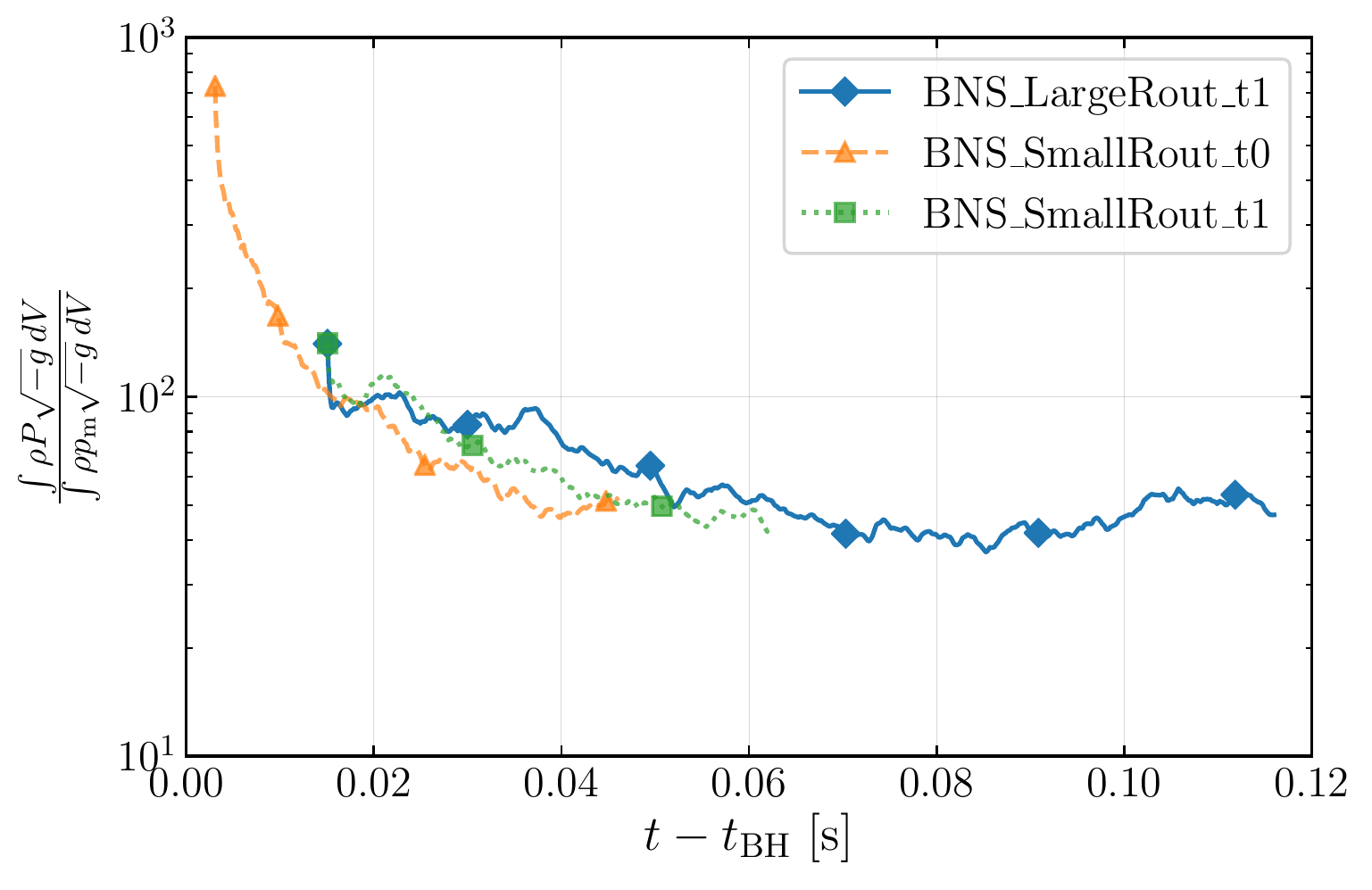}
	\caption{\textit{Top:} Magnetic energy of BNS postmerger
	runs (\textit{blue}), decomposed into toroidal (\textit{red})
	and poloidal (\textit{green}) components.
	\textit{Bottom:} Ratio of the $\rho$-weighted 
	integrals of thermal and magnetic pressure,
	as a measure of the magnetization of the disk for
	BNS postmerger runs.}
  \label{fig:BNS_Emag_t}
\end{figure}

	In Fig.~\ref{fig:BNS_Emag_t} (\textit{top})
	we plot the evolution of the magnetic energy for
	our postmerger simulations,
	decomposed into toroidal and poloidal components 
	(see Eqs.~\ref{eq:Emag}-\ref{eq:Emag_perp}),
	integrated out of the BH horizon.
	For our fiducial run, 
	\BNSLRto, we find 
	$E_{\mathrm{mag}}~\sim~10^{49}~\mathrm{erg}$, 
	with a dominant toroidal component.
	Our results in this respect are not 
	comparable with the reference simulations of Refs. 
	\cite{Rezzolla+2011} or \cite{Kawamura+2016},
	because their initial magnetic field was weaker than ours by three
	orders of magnitude.
	Instead, we consider the simulation 
	\texttt{H4B15d150} of Ref. 
    \cite{Kiuchi+2014},
	that has an initial field of
	similar strength, makes use of
	comparable resolution during the merger,
	and, though it uses a different EOS,
	the HMNS collapses to a BH rather promptly.
	With respect to this simulation,
	we find a consistent order of magnitude for
	$E_{\mathrm{mag}}$.
	Further, the dominance of the toroidal component after
	merger is a well-known result from the literature
	\citep[see, for instance, Ref. ][]{Kawamura+2016}.
	From these results, we conclude the \handoff
	captures the correct strength and topology of
	the magnetic field after merger.
	In Appendix~\ref{sec:resolution} we present resolution
	diagnostics that demonstrate that the MRI is properly
	resolved during our runs.

	To better understand the effect of
	transitioning between 
	codes on the magnetization of the disk, in
	Fig.~\ref{fig:BNS_Emag_t} (\textit{top})
	we also plot the
	curves for the earlier transition 
	\BNSSRtz, and for \BNSSRto.
	For every simulation, we
	notice the magnetic energy grows
	initially, as expected
	from magnetic winding and the MRI,
	but we notice that the curves for the later 
	transitions, \BNSSRto and \BNSLRto,
	do not match the curves of \BNSSRtz at $t_1$.
	Instead, the later transitions are initialized
	to a lower value than \BNSSRtz at $t_1$,
	indicating that	the magnetic growth is faster
	in \harm than in \igm, plausibly because
	the grid of the former has higher resolution 
	and spherical topology, better resolving the
	magnetic winding effects and the MRI.
	In fact, we notice that the magnetic energy
	of the later transitions
	have a very steep initial growth, until they
	catch up with the evolved values in \BNSSRtz.

	To measure the dynamical relevance of the
	magnetic field in the disk, in 
	Fig.~\ref{fig:BNS_Emag_t} (\textit{bottom})
	we plot the evolution of the
	ratio of the $\rho$-weighted integrals 
	of thermal and magnetic pressure (see
	Eq.~\eqref{eq:int_beta_rho}).
	We find that the dynamical relevance of the
	field grows in time, until saturation at
	${\sim}50$.
	As a comparison,
	the FM evolved in the last section,
	has ${\sim}30$ for this ratio, 
	at $t=9000M$.
	
\begin{figure*}[htb!]
        \centering
        \includegraphics[width=\columnwidth]{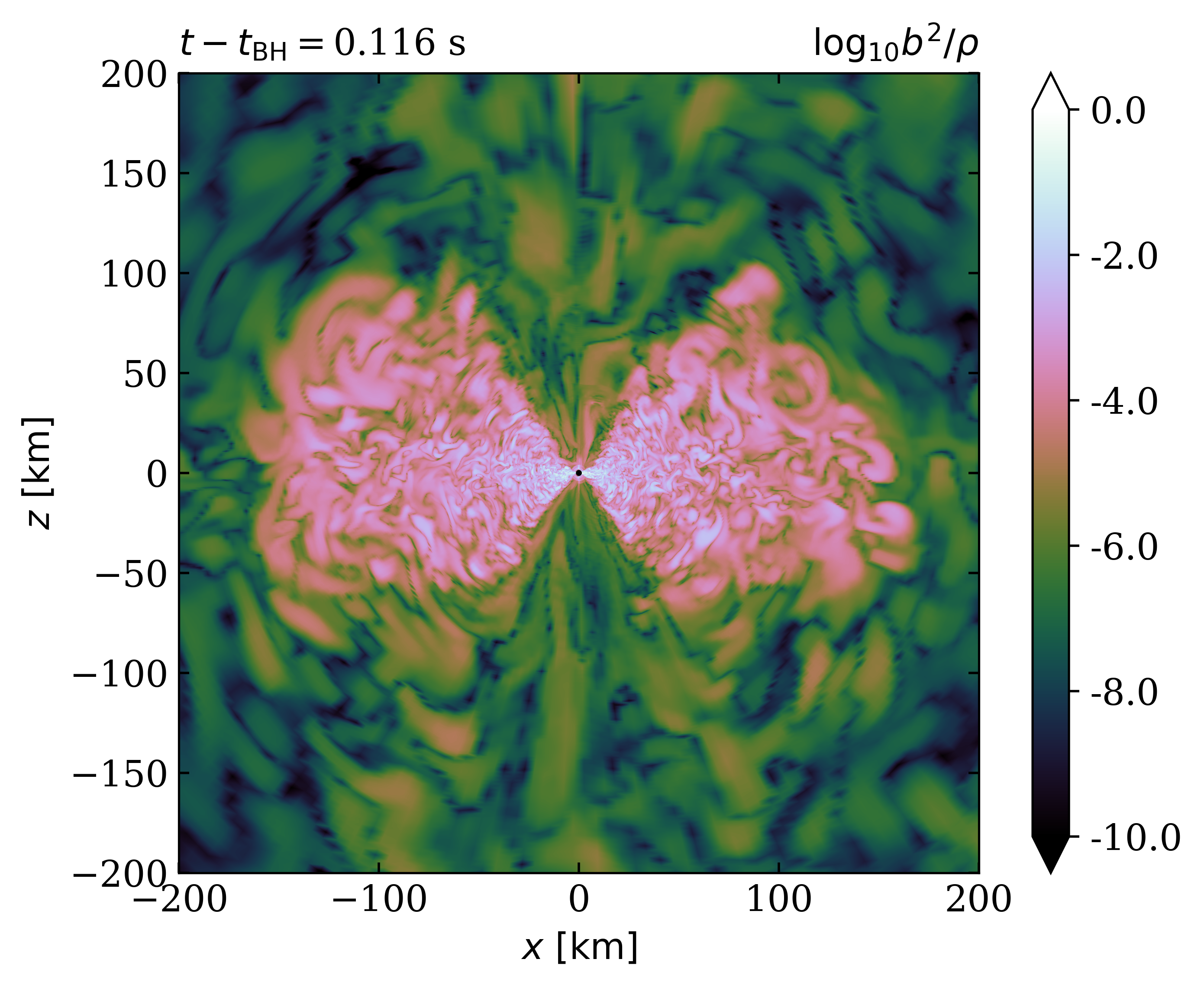}
        \includegraphics[width=\columnwidth]{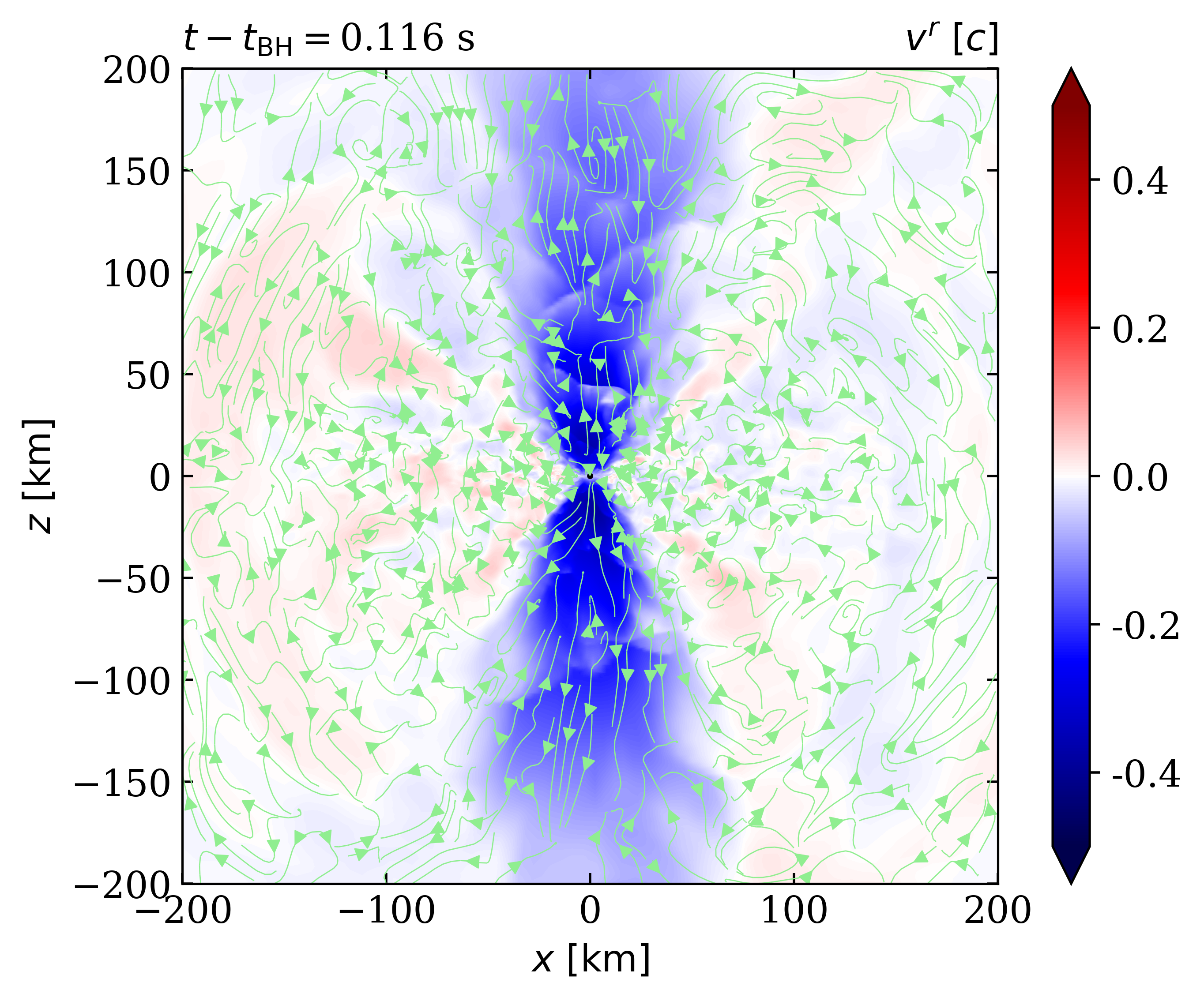}
	\caption{\textit{Left}: Magnetization $b^2/\rho$
	at the end of \BNSLRto. We notice that, although the
	disk is fairly magnetized,
	the funnel is not magnetically dominated.
	\textit{Right}: Radial velocity $v^r$ and magnetic
	field lines (\textit{green}) at the end of \BNSLRto.
	We notice the absence of outflows at the funnel, 
	and that magnetic field lines are poorly ordered.}
    \label{fig:outflow} 
\end{figure*}
	In Fig.~\ref{fig:outflow} we show a poloidal
	plot of the magnetization $b^2/\rho$ (\textit{left}),
	and of the radial velocity of the fluid (\textit{right})
	with magnetic field lines (\textit{green arrows}), 
	at the end of our fiducial run.
	Although the magnetization of the disk and the
	spin of the BH are rather high, we notice that
	the material in the funnel is not magnetically 
	dominated and a jet has not developed.
	This is in agreement with results
	from \cite{Kawamura+2016}.
	Several reasons might help to explain the
	absence of magnetic outflows:
	accretion of predominantly toroidal
	fields disfavours the formation
	of large-scale magnetic fields at the funnel 
	and its subsequent amplification 
	\citep[see, for instance, ][]{DeVilliers+2005,Beckwith+2008,McKinney+2012};
	while a long-lived NS remnant can build 
	a strong helical field at the funnel,
	prompt collapse prevents this scenario 
	\citep{Ciolfi2020a, Ciolfi2020b}; 
	we ignore the drag from neutrino winds, 
	which might help lower the degree of baryon
	pollution in the polar regions 
	\citep{Mosta+2020}.
	Still, in Fig.~\ref{fig:outflow} (\textit{right})
	we notice a moderate ordering of the magnetic
	field lines at the funnel, plausibly
	stretched by the inflowing material.
	A mildly relativistic jet might arise
	in the longer-term  \citep[see][and refs.~therein]{Ciolfi2020b}.
	While the magnetic outflows are supressed,
	Fig.~\ref{fig:outflow} (\textit{right}) 
	shows outgoing parcels of fluid,
	or \textit{winds}, 
	at the boundary of the funnel and the
	disk \citep{DeVilliers+2005}.

\begin{figure}[htb!]
	\centering
	\includegraphics[width=\columnwidth]{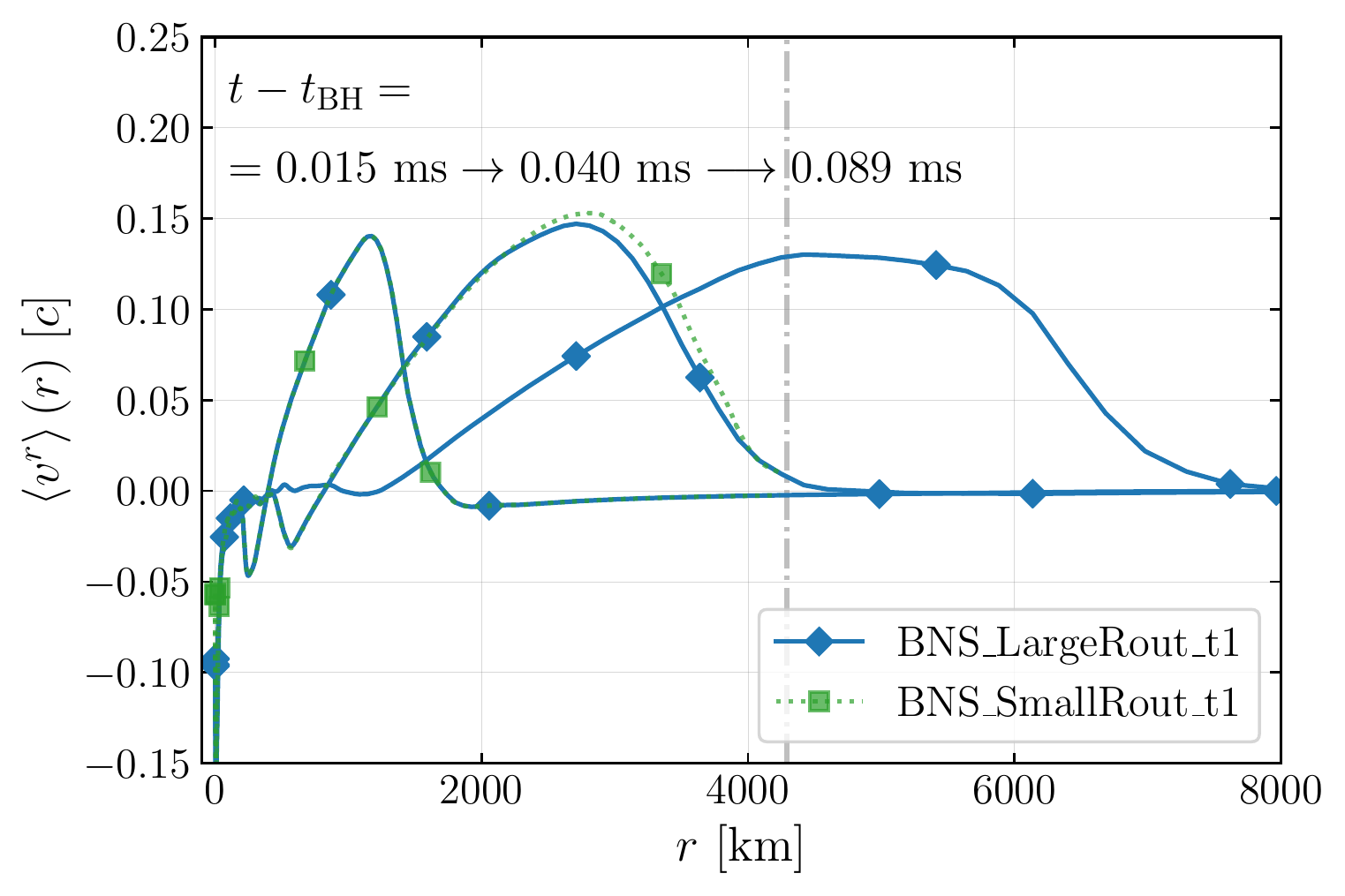}
	\caption{Radial velocity of the ejecta, averaged in
	$\theta$ and $\phi$, at three different times
	$t-t_{\mathrm{BH}}=0.015,\, 0.04,\, 0.089\, \mathrm{ms}$ (\textit{left to right}).
	The \textit{gray,dotted-dashed} line 
	represents the outer boundary of
	\BNSSRto.
	We notice that the ejecta expand 
	freely through the domain, indicating
	that the numerical
	atmosphere and extrapolated metric (\BNSLRto) introduced
	in the \handoff are physically valid.
	The qualitative agreement between the curves of
	\BNSSRto and \BNSLRto reflects the near equivalence
	between the interpolated numerical metric and
	extrapolated metric.}
 	\label{fig:BNS_vr_far} 
\end{figure}	
	Finally, we demonstrate the physical validity 
	of the extrapolated metric.
	In Fig.~\ref{fig:BNS_vr_far} we
	plot the radial velocity $v^r$
	for \BNSLRto and \BNSSRto, 
	averaged in $\theta$ and $\phi$:
	\begin{equation}
	    \left \langle v^r \right \rangle (r)= 
	    \frac{\int v^r \sqrt{-g}\, d\theta\, d\phi}{\int \sqrt{-g}\, d\theta\, d\phi}.
	\end{equation}
	We plot $\left \langle v^r \right \rangle (r)$ far from the BH, at the boundary of the ejecta with the numerical
	atmosphere, at three different times:
	$t-t_{\mathrm{BH}}=0.015,\, 0.04,\, 0.089\, \mathrm{s}$ (\textit{left to right}).
	The \textit{dotted-dashed} line in Fig.~\ref{fig:BNS_vr_far}
	represents	the outer boundary of
	\BNSSRto ($\sim 4300\ \mathrm{km}$).
	First, we notice that in both runs the ejecta
	expand freely through the domain, 
	showing the continuity of 
	the handed-off transition.
	Next, we notice the curves for \BNSLRto
	and \BNSSRto at $t-t_{\mathrm{BH}}=0.015,\, 0.04\, \mathrm{s}$
	are qualitatively the same.
    Since \BNSLRto makes use of the extrapolated
    metric, and \BNSSRto makes use of
    the interpolated numerical metric, we
    conclude these metrics are physically equivalent.
    We include a later curve for \BNSLRto 
    at $t-t_{\mathrm{BH}}=0.089\, \mathrm{s}$
    to show that the
    \handoff allows for the expansion of
    the ejecta even further out 
    than the outer boundary of the merger
    simulation ($\sim 5700\ \mathrm{km}$).
    We do not include this curve for 
	\BNSSRto because we did not evolve this run
	for long enough.

\section{Discussion}
    The key motivations behind the \handoff package are the advantages of
    using spherical coordinates over Cartesian to model azimuthal flows.
    In this article, we argued that spherical coordinates
    are preferable regarding both 
    physical precision and computational performance.
    In this section, we provide proof and discussion around such motives.
    
    \subsection{Numerical dissipation}
    \label{sec:dissipation}
    
    In the lore of numerical simulations of fluid dynamics there is
    a general conviction that momentum is evolved with higher precision
    if it is aligned with the direction of coordinate lines
    \citep[for similar comments, see ][]{Call+2010, Mignone+2012, Byerly+2014, Mewes+2020}.
    On the contrary, if the fluid momentum crosses the coordinate
    faces obliquely, it will be subject to severer numerical errors.
    In this sense, simulations where linear momentum is most
    relevant are preferably simulated using Cartesian coordinates, but those that care
    on angular momentum conservation and transport are preferably simulated using coordinates with spherical topology.
    These numerical errors usually fall under the name of
    \textit{numerical dissipation}---not to be confused with explicit numerical dissipation schemes that damp modes in the solution with wavelengths equal to the grid spacing \cite{KreissOliger1973}--- and avoiding them in the case
    of postmerger accretion disks is one of the main motivations behind the \handoff.
    
    A general proof and measure of this phenomenon
    remains elusive, as numerical errors in the fluxes are also affected 
    by the grid resolution, the particular physical
    processes involved, and the numerical methods adopted.
    At the same time, fiducial exact solutions to turbulent MHD are
    scarce. 
    Numerical experiments that capture the effects of
    numerical dissipation are usually limited to equilibrium
    solutions with spacetime symmetries, 
    where some component of the momentum is to be conserved.
    For the case of a rotating star or an orbiting torus around a BH, 
    one of the causes of Cartesian numerical dissipation has proven to
    be the poor choice
    of the coordinate vector base for the momentum decomposition; 
    the best choice being 
    a vector base aligned with the conserved component $l_{\phi}$, 
    like spherical coordinates.
    This simplifies the source terms in the evolution equations,
    and  reduces the propagation of numerical errors \cite{Call+2010, Mignone+2012, Byerly+2014}.
    Plausibly there are other causes behind the Cartesian numerical dissipation
    in azimuthal flows,
    like truncation errors introduced when interpolating from the coarsest to finest grid at AMR boundaries.
    In this section we provide further numerical experiment to demonstrate
    the effects of Cartesian numerical dissipation in an orbiting torus.

    \subsubsection{Fishbone-Moncrief disk in hydrostatic equilibrium}
    \begin{figure*}[htb!]
      \label{fig:hydro_rho}
      Hydrodynamical Fishbone-Moncrief Test\par\medskip
      \centering
      \includegraphics[width=\columnwidth]{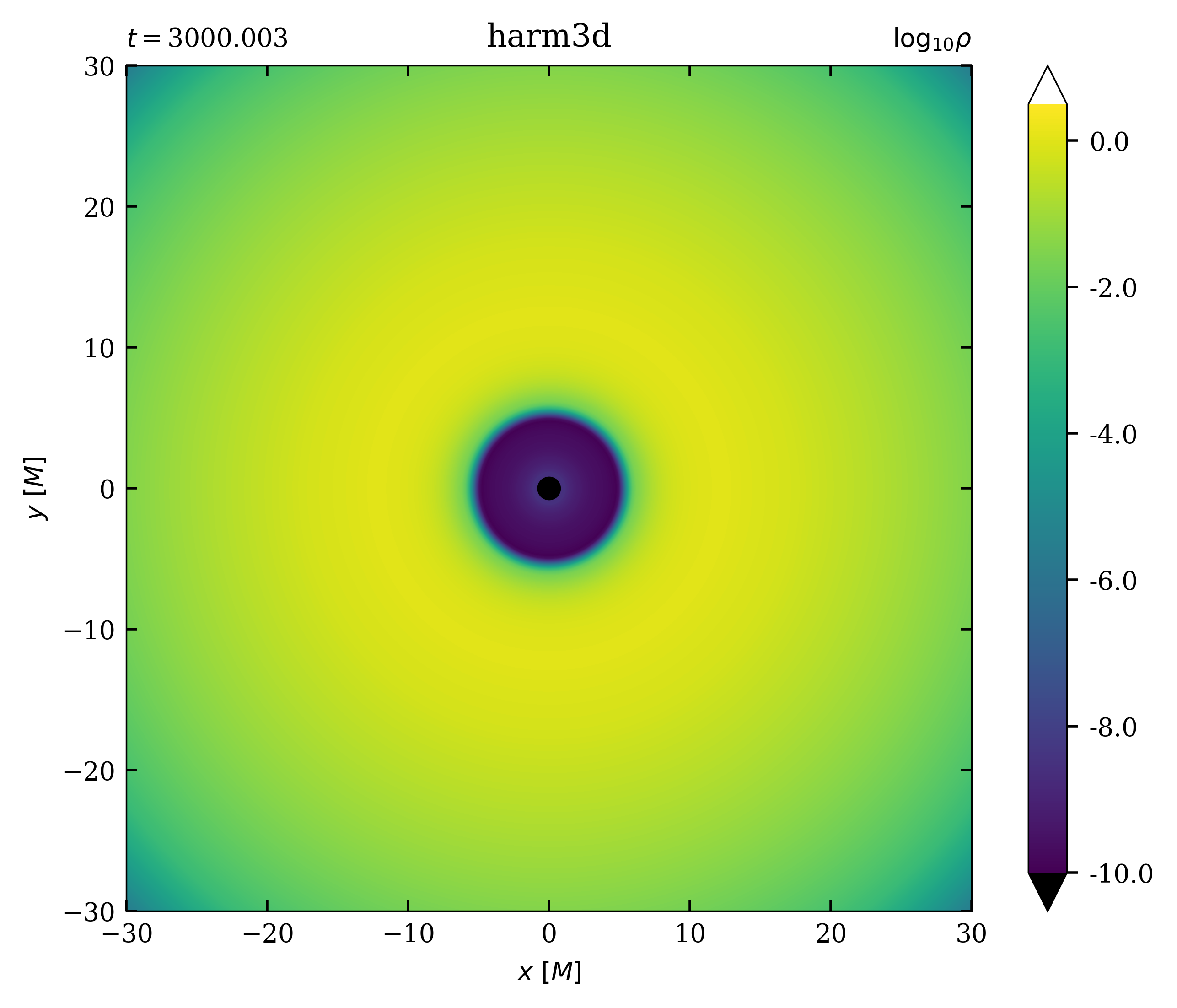}
      \includegraphics[width=\columnwidth]{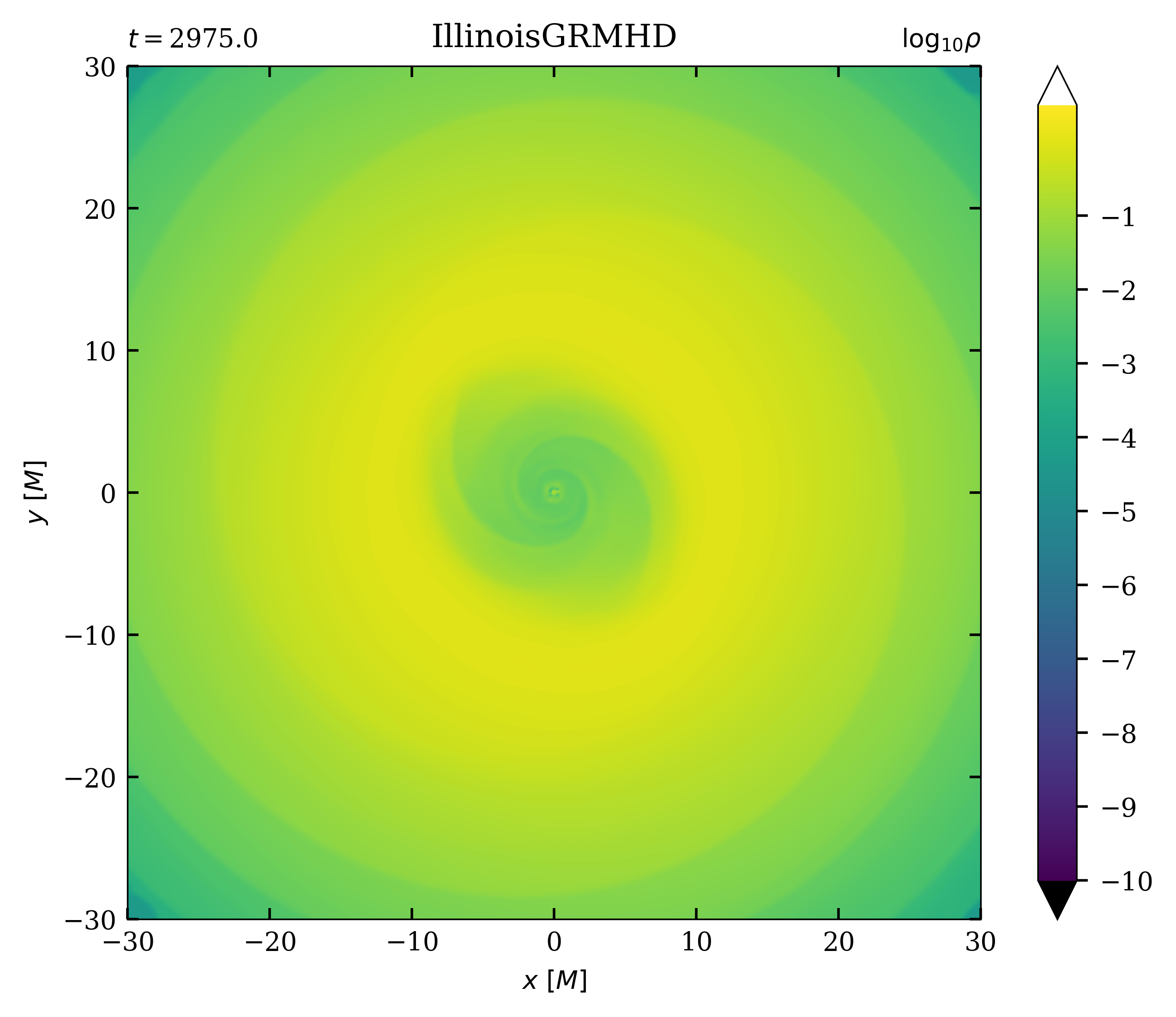}
      \caption{Rest-mass density $\rho$ at equatorial plane for hydrodynamical
      FM tests in \harm (\textit{left}) and \igm (\textit{right}), at
      time $t\sim{}3000M$. As a consequence of Cartesian numerical dissipation, \igm fails to maintain the initial hydrostatic equilibrium of the torus.}
      \label{fig:hydroRho}
    \end{figure*}
    \begin{figure}[htb!]
      \label{fig:hydro_mdot}
      \centering
      \includegraphics[width=\columnwidth]{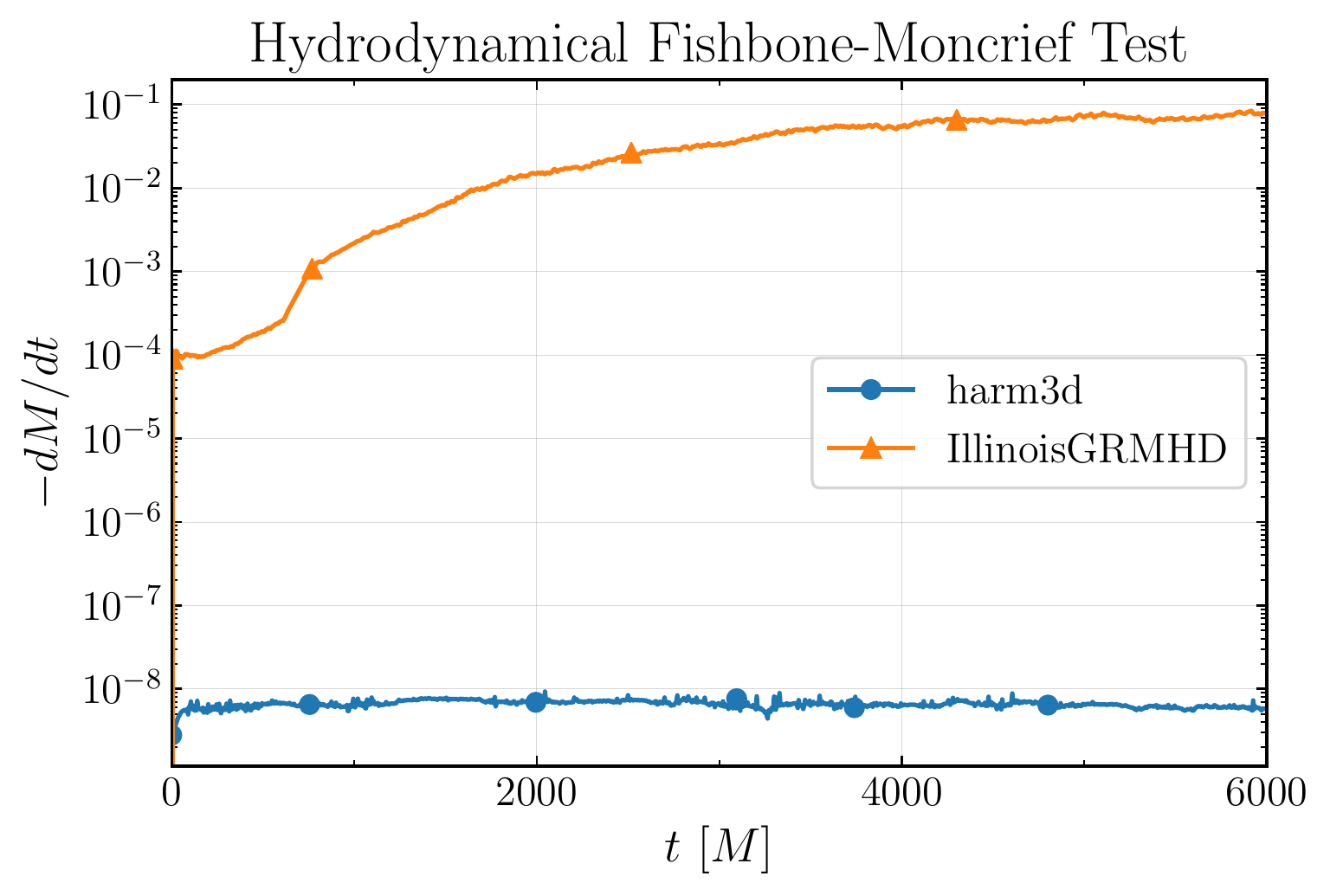}
      \caption{Accretion rate at the BH horizon as a function of time 
      for hydrodynamical FM tests in \harm (\textit{blue, circle}) 
      and \igm (\textit{orange, triangle}). Being the formal
      solution $-dM/dt = 0$, we notice that \harm with a
      spherical grid performs better than \igm by 7 orders of magnitude.}
      \label{fig:hydroRho}
    \end{figure}
    Following Sec.~\ref{sec:validation}, we initialize a FM disk in both \harm and \igm, with the same coordinate systems described in 
    such Section. Here, however, we nullify the initial magnetic field 
    and the initial perturbations to the gas pressure, so the ID is in hydrostatic equilibrium. 
    We evolve the system with both codes
    for $6000M$ in time, and proceed to analyze the deviations
    from equilibrium as a measure of numerical dissipation.
    
    In Fig.~\ref{fig:hydro_rho} we plot the density around the inner edge
    of the disk at the equatorial plane in both 
    \harm (\textit{left}) and \igm (\textit{right}) at
    $t=3000M$.
    We notice that \harm 
    manages to maintain the initial cylindrical symmetry of the torus, but
    \igm breaks the symmetry of the system developing spiral waves and
    angular momentum transport that causes accretion.
    Pressure-supported tori with constant specific angular
    momentum and entropy are unstable to nonaxisymmetric instabilities 
    \citep[see, for instance, ][]{Zurek+1986} and Cartesian
    numerical dissipation seems to be triggering those.
    To better quantify the physical relevance of such instability,
    in Fig.~\ref{fig:hydro_mdot} we plot the accretion rate through 
    the BH horizon in each case.
    Since the analytical expectation is zero, 
    we notice that
    \harm performs better than \igm by seven orders of magnitude.

    \subsubsection{Magnetized Fishbone-Moncrief disk}
    \begin{figure}[htb!]
      \centering
      \includegraphics[width=\columnwidth]{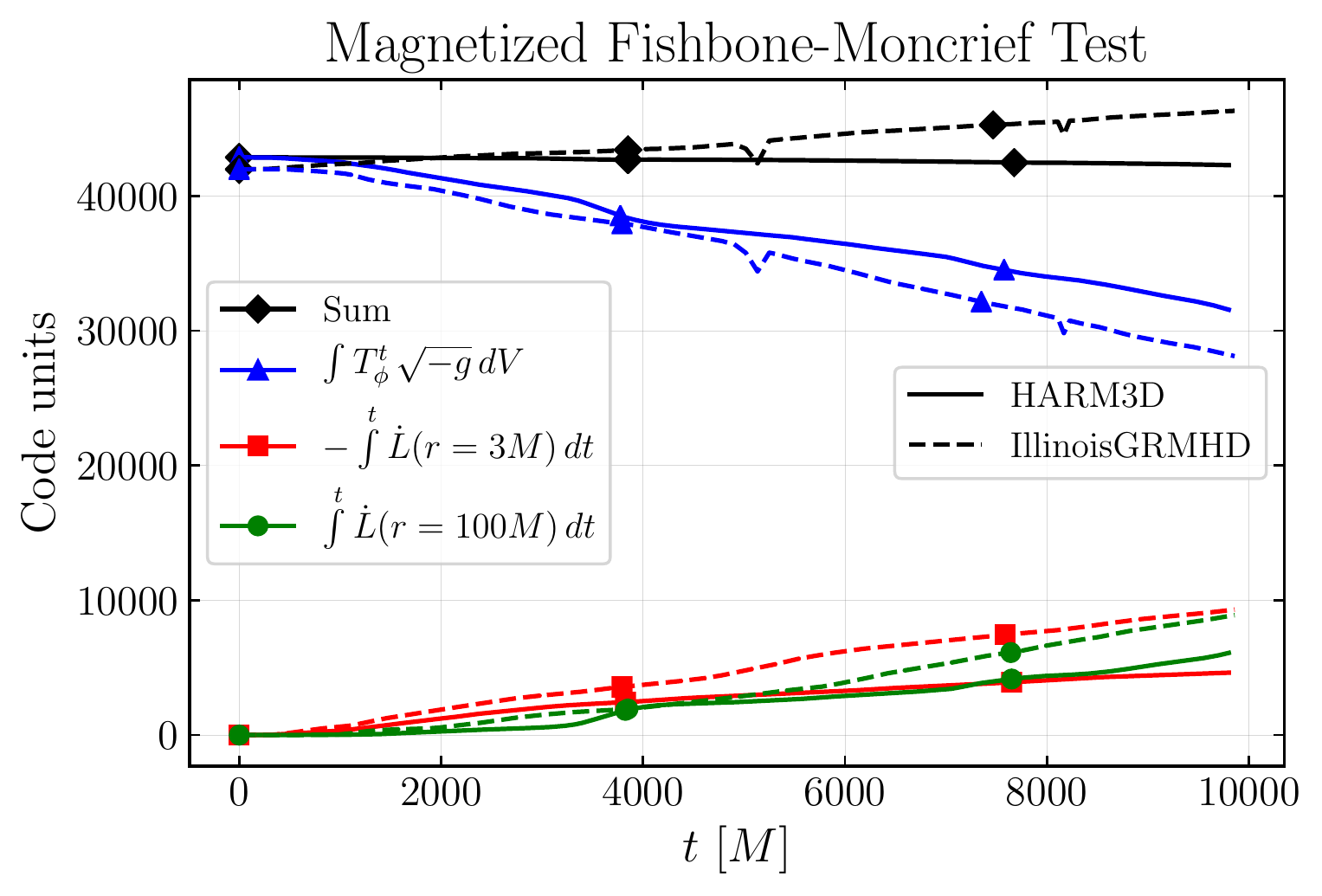}
      \caption{Angular momentum conservation as a function
      of time for a magnetized 
      FM test in \harm (\textit{solid, black, diamond}) 
      and \igm (\textit{dashed, black, diamond})-- the other
      curves represent the different terms in Eq.~\eqref{eq:angmomcons}.
      While \harm manages to conserve angular momentum by $1\%$
      with a spherical grid,
      Cartesian coordinates with block-structured AMR in \igm
      conserves angular momentum within $\%10$.}
      \label{fig:l_t}
    \end{figure}
    We showed that the effects of
    Cartesian numerical dissipation are drastic in the case of an
    orbiting torus in hydrostatic equilibrium. 
    The generalization of this result to more general turbulent MHD
    flows, however, has its caveats.
    Indeed, in Sec.~\ref{sec:validation} we demonstrated 
    that Cartesian coordinates with AMR can
    give qualitatively equivalent results than spherical coordinates for the
    case of a magnetized torus---although still inefficiently
    since the number of Cartesian cells was more than 80 
    times larger than of its spherical counterpart.
    Magnetic-induced turbulence 
    break the laminarity of the azimuthal flow and lessen the numerical
    convenience of using a spherical grid.
    But still, a more in-depth analysis will show 
    the superiority of spherical coordinates in modeling azimuthal flows.
    
    In the magnetized FM evolved in Sec.~\ref{sec:validation},
    given that the spacetime is isometric along the
    vector field $\partial_{\phi}$, the angular momentum of
    the plasma $\int T^t_{\mu} \partial^{\mu}_{\phi} \sqrt{-g} \, dV$ 
    should be conserved.
    However, in these simulations, we lfose track of
    some angular momenta, as parcels of the plasma are 
    accreted by the BH or expelled out of the outer boundary.
    We expect the following quantity to be conserved
    instead:
    \begin{equation}
    \label{eq:angmomcons}
    \begin{split}
        \Delta L(t) &= \int_{3M < r < 100M} T^t_{\phi} \sqrt{-g} \, dV + \\
        &+ \int^t \dot{L}(r=3M) \, dt + \int^t \dot{L}(r=100M) \, dt,
    \end{split}
    \end{equation}
    where the first term contains the angular momentum in the region
    $3M < r < 100M$, and the second and third terms track the loss of angular
    through the boundaries $r=3M$ and $r=100M$, 
    being $\dot{L}(r)$ the radial angular momentum flux:
    \begin{equation}
        \dot{L}(r) = \int -T^r_\phi \sqrt{-g} \, d\theta \, d\phi.
    \end{equation}
    In Fig.~\ref{fig:l_t} we plot these terms and their sum for both codes,
    \harm and \igm. 
    We find that \harm conserves angular momentum by $1\%$ but 
    \igm by $10\%$, demonstrating the higher fidelity of
    spherical coordinates in capturing the angular momentum
    conservation and transport.
    
    \subsection{Computational performance}    
    There are many reasons why using \harm instead
    of \igm for the postmerger evolution is
    computationally convenient:
    Flexible coordinate systems with spherical topology 
    can sample the domain optimally and,
    for instance, do not over-resolve
    the angular coordinates at large distances like
    AMR-structured Cartesian coordinates. 
    The absence of AMR boundaries
    in \harm reduces the number of buffer regions, improving
    memory usage and scaling.
    Freezing the spacetime evolution after  
    spacetime equilibration 
    significantly lessen the computational workload. 
    For azimuthal flows, fewer numerical cells are 
    required in spherical coordinates to capture the relevant physical processes (a factor of $\sim{}80$ for the FM test in Sec.~\ref{sec:validation}).
    The infrastructure around \harm is highly optimized
    and lightened for accretion disk simulations.
    
    To demonstrate the computational efficiency of \harm,
    we compare the performance of 
    \igm when evolving the spacetime and 
    matter fields of a stable neutron star, with the
    performance of \harm when evolving
    a magnetized torus in a static spacetime for a rotating
    BH.
    We submit both tests in a unigrid with $64^3$ cells, 
    and make use of a single computational process.
    We find that \igm performs $37,000$ cell updates per second per processor and \harm outperforms this by more than $\times 10$, performing
    $415,000$ cell updates per second per processor \footnote{Since \igm uses RK4 for
    time-integration and \harm uses RK2, we consider
    the number of cell updates per timestep
    to be the number of cells $\times 4$ and $\times 2$, respectively.}.
    
\section{Summary and conclusions}
\label{sec:conclusions}

    We presented the tools to transition a
    GRMHD simulation between two
    numerical codes, \igm and \harm,
    that make use of numerical grids
    with different resolutions and topologies.
    Moreover, we presented techniques
    to extrapolate the spacetime metric
    and MHD primitives to arbitrarily large
    radii, allowing us to extend the numerical
    domain of the destination grid.
    This set of methods, enclosed under
    the name of the \handoff, 
    are particularly interesting for 
    transitioning the outcomes
    of BNS or BH-NS mergers to a grid
    adapted to the geometry
    and requirements 
    of the postmerger.
    In Appendix~\ref{sec:grid} we give the
    implementation details of this grid.
    
    We validated the \handoff with the well-known
    case of a magnetized torus around a rotating
    BH. We applied the \handoff at different
    stages of the accretion process and
    confirmed the package successfully 
    captures the state of the plasma and
    spacetime metric and transforms them
    to a general grid.

    We applied the \handoff to a BNS postmerger,
    after the BH had formed.
    We modeled matter as a magnetized ideal fluid
    with adiabatic index $\Gamma = 2$.
    To test the reliability of the \handoff,
    we applied it at different times and
    to different destination grids.
    Our fiducial simulation, \BNSLRto, 
    starts from ID provided by the \handoff
    at $t-t_{\mathrm{BH}}=0.015 \mathrm{s}$, 
    makes use of the large grid described
    in Appendix~\ref{sec:grid}, and lasts for
    ${\sim}~0.1\mathrm{s}$.
    Since the destination grid is larger than the initial
    grid for the merger, 
    we needed to extrapolate
    the MHD primitives and spacetime metric to
    the complementary cells.
    After a careful analysis, we demonstrated
    our results from the \handoff are
    independent of the time of transition,
    and of the destination grid.
    Furthermore, we showed these results
    are in agreement with reference simulations
    in the literature.
    
    In Sec.~\ref{sec:dissipation} we demonstrated that the effects
    of Cartesian numerical dissipation are drastic 
    for an orbiting torus in hydrostatic equilibrium, and moderate
    for a highly magnetized and turbulent torus.
    This implies that long-term simulations of accretion disks
    in BNS postmergers with block-structured AMR can be spoiled 
    by Cartesian numerical dissipation, particularly in the absence
    of magnetic fields or even for realistic magnetic
    fields in the initial stars ($\sim 10^{12} \mathrm{G}$) that
    lead to mildly magnetized postmerger disks.
    For this reason, and the significant computational convenience
    of spherical grids to parameterize large computational domains,
    we consider that mapping the postmerger to a spherical
    grid is most convenient.
    
    We conclude the \handoff enables us to perform long-term,
    highly accurate and coordinate-optimized simulations
    of BNS postmergers, generating ID from the end of a 
    BNS merger simulation performed in full numerical relativity.
    Future work will focus on improving
    our models for the matter fields during
    the merger,
    adopting finite-temperature
    and tabulated EOS and taking
    neutrino effects 
    into account. Next, we will extend the \handoff
    to transition these realistic simulations
    to the modern code \harmnuc \citep{MurguiaBerthier+2021b}.

\begin{acknowledgments}

F.L.A. would like to thank Luciano Combi for helpful
comments and discussion.
We would also like to thank the referee and editor
for suggestions and comments that helped improve this manuscript.
This work was primarily funded through NASA Award No. TCAN-80NSSC18K1488, which provides support to all authors.
Additionally, F.L.A., M.C., L.E. L.J., and Y.Z. thank the NSF for support on Grants No. PHY-2110338, No. AST-2009330, No. OAC-2031744 and No. OAC-2004044.
Z.B.E. and L.R.W. gratefully acknowledge support from NSF Awards No.  
PHY-1806596, No. PHY-2110352, No. OAC-2004311, as well as NASA
Award No. ISFM-80NSSC18K0538. R.O’S. is supported by
NSF Awards No. PHY-2012057, No. PHY-1912632 and 
No. AST-1909534. B.J.K. is supported by NASA under Award No. 80GSFC21M0002. A.M-B is supported by the
UCMEXUS-CONACYT Doctoral Fellowship, and NASA
through the NASA Hubble Fellowship Grant No.  
HST-HF2-51487.001-A awarded by the Space Telescope Science Institute, which is operated by the Association of 
Universities for Research in Astronomy, Inc., for NASA, under
Contract No. NAS5-26555. V.M. is supported by the Exascale
Computing Project (17-SC-20-SC), a collaborative effort
of the U.S. Department of Energy (DOE) Office of 
Science and the National Nuclear Security Administration.
Work at Oak Ridge National Laboratory is supported
under Contract No. DE-AC05-00OR22725 with the U.S. 
Department of Energy.
Computational resources were provided by the TACC’s 
Frontera supercomputer Allocations No. PHY-20010 and 
No. AST-20021. Additional resources were provided by 
the RIT's BlueSky and GreenPrairies and Lagoon Clusters 
acquired with NSF Grants No. PHY-2018420,
PHY-0722703, PHY-1229173, and PHY-1726215.

\end{acknowledgments}

\appendix

\section{Design of postmerger grid}
\label{sec:grid}
\newcommand{\rin}{r_\mathrm{in}}
\newcommand{\rout}{r_\mathrm{out}}
\newcommand{\xx    }[1]{x^{\tiny \left(#1\right)}}
\newcommand{\QQ    }[1]{Q^{\left(#1\right)}}
\newcommand{\QQa   }[1]{\langle \QQ{#1} \rangle_{\rho}}
\newcommand{\xone  }{\xx{1}}
\newcommand{\xtwo  }{\xx{2}}
\newcommand{\xthree}{\xx{3}}
\newcommand{\rhe}{r_\mathrm{he}}
\newcommand{\xhe}{x_\mathrm{he}}
\newcommand{\nhe}{{n}}
\newcommand{\mbh}{M_\mathrm{BH}}
\newcommand{\ain}{a_{2 \mathrm{1}}}
\newcommand{\aout}{a_{2 \mathrm{2}}}
\newcommand{\Tau}{\mathcal{T}}

The  distorted spherical grid we often use in \harm, one that 
uses a uniform spacing in $\log(r)$ and is concentrated in the poloidal direction near the equator, is not particularly well-suited 
for long-term evolutions of ejecta launched closer to the equator \emph{and} relatively narrow jets along the poles. In order to resolve the jet at large distances and keep the number of poloidal cells fixed, we must increase  $\Delta\theta$ in the equator and decrease it near the poles as $r$ grows.  Using larger $\Delta \theta$ 
at the smallest radii also allows us to use larger time steps, as they are set by the smallest cell-crossing time of the fastest MHD wave in the domain, which is usually determined by the innermost azimuthal extent at the poles, $\simeq \rin \left(\Delta\theta/2\right) \, \Delta\phi$, where $\rin$ is the innermost radial coordinate on the grid.   This means that our conventional grid is adequate at smaller radii, but 
we must transition to one that is more focused along the poles at larger radii.  

Since the goal is to capture the ejected material out to $\mathcal{O}(1\mathrm{s})$ of time or $\mathcal{O}(1~\mathrm{light-second}) \sim \mathcal{O}(10^5 G \mbh / c^2)$ of distance, 
our typical $\log(r)$ grid would require so many grid points it would be computationally prohibitive.  Since the ejected material is expected to be nearly ballistic beyond $r\simeq 10^3 \mbh$, increasing $\Delta r(r)$ in a hyperexponential way provides an effective solution at covering large distances with fewer cells: 
\begin{equation}
\begin{split}
r(\xone)  =& r_0 + \left( \rin - r_0 \right) \times \\ &\exp{\left[ c_1 \left(\xone/\xhe\right)
+ c_2 \left(\xone/\xhe\right)^{\nhe} \right]} \quad ,  
\end{split}
\label{hyper-exp-radial-grid}  
\end{equation}
where the coefficients in the exponent ensure that $r(0) = \rin$, $r(\xhe) = \rhe$, $r(1)=\rout$ ($\rout$ is the outermost radial coordinate):
\begin{equation}
c_1 = \frac{d_1 - d_2 \, \xhe^{\nhe}}{1-\xhe^{\nhe-1}} \quad , \quad 
c_2 = \frac{d_2 \, \xhe^{\nhe} - d_1 \, \xhe^{\nhe-1}}{1-\xhe^{\nhe-1}} \quad , \label{c1-and-c2}
\end{equation}
and
\begin{equation}
d_1 = \ln{\left( \frac{\rhe - \rin }{\rin - r_0} \right) } \quad , \quad 
d_2 = \ln{\left( \frac{\rout - \rin }{\rin - r_0} \right) }
\quad . \label{d1-and-d2} 
\end{equation}
For the run \BNSLRto we use $n=10$, $r_0=0$, $\rin=1.31~\Msun$ 
, $\rhe=10^4~\Msun$, $\rout=7.2\times 10^4\Msun$, $d\xone = 1/1024$, and $0 \le \xone \le 1$. 

The poloidal discretization joins two regions: an inner 
one with cells focused near the equator and an outer region 
with more cells near the poles.  The two regions are smoothly 
and continuously connected through use of a transition function, similar to 
the strategy used in our ``dual fisheye" grids \citep{Zilhao+2014}. The poloidal coordinate in terms of the numerical coordinate $\xtwo$ is: 
\begin{equation}
\begin{split}
    \theta(\xtwo) = &\pi \left\{
    \xtwo - a_2 \left[ 
    \Tau(\xtwo) - \Tau(1/2) - \right.\right. \\ 
    &\left.\left.- \left(\xtwo - 1/2\right) \left( \Tau(1) - \Tau(0) \right)
    \right]
    \right\} \quad , \label{th-of-x2}
\end{split}
\end{equation}
where $\Tau(x)$ is the integral of the approximate boxcar function from 
\cite{Zilhao+2014}:
\begin{equation}
    \Tau(\xtwo) \equiv \frac{1}{2} \left[ \Sigma(\xtwo - 1/2 + \delta_2) - \Sigma(\xtwo - 1/2 - \delta_2) \right]  . \label{Tau-func}
\end{equation}
and
\begin{equation}
    \Sigma(x) \equiv \frac{1}{h_2} \ln \cosh \left( h_2 \ x  \right) \quad . \label{Sigma-func}
\end{equation}
Here, $h_2$ controls the steepness of the transition, and $\delta_2$ controls the fraction of cells that are in the equatorial portion of the poloidal grid. We use $h_2=20$, $\delta_2=0.3$, $d\xtwo = 1/200$, and $0 \le \xtwo \le 1$.

The transition between the inner and outer regions is handled by 
changing the amplitude of the grid distortion, $a_2$, with respect to $r$:
\begin{equation}
    a_2 = a_2(r) \equiv \ain f(r, r_j, h_j) + \left[1 - f(r, r_j, h_j) \right] \aout \quad , \label{a2-transition}
\end{equation}
where the transition function is 
\begin{equation}
    f(r, r_j, h_j) \equiv \frac{1}{2} \left\{ 1 + \tanh{\left[h_j \left(r - r_j \right) \right]} \right\} \quad . \label{transition-function}
\end{equation}
 The two amplitudes, $\ain$ and $\aout$, are set by different criteria.  The amplitude for the inner portion is set so that the spacing near the equator is such that the number of cells per vertical disk scale height, $N_{H/r}$, is equal to the 32 cells recommended to adequately resolve the MRI \citep{Sorathia+2012}:
\begin{equation}
    \ain = \frac{1 - N_2\left(H/r\right)/\left(\pi N_{H/r}\right)  }{1 - 2 \delta_2} \quad , \label{ain-def}
\end{equation}
where $N_2$ is the number of cell extents in the $\xtwo$ direction.
The outer amplitude is set so that the poloidal spacing near the poles follow the parabolic flow contours often found in the jets of GRMHD simulations \citep{Nakamura+2018}: 
$z=R^{n_1}$, where $R\equiv r \sin{\theta}$ is the cylindrical radius, and $1 < n_1 \lesssim 2.67$ for general outflows.  Using the small angle approximation, 
$\theta \simeq r^{n_j}$, where $n_j \equiv \left( 1/n_1 - 1 \right)$.  The funnel wall shape arising in GRMHD simulations with spinning black holes closely follows the curve with $n_1 \simeq 3/2$ or $n_j \simeq -1/3$ \citep{Nakamura+2018}, and is the value used here.  The amplitude, $\aout$, must vary with radius such that $d\theta_j$, the approximately $\xtwo$-independent spacing local to the jet, follows these contours at $r>r_j$, where $r_j$ is the radius at which we specify this transition occurs.  At $r=r_j$, we set $d\theta_j$ to be $90\%$ of what it would be if the grid was uniform in $\theta$. 
The amplitude is finally calculated using this $r$-dependent $d\theta_j$: 
\begin{equation}
    \aout = - \frac{1 - N_2 d\theta_j(r) / \pi }{ 2 \delta_2} \quad , \label{aout-def}
\end{equation}
where $d\theta_j(r) = \left(0.9 \pi / N_2\right) \left(r/r_j\right)^{n_j}$, 
$n_j = -1/3$, and $r_j = 300\Msun$. 

The azimuthal grid spacing for the new grid remains uniform, with $\phi\in\left[0,2\pi\right]$ and $d\phi = 2 \pi / N_3$, where $N_3$ is the number of cells in the azimuthal extent. 

In Fig.~\ref{fig:BNS_resolutions} (\textit{right}) 
we plot the cell lengths of this grid, as a function of 
radii and $\theta \in (0,\pi)$.
Regarding $\Delta r$, we notice the transition
of its slope from exponential to hyperexponential at
$r{\sim} 10^3~\mathrm{km}$.
Although $\Delta \phi$ is uniform, the cell length
$r\sin\theta\Delta\phi$ grows with $r$
because of its explicit radial dependence, and spans
different resolutions at a given radius because of  
the different values of $\theta \in (0,\pi)$.
Finally, $r \Delta \theta$ also grows with $r$
because of its explicit radial dependence, 
and spans different values at a given radius because,
as described above,
$\Delta \theta$ is not uniform but focuses on the
equator at small radii, and on the axis at larger radii.
We notice this transition at $r{\sim} 10^3~\mathrm{km}$.


\section{Boundary conditions at the axis}
\label{sec:BC}
\begin{figure}[htb!]
	\centering
	\includegraphics[width=.9\columnwidth]{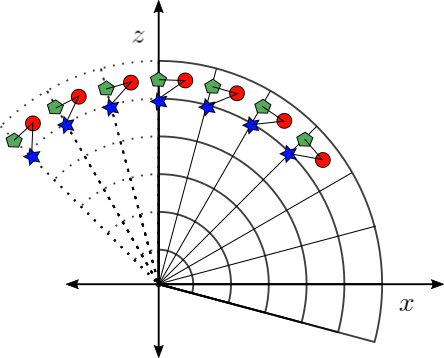}  
	\caption{Sketch of grid cells next to the positive $z$-axis. \textit{Red circles} 
	represent the positions of the cell centers, \textit{green polygons} represent the position of the lower
	$\theta$-face of each cell, and 
	\textit{blue stars} represent the lower corners.
	\textit{Dotted} cells represent ghosts cells. 
	The cell center of these ghost cells are 
	mapped to the cell center of 
	physical cells across the axis.
	The lower faces and corners of the ghost cells, however, 
	are shifted by $+\Delta \theta$ and 
	map to the upper-face of the corresponding physical cells.}
	\label{fig:BC}
\end{figure}
    Previous simulations of accretion disks in \harm
    excise a portion of the domain around the
    polar axis \citep[see, for instance,][]{Noble+2009, Noble+2012, Bowen+2019}.
    In these simulations, the polar coordinate $\theta$
    is restricted to an interval
    $(\theta_{\mathrm{c}},\pi - \theta_{\mathrm{c}}$),
    where $\theta_{\mathrm{c}}>0$ is the extent of the
    \textit{cutout} in $\theta$ from the axis.
    This is convenient for two reasons:
    First, the coordinate singularity 
    at the axis is
    excised from the domain. 
    Second, the innermost azimuthal
    extent at the poles is increased from 
    $ r_{\mathrm{in}} \sin (\Delta \theta /2) \Delta \phi$ to
    $ r_{\mathrm{in}} \sin ( \theta_{\mathrm{c}}+\Delta \theta /2) \Delta \phi$ and, since the time 
    step in spherical coordinates is restricted to the
    cell-crossing time of this extent, the time step
    is increased accordingly.
    However, cutting out the domain has its
    drawbacks: The simulation lacks a portion of the
    the funnel region and therefore of the magnetic
    outflows, and even
    the rest of the funnel might
    be affected by the boundary
    conditions imposed at the cutout, which are 
    usually reflective or outflow.

    These drawbacks are not severe in previous applications,
    where the funnel is mostly evacuated and the
    magnetic outflows expand through the numerical
    atmosphere, but that is not the case for a BNS
    postmerger.
    In the latter, the funnel is baryon polluted, with
    a complex magnetic field structure,
    and the magnetic outflows
    propagate through the dynamical ejecta creating
    a complex jet-cocoon system
    that eventually breaks-out from the debris.
    A cutout in $\theta$ 
    would prohibit the appropriate modeling of
    these phenomena.
    For this reason, we remove the cutout
    and implement appropriate boundary conditions
    at the polar axis, as we describe below.
    
    Formally, the cutout is still present to avoid the
    coordinate singularity at the axis, 
    but it is reduced
    to a small value. Specifically,
    the cell centered coordinates of the cells  
    next to the positive 
    [negative] $z$-axis have coordinates $(r,\Delta \theta/2+\theta_{\mathrm{c}},\phi)$
    [$(r,\pi - \Delta \theta/2 - \theta_{\mathrm{c}},\phi)$], with 
    $\theta_{\mathrm{c}}\sim 10^{-14}$.
    The three ghost cells in $\theta$
    before [after] these cells
    are mapped to the physical coordinates across the axis, 
    with cell centered coordinates $(r,\theta_{\mathrm{c}}+j\Delta \theta + \Delta \theta/2+\theta_{\mathrm{c}},\phi+\pi)$
    [$(r,\pi-\theta_{\mathrm{c}}-j \Delta \theta - \Delta \theta/2-\theta_{\mathrm{c}},\phi+\pi)$],
    with $j=0,1,2$.
    In Fig.~\ref{fig:BC} we sketch the cell centers
    of physical and ghost cells (\textit{dashed}) around the positive
    $z$-axis with \textit{red circles}.
    
    Although the cell centers of the ghost cells in
    $\theta$ match the cell centers of physical
    cells across the axis, that is not 
    the case for the lower corners and
    lower $\theta$-faces.
    The cell-corners and lower $\theta$-faces of
    the ghost cells are shifted by 
    $+\Delta \theta$, so they stand at
    a cell length distance from the corresponding
    cell positions at the boundary.
    In other words, the lower $\theta$-face of
    the ghost cells maps to the upper $\theta$-face 
    of the physical cells across the axis.
    In Fig.~\ref{fig:BC} we sketch the positions
    of the lower $\theta$-faces for ghosts (\textit{dashed}) 
    and physical cells with \textit{green polygons}, and the
    corners with \textit{blue stars}.
    This is important, for instance, when
    calculating finite differences
    across the axis for the spacetime 
    connections, or for the curl of
    $A_{\mu}$.
    
    Having set the coordinates of the ghost 
    cells in $\theta$, we evaluate
    the spacetime metric at these cells, and multiply
    its components $g_{\mu \nu}$ by parity factors that
    ensure continuity across the axis.
    In particular, we follow Refs. \cite{Mewes+2020} 
    (see Table 1 of that reference) and change the sign
    of $g_{t\theta}, g_{r\theta}, g_{\phi\theta}$
    and their symmetric permutations.
    The metric component $g_{\phi \phi}$ tends to
    zero towards the axis, so we enforce
    $g_{\phi \phi}=1\e{-14}$ if this component takes
    a smaller value at the $\theta$-face
    and corners of the physical cells next to
    the axis.
    We fill the ghost cells
    with MHD primitives from the matching 
    physical cells, and take parity factors into account
    by changing the sign of the tensor components 
    $B^{\theta}, v^{\theta}$ \citep{Mewes+2020}.
    We use the geometry and primitives in
    these ghost regions as boundary conditions 
    to evolve the conserved
    variables in the physical cells
    next to the axis. Subsequently, the MHD primitives
    in the ghost cells are updated after each time step 
    with the evolved values at the matching physical cells,
    and multiplied by the corresponding 
    parity factors.
    
    The evolution of the magnetic field requires further care.
    In the \texttt{FluxCT} algorithm \citep{Toth2000}, 
    the electric field $\mathbf{E}$
    is calculated at the cell-edges, and the extent or definition
    of those edges is singular
    for cells next to the axis.
    As first described by Refs. \cite{Liska+2017} 
    (see the supplemental material of that reference), 
    we find that a nonzero value of $E_{\phi}$ at the axis 
    gives raise to an artificial growth of the magnetic
    flux in the radial direction.
    We apply the same techniques of Ref. \cite{Liska+2017} to
    circumvent this problem.
    Since the extent of the cell-edge in $\phi$ is zero
    at the axis, we set the electric field component $E_{\phi}$ 
    to zero in those cells. Further, since the cell-edge
    in $r$ is the same for every cell around the axis at a
    given height $z$, we set $E_r$ to a unique value
    in these cells.
    This value results from the $\phi$-average of $E_r$
    around the axis at height $z$.

\section{Resolution Diagnostics}
\label{sec:resolution}

    In this appendix we calculate resolution diagnostics to
    demonstrate the convergence of the MRI-driven turbulence
    in the postmerger disk.
    We follow the analyses of Refs. \cite{Hawley+2011, Noble+2012, Hawley+2013, Kiuchi+2018}.

\begin{figure*}[htb!]
        \centering
        \includegraphics[width=\columnwidth]{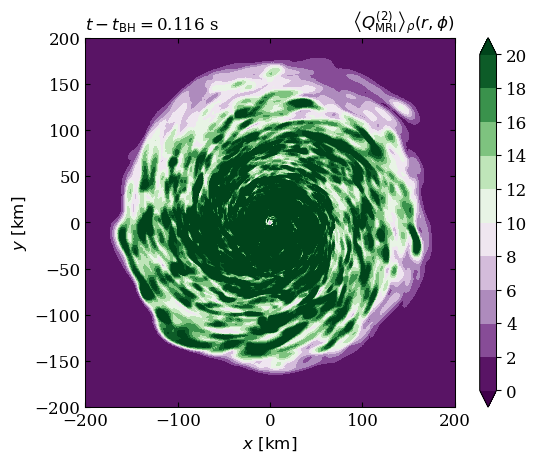}
        \includegraphics[width=\columnwidth]{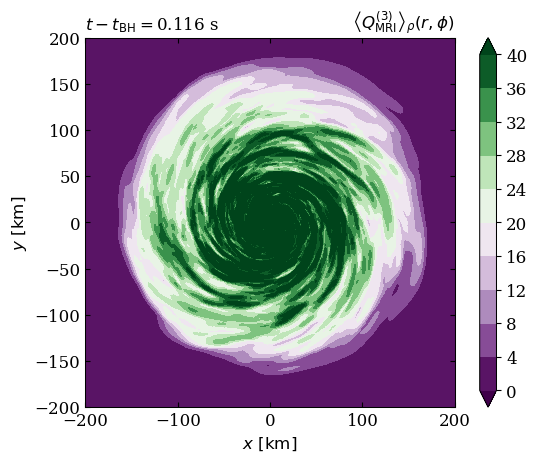}
	    \caption{Quality factors of MRI resolution, weighted by $\rho$ and averaged in $\theta$. 
	    We emphasize they satisfy $\left \langle Q^{(2)} \right\rangle_{\rho} > 10$ (\textit{left}) and 
	    $\left \langle Q^{(3)} \right \rangle_{\rho} > 20$ (\textit{right}), 
	    adequately resolving
	    the MRI wavelength.}
	    \label{fig:Qmri}
\end{figure*}
    We start by calculating the
    number of cells within the 
    linear MRI wavelength, 
    or quality factors:
\begin{equation}
    Q^{(i)} = \frac{2 \pi \left| b^{(i)} \right|}{\Delta x^{(i)} \Omega_{\mathrm{K}}(r)  \sqrt{\rho h + 2 p_{\mathrm{m}}}}.
\end{equation}
    We  multiply the latter 
    by $\rho$ to focus on the disk,
    and average over the polar
    direction:
\begin{equation}
    \left \langle Q^{(i)} \right \rangle_{\rho}(r,\phi) = \frac{\int_0^1 Q^{(i)} \rho \sqrt{g'} \, dx^{(2)} }{\int_0^1 \rho \sqrt{g'} \, dx^{(2)}}.
\end{equation}
    In Fig.~\ref{fig:Qmri} we plot
    these quantities at the end of 
    \BNSLRto, and notice
    they satisfy 
    $\left \langle Q^{(2)} \right\rangle_{\rho} > 10$ and 
    $\left \langle Q^{(3)} \right \rangle_{\rho} > 20$, the conditions for converged MRI behavior from resolution studies.

\begin{figure*}[htb!]
  \centering
  \includegraphics[width=\columnwidth]{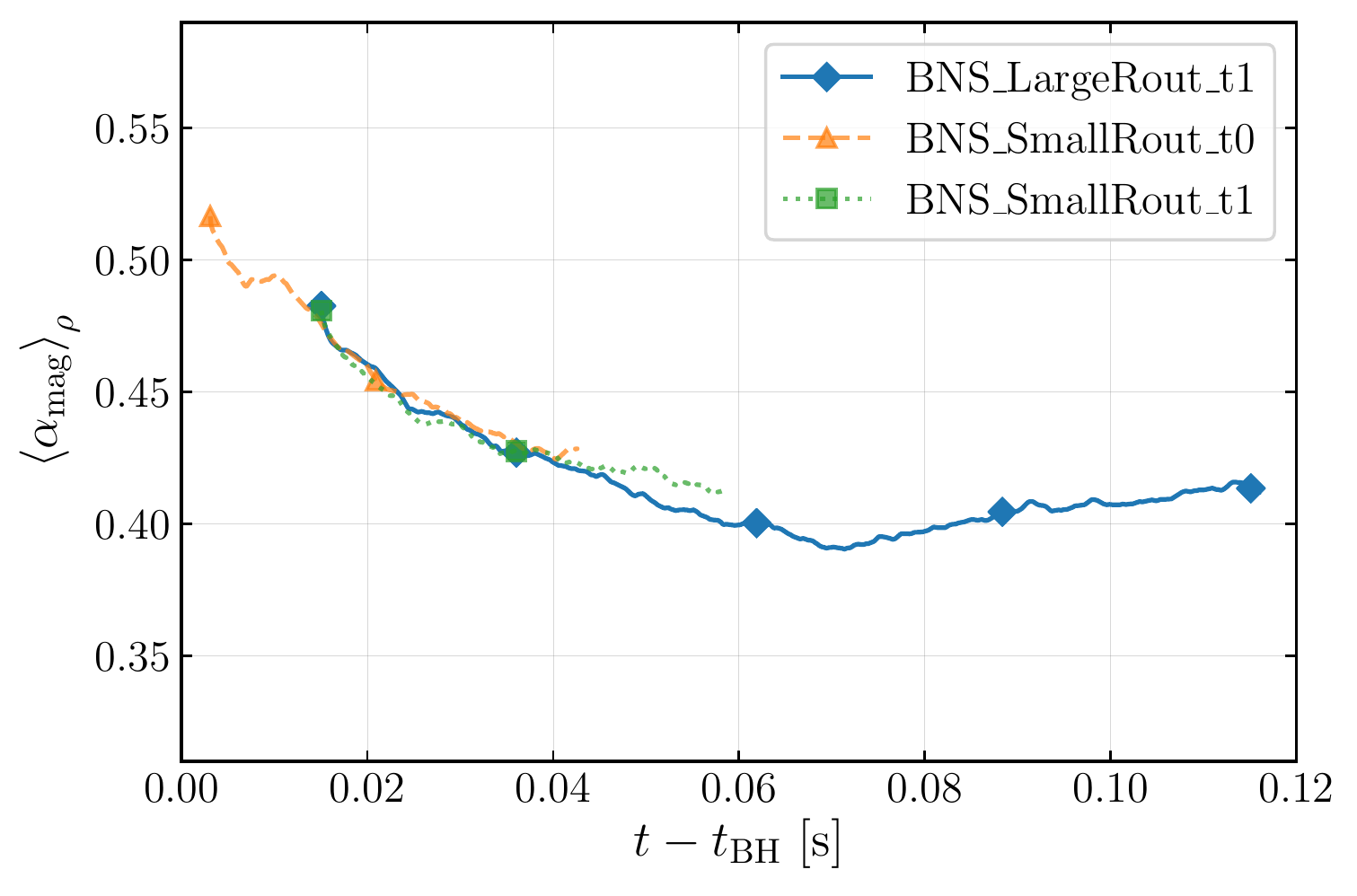}
  \includegraphics[width=\columnwidth]{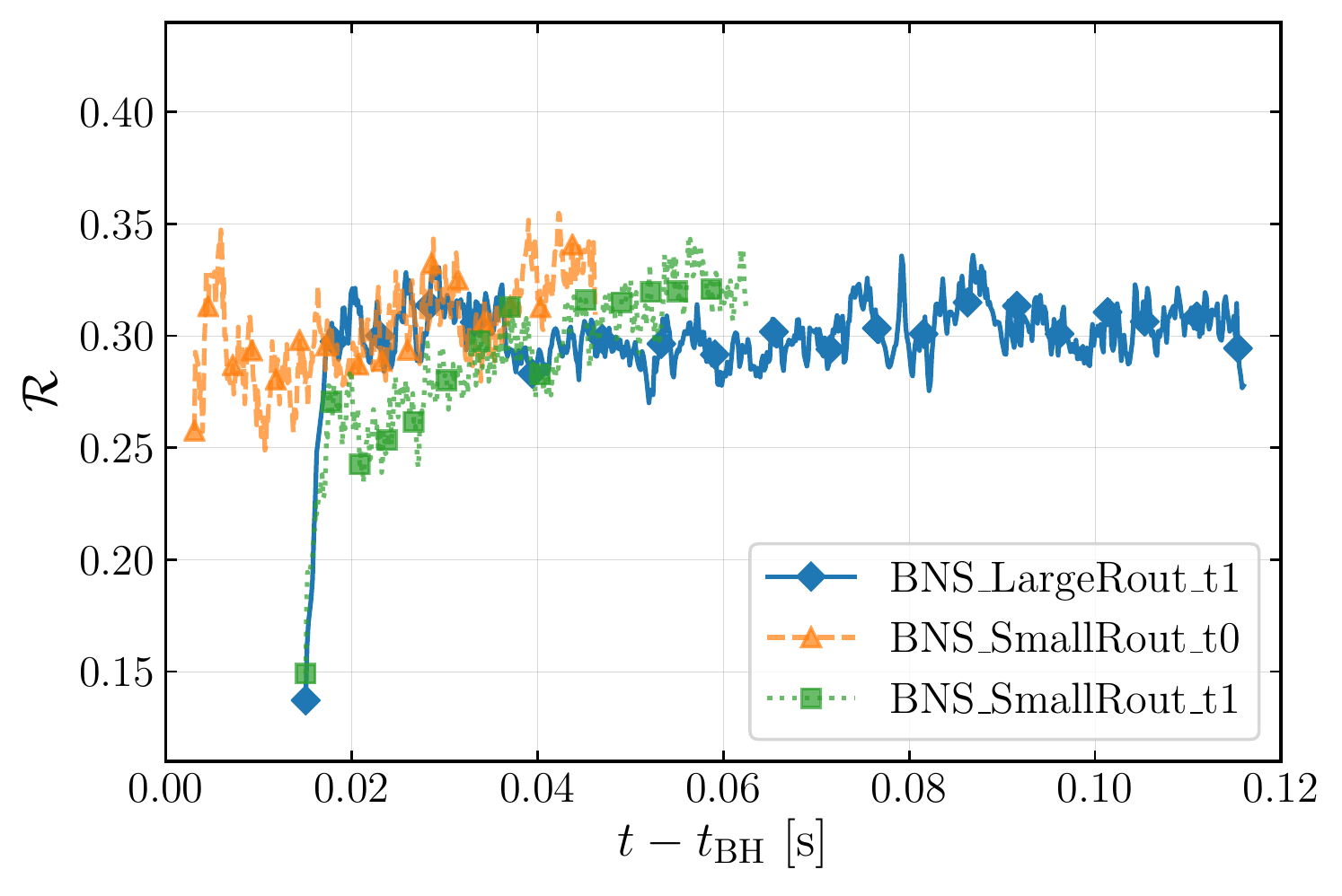}
  \caption{Nonlinear diagnostics of MRI resolution
  $\alpha_{\mathrm{mag}}$ (\textit{left}) and
  $\mathcal{R}$ (\textit{right}) (see Eqs.~\ref{eq:alphamag} and \ref{eq:R}).
  They converge to $\sim 0.4$ and $\sim 0.3$, respectively, 
  implying proper resolution of the MRI modes \citep{Hawley+2011, Hawley+2013}.}
  \label{fig:alpha_R}
\end{figure*}
    We also calculate nonlinear diagnostics:
    The weighted ratio of the Maxwell stress to 
    the magnetic pressure
\begin{equation}
    \label{eq:alphamag}
    \left< \alpha_{\mathrm{mag}} \right>_{\rho} = \frac{ \int \alpha_{\mathrm{mag}} \rho \sqrt{g} \, dV}{\int \rho \sqrt{g} \, dV }, \ \
    \alpha_{\mathrm{mag}} = \frac{ 2 \sqrt{ b_rb^r} \sqrt{b_{\phi}b^{\phi} }}{b^2},
\end{equation}
    and the weighted ratio of energies between these components
\begin{equation}
    \label{eq:R}
    \mathcal{R} = \frac{ \int B_r B^r \rho \sqrt{-g} \, dV}{ \int B_{\phi} B^{\phi} \rho \sqrt{-g} \, dV}.
\end{equation}
In Fig.~\ref{fig:alpha_R} we plot these quantities
as a function of time.
We find $\alpha_{\mathrm{mag}}$ becomes approximately constant, as expected from the correlation induced
 on $b^r$ and $b^{\phi}$ by the MRI, and converges to $\sim 0.4$, the saturation value of shearing-box
simulations of Ref. \cite{Hawley+2011}. 
On the other hand, we find $\mathcal{R}\sim 0.3 > 0.2$,
satisfying the criteria of Ref. \cite{Hawley+2013}.
From this plot we also notice that
the later transitions, \BNSSRto and
\BNSLRto, are initialized to a smaller value
than \BNSSRtz at such time, and then they  
rapidly catch up with the convergent value. 
This suggests the MRI was not properly resolved in \igm, proving the need for
the transition to a higher-resolution grid.
We conclude the MRI is well resolved in the run \BNSLRto that makes
use of the grid described in Appendix~\ref{sec:grid}.


\section{Magnetic field decomposition}
\label{sec:bfield_decomp}
Along this article, we decomposed the magnetic field in the direction
of the fluid velocity $\tilde{v}^i$ and its orthogonal frame, and
interpreted these as toroidal and poloidal components of the field,
respectively 
(see Eqs.~\ref{eq:Emag}-\ref{eq:Emag_perp}).
This interpretation is supported by the reduction of this decomposition
to poloidal-toroidal in the case of axisymmetric axial flows, and by
Ref.~\cite{Fanci+2013} that successfully applied this decomposition 
and interpretation to rotating NSs. In this Appendix we demonstrate
that such decomposition and interpretation are also valid for the 
case of magnetized accretion disks.

We decompose the magnetic field in terms of the
poloidal coordinate vector $\phi^i$:
\begin{equation}
    B^i = B_{\mathrm{tor}} \phi^i + B_{\mathrm{pol}}^i,
\end{equation}
where
\begin{equation}
    \phi^i = \left[ 0,0,\frac{1}{\sqrt{g_{\phi \phi}}} \right],
\end{equation}
\begin{equation}
    B_{\mathrm{tor}} = B^i \phi_i.
\end{equation}
It can be proved that the magnetic energy (Eq.~\ref{eq:Emag}) takes the form:
\begin{equation}
    E_{\mathrm{mag}} =  E^{\phi-\mathrm{tor}}_{\mathrm{mag}} +
    E^{\phi-\mathrm{pol}}_{\mathrm{mag}} + E^{\mathrm{\times}}_{\mathrm{mag}},
\end{equation}
where
\begin{equation}
    \label{eq:phiPol}
    E^{\phi-\mathrm{pol}}_{\mathrm{mag}} = \frac{1}{2} \left[ B^2_{\mathrm{pol}} \left( 1 + \tilde{v}^2 \right) - \left( B^i_{\mathrm{pol}} \tilde{v}_i \right)^2\right],
\end{equation}
\begin{equation}
    \label{eq:phiTor}
    E^{\phi-\mathrm{tor}}_{\mathrm{mag}} = \frac{1}{2} \left[ B_{\mathrm{tor}}^2 \left( 1 + \tilde{v}^2 - \frac{\tilde{v}_{\phi} \tilde{v}_{\phi}}{\left| g_{\phi \phi} \right|}\right) \right],
\end{equation}
\begin{equation}
    \label{eq:phiCross}
    E^{\mathrm{\times}}_{\mathrm{mag}} = - \left( B^i_{\mathrm{pol}} \tilde{v}_i \right) B_{\mathrm{tor}} \frac{\tilde{v}_{\phi}}{\sqrt{g_{\phi \phi}}}.
\end{equation}
As we can see from the latter expressions, 
the decomposition in terms of the coordinate vector $\phi^i$ includes a cross term
$E^{\mathrm{\times}}_{\mathrm{mag}}$ in the expansion of the magnetic energy.

Calculating the terms \eqref{eq:phiPol}-\eqref{eq:phiCross} 
for \BNSLRto and comparing them 
with the respective terms in the decomposition in terms of the fluid
velocity, we found that 
$E^{\phi-\mathrm{pol}}_{\mathrm{mag}}, E^{\phi-\mathrm{tor}}_{\mathrm{mag}}$
are of the same order of magnitude as 
$E^{\mathrm{pol}}_{\mathrm{mag}}, E^{\mathrm{tor}}_{\mathrm{mag}}$, 
respectively, and that the cross term $E^{\mathrm{\times}}_{\mathrm{mag}}$
is negligible.
This supports the interpretation of the terms 
$E^{\mathrm{pol}}_{\mathrm{mag}}, E^{\mathrm{tor}}_{\mathrm{mag}}$
as poloidal and toroidal contributions to the magnetic energy
in the case of accretion disks.
This interpretation, however, might need to be revised in the presence of
a relativistic jet, where fluid lines align with
poloidal magnetic field lines at the funnel.

\bibliography{apssamp} 

\end{document}